\documentclass[12pt]{article}
\pdfoutput=1

\usepackage{jheppub}
\usepackage{amsmath,amssymb,euscript,array,mathrsfs}
\usepackage{slashed}

\usepackage{epsfig}

\usepackage{tikz}
\usetikzlibrary{matrix,arrows,positioning,shapes,snakes,fadings,decorations.pathmorphing,
decorations.pathreplacing,automata,shadows,decorations.markings,patterns}
\usetikzlibrary{calc,shapes.callouts,shapes.arrows}

\def\a{\alpha}
\def\b{\beta}
\def\c{\gamma}
\def\d{\delta}
\def\e{\epsilon}

\def\l{\lambda}
\def\m{\mu}
\def\n{\nu}

\def\r{\rho}
\def\s{\sigma}

\def\w{\omega}

\def\D{\Delta}

\def\S{\Sigma}

\def\tr{{\rm tr}}

\def\RR{{\mathfrak R}}

\def\Dbarslash{\,\,{\raise.15ex\hbox{/}\mkern-12mu {\bar D}}}
\def\Dslash{\,\,{\raise.15ex\hbox{/}\mkern-12mu D}}
\def\delslash{\,\,{\raise.15ex\hbox{/}\mkern-9mu \partial}}
\def\delbarslash{\,\,{\raise.15ex\hbox{/}\mkern-9mu {\bar\partial}}}

\def\RR{{\mathfrak R}}

\def\rta{\rightarrow}
\def\tr{{\rm tr}}

\def\Im{{\rm Im}}
\def\Re{{\rm Re}}

\title{\begin{center}Dynamical Evolution of Gravitational Leptogenesis\end{center}}

\author{Jamie I.~McDonald$^{1,\dagger}$ and Graham M.~Shore$^{2,*}$}
\emailAdd{$^\dagger$jamie.mcdonald@tum.de}
\emailAdd{$^*$g.m.shore@swansea.ac.uk}

\affiliation{\vskip0.2cm
$^1$Technische Universit\"at M\"unchen, Physik-Department, James-Franck-Stra{\ss}e, 
85748 Garching, Germany. 
\vskip0.1cm
$^2$Department of Physics, College of Science, Swansea University, Singleton Park, Swansea, SA2 8PP, UK.}

\date{\today}

\abstract{Radiatively-induced gravitational leptogenesis is a potential mechanism to 
explain the observed matter-antimatter asymmetry of the universe. 
Gravitational tidal effects at the quantum loop level modify the dynamics of the leptons
in curved spacetime and may be encoded in a low-energy effective action $S_{eff}$.
It has been shown in previous work how in a high-scale BSM theory 
the CP odd curvature-induced interactions in $S_{eff}$ modify the dispersion relations 
of leptons and antileptons differently in an expanding universe, giving rise to an 
effective chemical potential and a non-vanishing equilibrium lepton-antilepton asymmetry.
In this paper, the CP even curvature interactions are shown to break lepton number current
conservation and modify the evolution of the lepton number density as the universe expands. 
These effects are implemented in a generalised Boltzmann equation and used to trace
the dynamical evolution of the lepton number density in different cosmological scenarios.
The theory predicts a potentially significant gravitationally-induced lepton-antilepton asymmetry
at very early times in the evolution of the universe.
} 

\begin{document}

\maketitle
\flushbottom

\setlength{\parskip}{10pt}

\section{Introduction}\label{sect 1}

The origin of matter-antimatter asymmetry in the universe is one of the most important 
outstanding issues in cosmology. Radiatively-induced gravitational leptogenesis (RIGL)
\cite{McDonald:2015ooa, McDonald:2015iwt, McDonald:2016ehm}
is a particularly elegant and economical mechanism to generate a lepton number asymmetry
in the early universe with the potential to explain the presently
observed baryon-to-photon ratio $\eta = (n_B - n_{\bar{B}})/n_\c  \simeq 6 \times 10^{-10}$.

The fundamental idea is that tidal gravitational effects at the quantum loop level can induce 
an asymmetry in the propagation of leptons and antileptons. This may be interpreted as the
generation of an effective chemical potential for lepton number, which in the high temperature
environment of the early universe produces a non-vanishing lepton number asymmetry
in (quasi-)equilibrium.

The three basic conditions for the generation of a baryon or lepton number asymmetry have
been known since the early work of Sakharov \cite{Sakharov:1967dj}, {\it viz.} (i) a source of 
$B$ or $L$ violation; (ii) C and CP non-conservation;  
(iii) out-of-equilibrium $B$ or $L$ violating reactions. 
Gravitational leptogenesis \cite{Davoudiasl:2004gf}  circumvents the third condition by exploiting 
the time dependence of the gravitational background in the expanding early universe,
allowing a lepton asymmetry to be induced essentially in equilibrium.
This may subsequently be converted to a baryon asymmetry at lower temperatures through 
electroweak sphaleron processes \cite{Khlebnikov:1988sr}. Beyond the Standard Model (BSM) 
physics remains necessary in this scenario to satisfy the first two criteria.

The RIGL mechanism is most easily understood in terms of an effective action for the 
light neutrinos $\n_L$ in curved spacetime 
\cite{McDonald:2014yfg}. 
This extends the original effective action for tidal gravitational effects in QED
\cite{Drummond:1979pp} to the neutrino sector \cite{Ohkuwa:1980jx},
To leading order in the gravitational field strength,
this may be written as
\begin{align}
S_{eff}= \int d^4 x\sqrt{-g} \biggl[ 
\frac{1}{2} i\,\overline{\n_L} \c.\overleftrightarrow{D}\, \n_L
&\,+ \,\tilde{a}\, R_{\m\n} \, i \,\overline{\n_L} \c^\m \overleftrightarrow{D}^\n \,\n_L
\,+\, b \,\partial_\m R \,\,\overline{\n_L} \c^\m \,\n_L \nonumber \\
&~~~\,+ c \, R \, i \,\overline{\n_L} \c.\overleftrightarrow{D} \,\n_L
\,+\, \tilde{d}\, i\left(D_\m \overline{\n_L}\right) \c.\overleftrightarrow{D} \, 
D^\m \n_L \biggr] \ ,
\label{a1}
\end{align}
suppressing flavour labels.  The direct curvature couplings in this Lagrangian effectively 
violate the strong equivalence principle, allowing gravity to distinguish between matter 
and antimatter. They are generated perturbatively at one and two-loop level in the standard
model or, as required for realistic leptogenesis, a BSM theory characterised by a 
high mass scale $M$ and exhibiting CP non-conservation, and can be understood 
qualitatively as the result of tidal gravitational forces acting on the extended cloud of 
virtual particles in the quantum loops. The effective action (\ref{c1})
is a faithful description of the dynamics for weak gravitational fields $\RR/M^2 < 1$
(where $\RR$ denotes a typical curvature) and for sufficiently low energies, as we specify 
carefully later.

The physics implications of these new gravitational interactions is quite different
depending on whether the corresponding operators are CP even or CP odd.
The only CP odd operator is the one with coefficient $b$ and this is responsible for 
generating a non-vanishing equilibrium lepton number density $n_L^{eq}$.
Noting that $J^\m = \overline{\n_L}\c^\m \n_L$ is the lepton (neutrino) number current,
its $\m=0$ component is the corresponding charge and the operator in (\ref{a1})
may be understood as the introduction of an associated chemical potential
$\m = b \dot{R}$, which is non-vanishing in a time-varying gravitational field.
At finite temperature, this gives rise to an asymmetry 
$n_L^{eq} = \tfrac{1}{3} b \,\dot{R}\, T^2$.
The two-loop BSM calculation of this operator, its interpretation in a modified Boltzmann
equation, and its implications for cosmology have been described in detail in
\cite{McDonald:2015ooa, McDonald:2015iwt, McDonald:2016ehm}.

The new feature in this paper is an analysis of the role of the CP even operators
with coefficients $\tilde{a}$, $c$ and $\tilde{d}$ in driving the evolution of the lepton
asymmetry at very early times. One way to understand these effects 
is to note\footnote{Non-conservation of the lepton number current in a Lagrangian with an
arbitrarily included operator of type $R\, \overline{\n_L} \c.D \n_L$ was observed in an
interesting recent paper \cite{Antunes:2019phe}, which was an important motivation for the present
work. The application to leptogenesis proposed in \cite{Antunes:2019phe} is however very different
from the mechanism developed here.} 
that the effective action (\ref{a1}) implies that the lepton number current is not conserved.
In fact,
\begin{equation}
D_\m J^\m \,=\, - 2a \, R_{\m\n} D^\m J^\n \, - \, 2 \hat{b} \, \partial_\m R J^\m \ ,
\label{a2}
\end{equation}
where $a = \tilde{a} - \tfrac{1}{2}\tilde{d}$ and 
$\hat{b} = \tfrac{1}{2}\tilde{a} + c - \tfrac{1}{2}\tilde{d}$.
As we show in section \ref{sect 5}, this implies the following equation for the dynamical evolution
of the lepton number density $n_L$ in a FRW universe,
\begin{equation}
\frac{dn_L}{dt} \,+\, 3H n_L \,+\, 2a \left(-3 R^0{}_0 + R^i{}_i\right) H n_L \,+\, 
2 \hat{b} \,\dot{R} \,n_L \,=\, 0 \ ,
\label{a3}
\end{equation}
where $H$ is the Hubble parameter. 
These new curvature terms crucially modify the evolution of $n_L$ at very early times.
Whether this acts to amplify or suppress the magnitude of $n_L$ depends on the signs
and relative magnitudes of the coefficients, especially $a$, which are not arbitrary in RIGL
but are determined by the fundamental BSM theory.

The combined effect of these two distinct mechanisms -- the generation of an equilibrium asymmetry 
$n_L^{eq}$ by the CP odd $b$ operator and the modified evolution of $n_L$ by the CP even
$\tilde{a},\, c,\, \tilde{d}$ operators -- in a cosmological setting is clearly expressed in terms
of a generalised Boltzmann equation. This is traditionally written for the lepton-to photon
ratio $N_L = n_L/n_\c$ expressed as a function of inverse temperature $z = M/T$.
In section 5, we derive this new Boltzmann equation for gravitational leptogenesis as,
\begin{equation}
\frac{dN_L}{dz} \,=\, -\, W(z) \, \bigl(N_L\,-\, N_L^{eq}(z)\bigr) \,-\, \mathcal{W}(z)\, N_L \ .
\label{a4}
\end{equation}
Here, $\mathcal{W}(z)$ reflects the curvature-induced evolution terms in (\ref{a3}), while
$W(z)$ is determined by the $L$-violating interactions in the BSM theory. Normally interpreted 
(see for example \cite{Buchmuller:2004nz})
as a `washout' term in leptogenesis mechanisms for which $N_L^{eq}$ is zero, in our
theory it plays the quite different role of driving the lepton asymmetry to its equilibrium value.
Putting everything together, the entire evolution predicted by the RIGL Boltzmann equation
is shown in Fig.~\ref{Fig 5.1}. Essentially we find three stages: a very early high temperature 
phase in which the new evolution term $\mathcal{W}(z)$ keeps $N_L$ below equilibrium;
followed by a phase where $W(z) > \mathcal{W}(z)$ and $N_L$ is driven to $N_L^{eq}$;
and finally decoupling when $W(z)$ becomes too weak to hold
$N_L$ to $N_L^{eq}$ ({\it i.e.}~the $L$-violating reactions are too slow compared to the 
Hubble expansion to maintain equilibrium) and it decouples leaving the final constant 
asymmetry predicted by this theory.

Clearly the transitions between these stages depend on the dynamical balance between 
the curvature and temperature dependent rates $W(z)$  and $\mathcal{W}(z)$, and 
$N_L^{eq}(z)$, as the universe expands and cools. In particular, the sign of $\mathcal{W}(z)$,
controlled by the coefficient $a$ in the effective action, is key to whether the new
evolution term amplifies or suppresses the lepton asymmetry at early times.

We emphasise that these mechanisms for the generation and evolution of leptogenesis
are very general and depend only on the SEP violating effective action (\ref{a1}).
The philosophy of RIGL is that this effective action is generated automatically, 
and necessarily, in any given BSM theory incorporating CP violation. Nothing is added 
by hand, and all the coefficients are calculable in the fundamental theory.
The overall picture of gravitational leptogenesis presented here is, however, not dependent
on any particular choice of BSM theory.

The remainder of the paper is organised as follows. For clarity, we present our general
theory using a particularly well-motivated BSM theory\cite{Fukugita:1986hr} 
in which the standard model is
augmented with heavy right-handed sterile neutrinos $\n_R$. This model provides an
explanation, via the see-saw mechanism, for the light neutrino masses. As we see in
section \ref{sect 6}, where we discuss the implications for cosmology, the parameter 
bounds set by the experimental values for the neutrino masses severely constrain the
resulting gravitational leptogenesis predictions in this model.
The BSM Lagrangian is introduced in section \ref{sect 2}, where explicit calculations of the
one-loop diagrams determining the coefficients $\tilde{a}$, $c$, $\tilde{d}$ are presented.
The two-loop coefficient $b$ was previously calculated in \cite{McDonald:2015iwt}.
These Feynman diagrams for the neutrino self-energies are matched to the effective
action in section \ref{sect 3}.

The new dynamical evolution mechanism is presented in sections \ref{sect 4} and \ref{sect 5}.
First, in section \ref{sect 4}, the wave solutions of the equation of motion derived from the
effective action are found using the eikonal approximation \cite{McDonald:2014yfg}. 
At leading order, the {\it phase}
is determined by the CP odd $b$ operator -- this modifies the dispersion relation in a 
different way for the light neutrinos and antineutrinos, giving the essential asymmetry
which ultimately leads to a non-vanishing $n_L^{eq}$ in thermal equilibrium.
The sub-leading eikonal order describes the time dependence of the wave {\it amplitude},
which is interpreted as lepton number density. This provides an independent derivation
of the evolution equation (\ref{a3}). The derivation from current non-conservation is
given in section \ref{sect 5}. Section \ref{sect 5} also contains the derivation and initial
interpretation of the gravitationally modified Boltzmann equation.

Finally, in section 6, we return to the full BSM theory and study gravitational leptogenesis
quantitatively in different cosmological scenarios 
\cite{Davoudiasl:2004gf, McDonald:2016ehm} -- first in the standard radiation-dominated 
era where FRW cosmology is well-established, then during conventional reheating where
the effective equation of state is characterised by $0<w<1/3$, and finally
in a more speculative post-inflationary era in which the usual reheating phase 
is replaced by a period in which the expansion is dominated by inflaton dynamics 
with $w > 1/3$. We track the gravitationally-induced evolution of the lepton number 
density from very early times through to decoupling and discuss the conditions and
parameter choices for which the observed value of the baryon-to-photon ratio $\eta$
may be realised.

Analytic results complementing the numerical plots of the evolution of $N_L(z)$ 
in section \ref{sect 6} are given in Appendix \ref{Appendix A}.

\section{Fundamental BSM Theory and Leptogenesis}\label{sect 2}

The requirements for a fundamental theory which could realise our mechanism of radiatively-induced 
gravitational leptogenesis include CP violation active at a high energy scale 
where the spacetime curvature is sufficiently large, and a mechanism for lepton number violation.
These are naturally incorporated in a BSM theory with three sterile neutrinos ({\it i.e.}~having 
no interactions with the SM gauge fields) with a hierarchy of large Majorana masses, 
coupling to the usual left-handed leptons via the SM Higgs field. Independently of the question 
of leptogenesis, there are compelling reasons to augment the standard model in this way, 
with the see-saw mechanism then explaining the existence of the non-zero masses of the 
light neutrinos.

\subsection{BSM Lagrangian and lepton number violation} \label{sect 2.1}

The fundamental action, including the right-handed neutrinos, is then:
\begin{equation}
S\,=\, \int d^4 x \sqrt{-g} \biggl[{\cal L}_{\rm SM} + \biggl(
\frac{1}{2} i\,\overline{\n_R}\, \c.\overleftrightarrow{D} \n_R - \frac{1}{2} \overline{\n_R^{\,\,c}}\, M\,\n_R
- \overline{\ell_L} \,\l \,\phi\, \n_R + {\rm h.c.} \biggr)\biggr] \ .
\label{b1}
\end{equation}
Here, $\n_R^\a$ \,($\a = 1,\ldots 3$) are the right-handed neutrinos, with Majorana mass matrix
$M_{\a\b}$, which we take to be diagonal. $\ell_L^i$ \,($i = 1\ldots 3$) are the SM lepton doublets
and $\phi$ is the Higgs field.\footnote{Our notation here is $\phi_r = \e_{rs}\tilde\phi_s^*$ where 
$\tilde\phi$
is the usual Higgs doublet giving mass to the lower fields in the $SU(2)$ lepton doublets.}
The complex Yukawa couplings $\l_{i\a}$ introduce CP violation into the theory, 
as required by the second Sakharov condition for leptogenesis. 

With a Higgs VEV $v$, the usual see-saw mechanism gives rise to three light Majorana neutrinos
with mass matrix
\begin{equation}
(m_\n)_{ij} ~=~ \sum_\a \l_{i\a}\,\frac{1}{M_\a}\,\l_{\a j}^T\, v^2 \ ,
\label{bb1}
\end{equation}
along with the heavy sterile Majorana neutrinos with masses $M_\a$. This is diagonalised by the 
PMNS matrix $U$ such that $U^T m_\n U = m_{diag}$, with the corresponding relation between the
mass and flavour eigenstates. Except where explicitly mentioned, we neglect these light neutrino
masses in the rest of this paper.

The coupling to gravity is through the connection alone, as required for a Lagrangian satisfying the strong
equivalence principle, with the covariant derivative acting on spinors being 
$D_\m = \partial_\m - \tfrac{i}{4}\w_{\m a b}\s^{a b}$, where $\w_{\m a b}$ is the spin connection
and $\s^{ab} = \tfrac{i}{2} \left[\c^a,\c^b\right]$.\footnote{We use Greek indices $\m,\n,\ldots$ to refer 
to coordinates in the curved spacetime and Latin indices $a,b,\ldots$ to refer to the local Lorentz frame 
at each point. Our metric and curvature sign conventions are [S1][S2][S3] = $-++$ in the terminology
of \cite{Hobson:2006se} and we use the Dirac matrix definitions of \cite{Peskin:1995ev}. }

To see how the gravitational interactions induce an asymmetry in the propagation of leptons and
antileptons, it is convenient to write $S$ explicitly in terms of the fields $\ell_L, \n_R$ and their charge
conjugates $\ell_L^{\,\,c} \equiv (\ell_L)^c$ and $\n_R^{\,\,c}$. Note that $\ell_L^{\,\,c}$ is a right-handed field,
since in general $(\psi_L)^c = (\psi^c)_R$. We then have:
\begin{align}
S\,=\, \int d^4 x \sqrt{-g}&\biggl[{\cal L}_{\rm SM} + 
\frac{1}{4}\left(i\,\overline{\n_R} \,\c.\overleftrightarrow{D}\, \n_R  
+ i\, \overline{\n_R^{\,\,c}}\,\c.\overleftrightarrow{D}\,\n_R^{\,\,c}\right) 
- \frac{1}{2} \Bigl(\overline{\n_R^{\,\,c}}\, M\, \n_R  + \overline{\n_R} \,M \,\n_R^{\,\,c} \Bigr) \nonumber \\
&-\frac{1}{2} \left(\overline{\ell_L}\, \l \,\phi \,\n_R + \overline{\n_R}\, \l^\dagger \phi^\dagger\, \ell_L
+ \overline{\ell_L^{\,\,c}} \,\l^* \phi^*\, \n_R^{\,\,c} + \overline{\n_R^{\,\,c}} \,\l^T \phi^T \,\ell_L^{\,\,c} \right) 
\biggr] \ .
\label{b2}
\end{align}

The corresponding propagators are denoted by
\begin{align}
\langle\, \ell_L ~~\overline{\ell_L}\,\rangle ~&=~ \langle \, \ell_L^{\,\,c} ~~ \overline{\ell_L^{\,\,c}}\,\rangle 
~=~ \D(x,y) \ ,  \\
\langle\,\phi_r ~~ \phi_s^*\,\rangle ~&=~ G_{rs}(x,y)\ ,
\label{b3}
\end{align}
while for the right-handed fields we have both {\it `charge conserving'} and {\it `charge violating'}
propagators,
\begin{align}
\langle\, \n_R ~~ \overline{\n_R}\,\rangle ~&=~ \langle\,\n_R^{\,\,c} ~~ \overline{\n_R^{\,\,c}}\,\rangle 
~=~ S(x,y) \ , \\
\langle\, \n_R ~~ \overline{\n_R^{\,\,c}}\,\rangle ~&=~ \langle\,\n_R^{\,\,c} ~~ \overline{\n_R}\,\rangle 
~=~ S^\times(x,y) \ .
\label{b5}
\end{align}
Note that in curved spacetime, translation invariance is lost and the propagators are not simply
functions of the coordinate difference $|x-y|$ as in flat spacetime. This becomes crucial below.
In flat spacetime, the momentum space propagators are
\begin{equation}
\D(p) ~=~ \frac{i \,\c.p}{p^2}  \ , ~~~~~~~~~~~~~~~~~~~~~~~~ G(p) ~=~ \frac{i}{p^2 - m_H^2} \ ,
\label{b7}
\end{equation}
(neglecting the light neutrino masses, and writing $G(p)$ for the physical Higgs component only) and 
\begin{equation}
S_\a(p) ~=~ \frac{i\,\c.p}{p^2 - M_\a^2} \ , ~~~~~~~~~~~~~~~~~ 
S_\a^\times(p) ~=~ \frac{i\, M_\a}{p^2 - M_\a^2} \ .
\label{b8}
\end{equation}

\begin{figure}[h!]
\centering
{\includegraphics[scale=0.50]{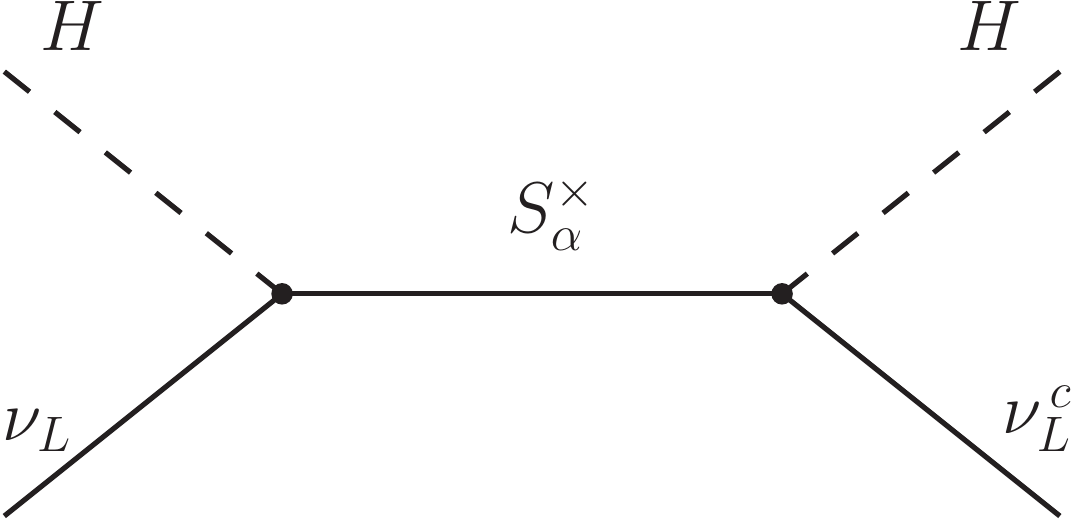}
\hskip1.9cm
\includegraphics[scale=0.50]{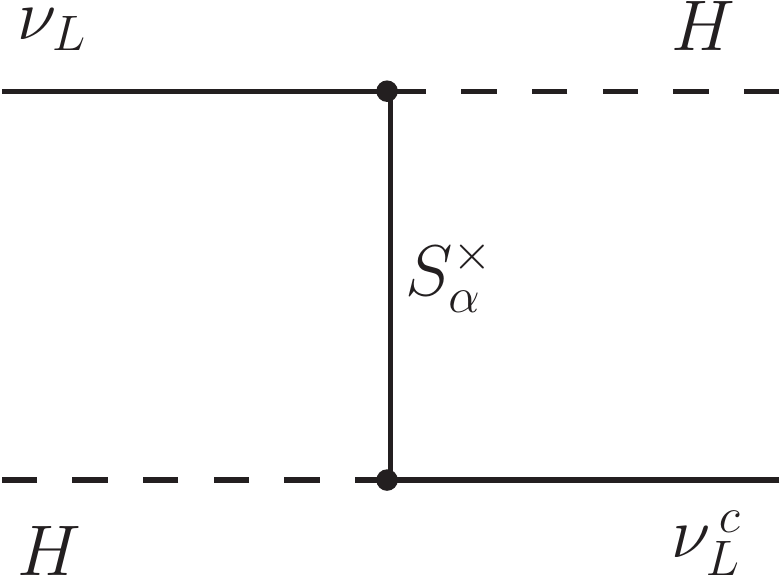}}
\vskip1cm
{\includegraphics[scale=0.50]{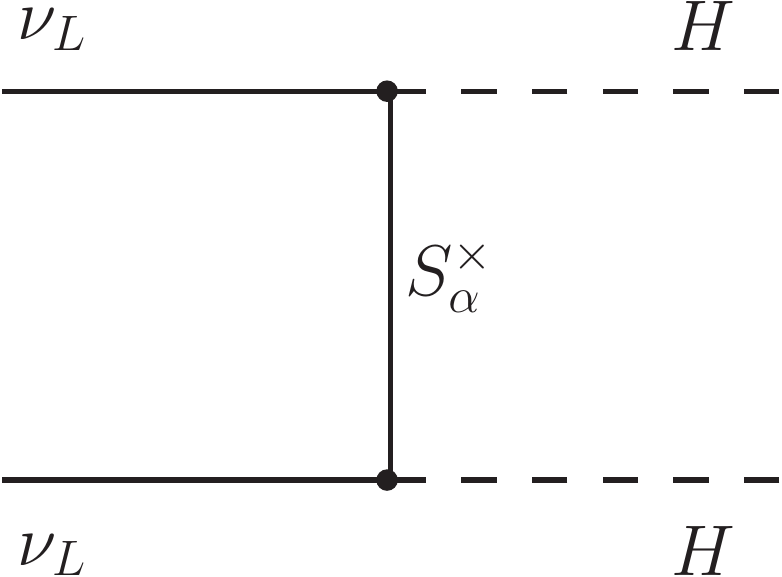}
\hskip3.4cm
\includegraphics[scale=0.50]{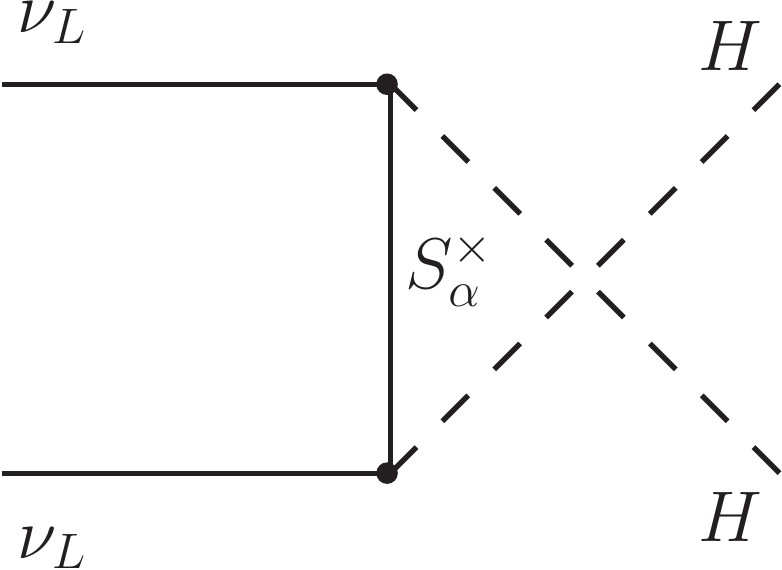}}
\caption{Feynman diagrams for the lepton number violating reactions 
$\n_L~H \leftrightarrow \n_L^{\,c}~ H$ and $\n_L~\n_L \leftrightarrow H ~H$, mediated by 
the `charge violating' $\n_R$ propagator $S_\a^\times$. }
\label{Fig 2.1}
\end{figure}
With a non-vanishing Majorana mass, the propagator $S_\a^\times$ allows reactions which violate 
lepton number by 2 units, {\it viz.} $\n_L~H \leftrightarrow \n_L^{\,\,c}~ H$ and 
$\n_L~\n_L \leftrightarrow H ~H$, illustrated in Fig.~\ref{Fig 2.1}. Both diagrams depend on the 
Yukawa coupling factor through $\l \,S^\times \l^T \,=\, \sum_\a \l_{i\a} S_\a^\times \l_{\a j}^T$.
This is the source of the lepton number violation which is required by the first Sakharov condition.

\subsection{Lepton-antilepton asymmetry in curved spacetime}\label{sect 2.2}

To implement the mechanism of radiatively-induced gravitational leptogenesis, we need to show
that the propagation of leptons and antileptons is different in a gravitational field.
Specifically, we find that at loop level, the self-energies $\S$ and $\S^c$ for the leptons and
antileptons differ when translation invariance no longer holds, leading to distinct dispersion
relations. Together with the lepton number violating reactions in Fig.~\ref{Fig 2.1},
this enables a lepton-antilepton asymmetry to be generated in thermal equilibrium in an
expanding universe.

\begin{figure}[h!]
\vskip0.5cm
\centering
\includegraphics[scale=0.52]{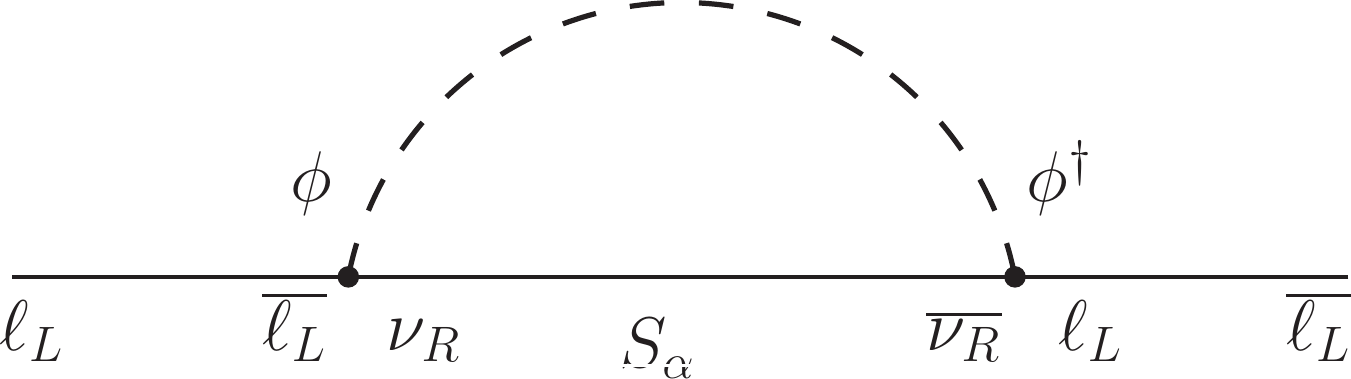}
\caption{ One-loop self-energy diagram for the light $\n_L$ neutrinos with an intermediate
charge-conserving $\n_R$ propagator $S$.  The fields at the vertices are shown explicitly for 
comparison with the Lagrangian (\ref{b2}).}
\label{Fig 2.2}
\end{figure}
At one loop, there is a single self-energy diagram involving the right-handed neutrinos,
shown in Fig.~\ref{Fig 2.2}. Evidently, at this order there is no corresponding diagram involving
the charge-violating propagator. For clarity, we explicitly show the fields at the vertices
in the diagram to allow the Yukawa couplings to be easily read off from the Lagrangian
in the form (\ref{b2}). The self-energy is therefore,
\begin{equation}
\S_{ij}(x,y) ~=~ \sum_\a \l_{i\a} \l_{aj}^\dagger\, G(x,y)\, S_\a (x,y) \ .
\label{b9}
\end{equation}
Note that depending on the Yukawa couplings, this can induce lepton flavour-changing processes
dependent on the gravitational field.

The corresponding self-energy diagram for the antileptons is evidently given by
\begin{equation}
\S_{ij}^c(x,y) ~=~ \sum_\a \l_{i\a}^* \l_{aj}^T \, G(x,y)\, S_\a (x,y) \ .
\label{b10}
\end{equation}
Since we are interested in the violation of total lepton number, we can trace over the light lepton
flavours, leaving\begin{align}
\tr \left(\S_{ij}(x,y) - \S_{ij}^c(x,y)\right) ~&=~ 
\sum_\a \left(\l^\dagger \l - \l^T \l^*\right)_{\a\a} \, G(x,y)\, S_\a (x,y)  \nonumber \\
&=~ 2i\, \sum_\a \Im \left(\l^\dagger \l\right)_{\a\a} \,  G(x,y)\, S_\a (x,y) ~=~ 0 \ .
\label{b11}
\end{align}
As we calculate below, these one-loop self-energies contribute to the CP even terms in the effective 
Lagrangian (\ref{c1}), though not to the CP odd term generating the matter-antimatter asymmetry.

\begin{figure}[h!]
\vskip0.5cm
\centering
\includegraphics[scale=0.44]{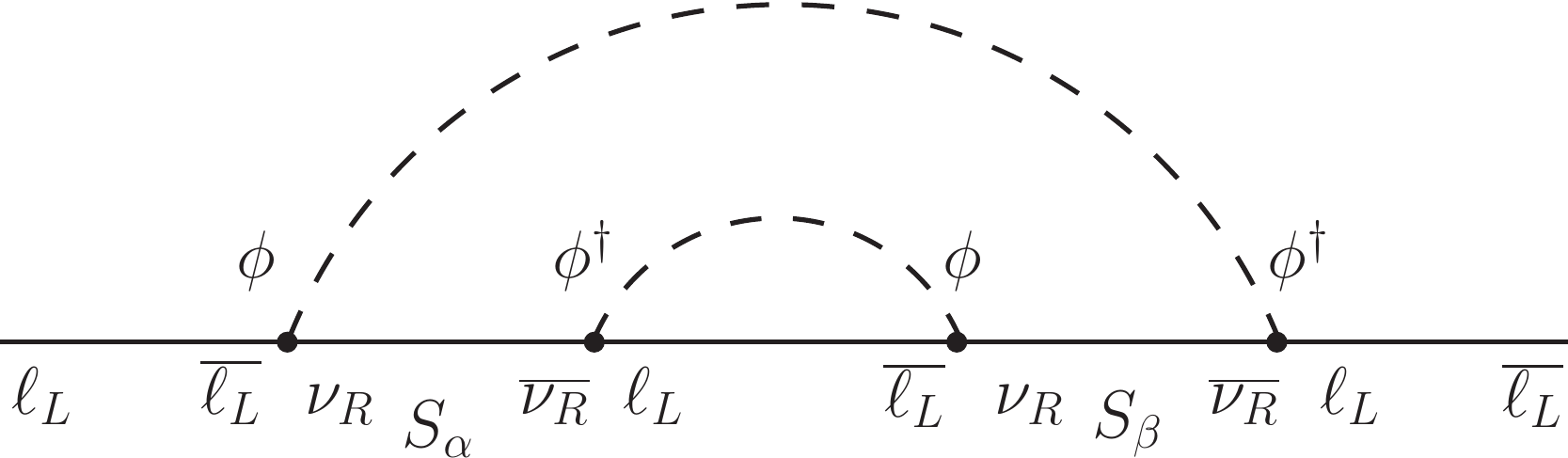}
\vskip1.2cm
\includegraphics[scale=0.44]{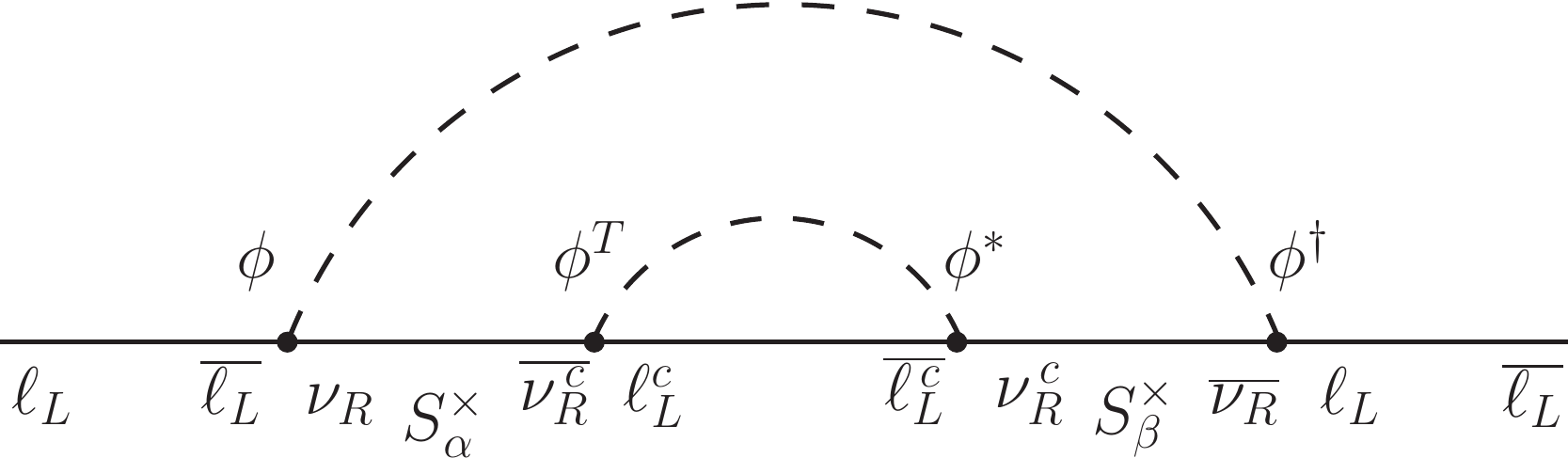}
\hskip0.5cm
\includegraphics[scale=0.44]{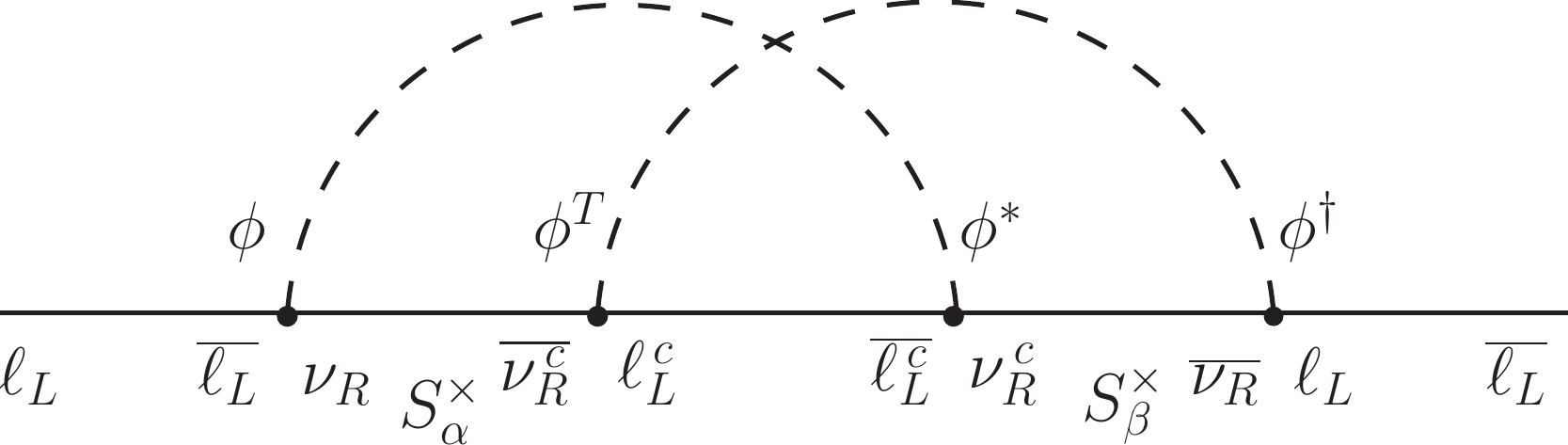}
\caption{Two-loop self-energy diagrams for the light $\n_L$ neutrinos giving rise to a 
lepton-antilepton asymmetry in curved spacetime.}
\label{Fig 2.3}
\end{figure}
The situation is different at two loops \cite{McDonald:2015ooa, McDonald:2015iwt}. 
Here, we have the three self-energy diagrams illustrated in Fig.~\ref{Fig 2.3}.
The corresponding self-energies are:
\begin{align}
\S_{ij}^{(1)}(x,y) ~=~ \int d^4 z \sqrt{-g} &\int d^4 z' \sqrt{-g}\, 
\sum_{\a,\b,k} \left(\l_{i\a} \l_{\a k}^\dagger \l_{k\b} \l_{\b j}^\dagger \right)  \nonumber \\
& \times ~G(x,y)\, G(z,z')\, S_\a(x,z) \,\D(z,z') \, S_\b(z',y) \ , 
\label{b12}
\end{align}
from the `nested' diagram with two $S$ propagators, and 
\begin{align}
\S_{ij}^{(2)}(x,y) ~=~ \int d^4 z \sqrt{-g} &\int d^4 z' \sqrt{-g}\, 
\sum_{\a,\b,k} \left( \l_{i\a} \l_{\a k}^T \l_{k\b}^* \l_{\b j}^\dagger \right)  \nonumber \\
& \times ~G(x,y) \, G(z,z') \, S_\a^\times(x,z) \,\D(z,z') \, S_\b^\times(z',y) \ , 
\label{b13}
\end{align}
and
\begin{align}
\S_{ij}^{(3)}(x,y) ~=~ \int d^4 z \sqrt{-g} &\int d^4 z' \sqrt{-g}\, 
\sum_{\a,\b,k} \left( \l_{i\a} \l_{\a k}^T \l_{k\b}^* \l_{\b j}^\dagger \right)  \nonumber \\
& \times ~G(x,z') \, G(z,y) \, S_\a^\times(x,z) \,\D(z,z') \, S_\b^\times(z',y) \ , 
\label{b14}
\end{align}
for the `nested' and `overlapping' diagrams with two $S^\times$ propagators.
Here, $G(x,y)$ generically denotes the appropriate component of the
Higgs field propagator. Note that there is no overlapping-type diagram with two $S$ propagators -- 
this follows from the complex $SU(2)_L$ doublet nature of the Higgs propagator.

In curved spacetime, all these diagrams produce an asymmetry in $\S_{ij} - \S_{ij}^c$ but, 
as we now show, only the diagrams with two $S^\times$ propagators
produce a {\it total} lepton-antilepton asymmetry,
unsurprisingly since these are the diagrams involving the charge-violating propagators $S^\times$.
Tracing over the light flavours as before, and focusing on the Yukawa couplings, we have,
with the integral factor read off from (\ref{b12}), 
\begin{equation}
\tr \left(\S_{ij}^{(1)} - \S_{ij}^{(1)c}\right) ~=~
2i \sum_{\a,\b} \,\Im \left[(\l^\dagger \l)_{\b\a} \, (\l^\dagger \l)_{\a\b} \right]\, I_{\a\b}^{(1)} ~=~ 0 \ ,
\label{b15}
\end{equation}
whereas
\begin{align}
\tr \left(\S_{ij}^{(2)} - \S_{ij}^{(2)c}\right) ~&=~
2i \sum_{\a,\b} \,\Im \left[(\l^\dagger \l)_{\b\a} \, (\l^T \l^*)_{\a\b} \right]\, I_{\a\b}^{(2)} 
\nonumber \\
&=~2i \sum_{\a,\b} \,\Im \left[(\l^\dagger \l)_{\b\a} \, (\l^\dagger \l)_{\b\a} \right]\, I_{\a\b}^{(2)}  \ ,
\label{b16}
\end{align}
and\begin{equation}
\tr \left(\S_{ij}^{(3)} - \S_{ij}^{(3)c}\right) ~=~
2i \sum_{\a,\b} \,\Im \left[(\l^\dagger \l)_{\b\a} \, (\l^\dagger \l)_{\b\a} \right]\, I_{\a\b}^{(3)}  \ .
\label{b17}
\end{equation}
Noting further that $\Im \left[(\l^\dagger \l)_{\b\a} (\l^\dagger \l)_{\b\a}\right]$ 
is {\it antisymmetric} in $\a,\b$, we can replace the integral factors above by 
$I_{[\a\b]}^{(2)}$ and $I_{[\a\b]}^{(3)}$.

This brings us to a key point \cite{McDonald:2015ooa, McDonald:2015iwt}. 
To obtain a lepton-antilepton asymmetry, the integral factors 
in (\ref{b16}), (\ref{b17}) must have a contribution which is antisymmetric in $\a,\b$.
Now, in flat spacetime, translation invariance implies that the propagators are all functions
of the difference of coordinates, {\it i.e.} $\D(x,y) \rightarrow \D(x-y)$, {\it etc}. 
But then, making a suitable change of variables on $z,z'$, we can readily show\footnote{For example,
for diagram (2), translation invariance implies
\begin{align*}
I_{\a\b}^{(2)}(x,y) ~&=~\int d^4 z \int d^4 z' \, G(x-y)\,G(z-z')\,S_\a^\times(x-z)\,\D(z-z')
\,S_\b^\times(z'-y)  \\
&=~\int d^4 u \int d^4 u' \, G(x-y)\,G(u-u')\,S_\b^\times(x-u)\,\D(u-u')
\,S_\a^\times(u'-y)  \\
&=~ I_{\b\a}^{(2)}(x,y) \ ,
\end{align*}
under the change of dummy variables $u = x+y-z'$ and $u'=x+y-z$. }
that the factors $I_{\a\b}^{(2)}(x,y)$ and $I_{\a\b}^{(3)}(x,y)$ are symmetric in $\a,\b$.

This is, however, special to flat spacetime. It accords with the general theorems presented in
\cite{McDonald:2015iwt} that CPT invariance together with translation invariance ensures 
that the propagation of particles and antiparticles is identical. 
It is no longer true in curved spacetime. In this case, we have found that at two loops there is an 
asymmetry in the self-energies of the light leptons and antileptons given by
\begin{equation}
\tr \left(\S_{ij} - \S_{ij}^c \right) ~=~ 2i \sum_{\a,\b} \Im \,[(\l^\dagger \l)_{\b\a} (\l^\dagger \l)_{\b\a}]\,
\left(I_{[\a\b]}^{(2)} + I_{[\a\b]}^{(3)}\right) \ .
\label{b18}
\end{equation}
Provided that the Majorana masses of the sterile neutrinos are non-degenerate, we therefore 
have a gravitational mechanism for establishing a lepton-antilepton asymmetry through 
two-loop radiative corrections to the light lepton propagators. In subsequent sections,
we show how this leads to a non-vanishing net lepton number density in thermal equilibrium 
in an expanding universe.

\subsection{Weak gravitational field expansion}\label{sect 2.3}

The gravitational dynamics of the light neutrinos in this model is captured by the effective action
$S_{eff}$ discussed in section \ref{sect 3}. The nature of the interactions in $S_{eff}$ is constrained by
general principles including local Lorentz invariance and hermiticity, with the leading terms at
low energy and first order in the curvature shown in (\ref{c1}). This includes both CP conserving 
and violating operators, reflecting the potential breaking of CP invariance by the Yukawa couplings
in the fundamental theory.

It remains only to determine the coefficients of these operators in terms of the parameters and couplings
of the fundamental theory (\ref{b1}).
The simplest method is to compare the predictions of the two theories for the light neutrino-graviton vertex
in the weak gravitational field limit. In fact, as explained in \cite{McDonald:2014yfg},
it suffices to consider only conformally flat metrics for this purpose, but note that this is {\it not} a
restriction on the validity of $S_{eff}$ for general curved spacetimes.

So, writing the metric in the weak-field approximation as
\begin{equation}
g_{\m\n} = \Omega^2 \eta_{\m\n} = (1+h) \eta_{\m\n}  \ ,
\label{b19}
\end{equation}
for which the expansion of the curvatures at $O(h)$ is
\begin{equation}
R_{\m\n} = - \partial_\m \partial_\n h - \frac{1}{2} \,\eta_{\m\n} \partial^2 h \ , ~~~~~~~~~~~
R = - 3\, \partial^2 h  \ ,
\label{b19a}
\end{equation}
together with conformally rescaled fields (in $n$ dimensions)
\begin{equation}
\ell_L \rta \Omega^{-(n-1)/2} \ell_L \ , ~~~~~~\n_R \rta \Omega^{-(n-1)/2} \n_R \ , ~~~~~~
\phi \rta \Omega^{-(n-2)/2} \phi \ ,
\label{b20}
\end{equation}
the neutrino-Higgs sector of the BSM Lagrangian is expanded at $O(h)$ as
\begin{align}
\mathcal{L}_h~=~ \biggl[&-\frac{1}{4} M\left(\overline{\n_R^{\,\,c}} \,\n_R \,+\, 
\overline{\n_R} \,\n_R^{\,\,c}\right) 
\,-\, m_H^2 H^2
\nonumber \\
&+ \frac{1}{8} (n-4)H \left(\overline{\n_L} \,\l \,\n_R \,+\, \overline{\n_R} \,\l^\dagger\, \n_L
\,+\, \overline{\n_R^{\,\,c}}\, \l^* \,\n_L^{\,\,c} \,+\, 
\overline{\n_L^{\,\,c}}\, \l^T\, \n_R^{\,\,c} \right)\,\biggr] \, h \ ,
\label{b21}
\end{align}
where for simplicity we have taken the Higgs field to be a conformal scalar, {\it i.e.}~$\xi = 1/6$ 
in the Lagrangian term $\xi R \phi^2$. In general there will be an additional $hHH$
vertex proportional to $(\xi -1/6)q^2$ which will carry through the diagram calculations leaving
a $\xi$ dependence in the final effective Lagrangian coefficient. 
\begin{figure}[h!]
\vskip0.5cm
\centering
\includegraphics[scale=0.5]{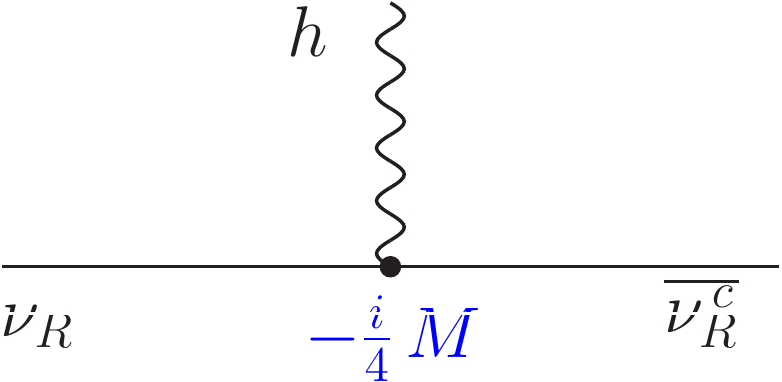}
\hskip1cm
\includegraphics[scale=0.5]{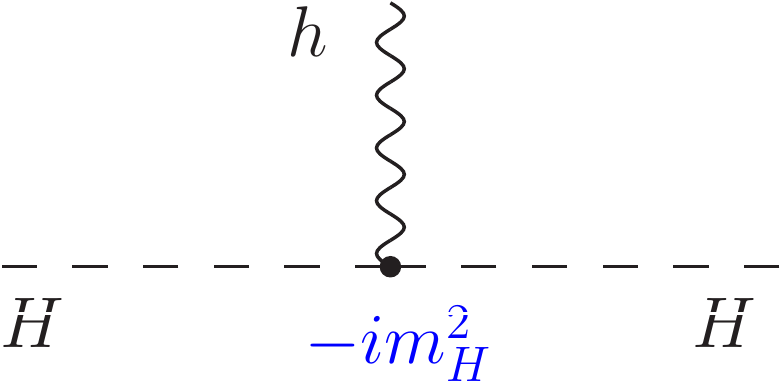}
\hskip1cm
\includegraphics[scale=0.5]{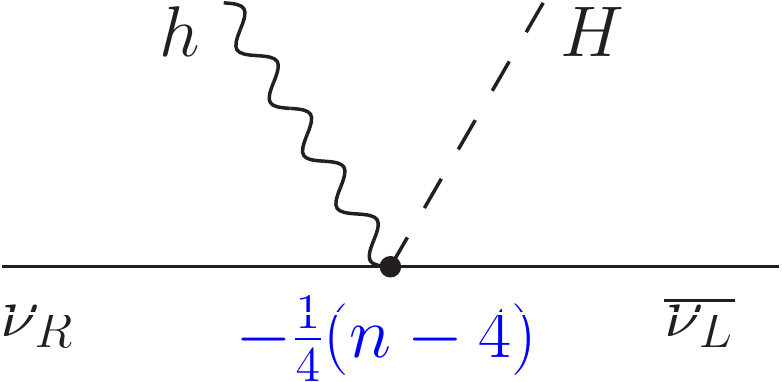}
\caption{Elementary gravitational couplings in the fundamental theory in the weak-field limit.}
\label{Fig 2.4}
\end{figure}

From this, we can read off the elementary vertices for the $h$ couplings to the left and right-handed
neutrinos and the Higgs field. A representative set are shown in Fig.~\ref{Fig 2.4}. 
Note particularly the factor of $O(n-4)$ entering all the Yukawa vertices -- these will give a finite
contribution when inserted into UV divergent diagrams.

The next step is to incorporate these into the one and two-loop self-energy diagrams for the light
neutrinos, effectively integrating out the heavy fields, and subsequently comparing to the equivalent 
vertices derived directly from the effective Lagrangian. This enables us to match coefficients 
and determine the values of the effective couplings in $S_{eff}$ in this model.

\subsection{Feynman diagrams}\label{sect 2.4}

\begin{figure}[h!]
\centering
\includegraphics[scale=0.55]{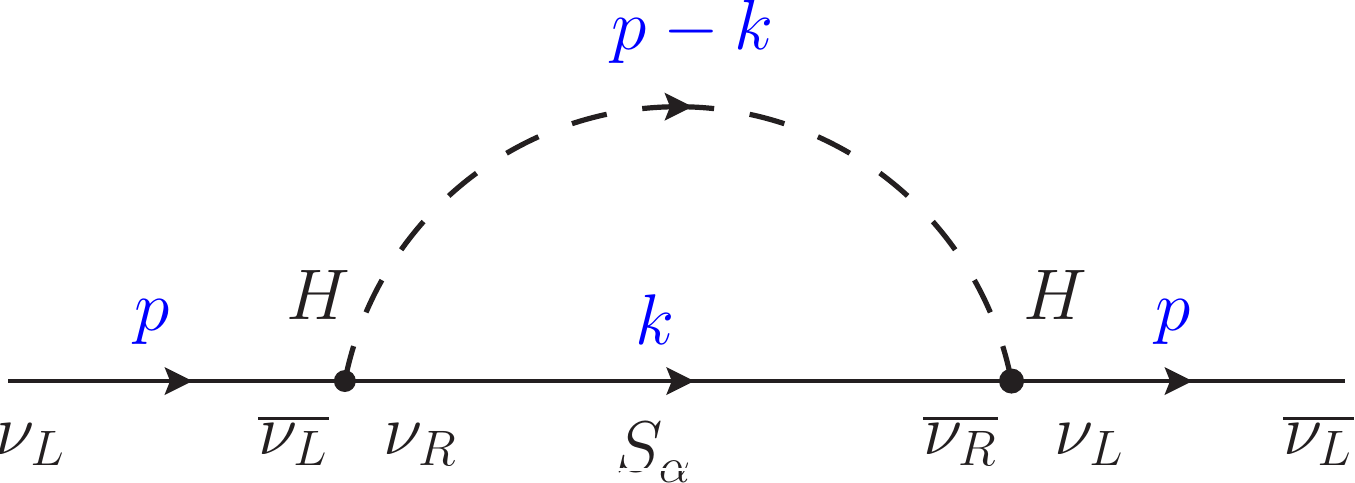}
\caption{One-loop self-energy diagram for the light neutrinos in flat spacetime.}
\label{Fig 2.5}
\end{figure}
Consider first the one-loop self-energy diagram in Fig.~\ref{Fig 2.5}, without an $h$ insertion.
Evaluating in flat spacetime, the self-energy is\footnote{
Evaluating the finite part of the self-energy diagram gives
\begin{align*}
\S_{ij} &=-\frac{i}{(4\pi)^2} \left(\l \l^\dagger\right)_{ij} \,\slashed{p}\,\biggl[
\frac{1}{n-4} - \log{4\pi} + \c \\
&~~~~~~~~~~~~~~~~~~~~~~~~~~~~~~~~~~~~~+\frac{1}{2p^4}\biggl(
-p^2\left(M_\a^2 + 2p^2\right) + M_\a^4 \log{\frac{M_\a^2}{\m^2}}
-\left(M_\a^4 - p^4 \right) \log{\frac{\left(M_\a^2 -p^2\right)}{\m^2}}\biggr) \biggr] \\
&= -\frac{i}{(4\pi)^2} \left(\l \l^\dagger\right)_{ij} \,\slashed{p}\,\left[
\frac{1}{n-4} + \frac{1}{2}\log{\frac{M_\a^2}{\m^2}} - \log{4\pi} + \c -\frac{3}{4} - \frac{1}{3}\frac{p^2}{M_\a^2}
+ O\left(\frac{p^4}{M_\a^4}\right)  \right]\ ,
\end{align*} 
where we have expressed the result in terms of the dimensionally renormalised couplings and $\m$ is the
corresponding mass scale. 
\label{fn2.4}}
\begin{align}
\S_{ij}(p,p) ~&=~ \sum_\a \l_{i\a}\, \l_{\a j}^\dagger \int \frac{d^n k}{(2\pi)^n}\, \frac{\slashed{k}}{k^2 - M_\a^2}\,
\frac{1}{(p-k)^2 - m_H^2} \nonumber \\
&=~ - \frac{i}{n-4} \, \left(\l \l^\dagger\right)_{ij} \, \frac{1}{n-4} ~\slashed{p}~~+~~~{\rm finite} \ .
\label{b22}
\end{align}

\begin{figure}[h!]
\centering
\includegraphics[scale=0.50]{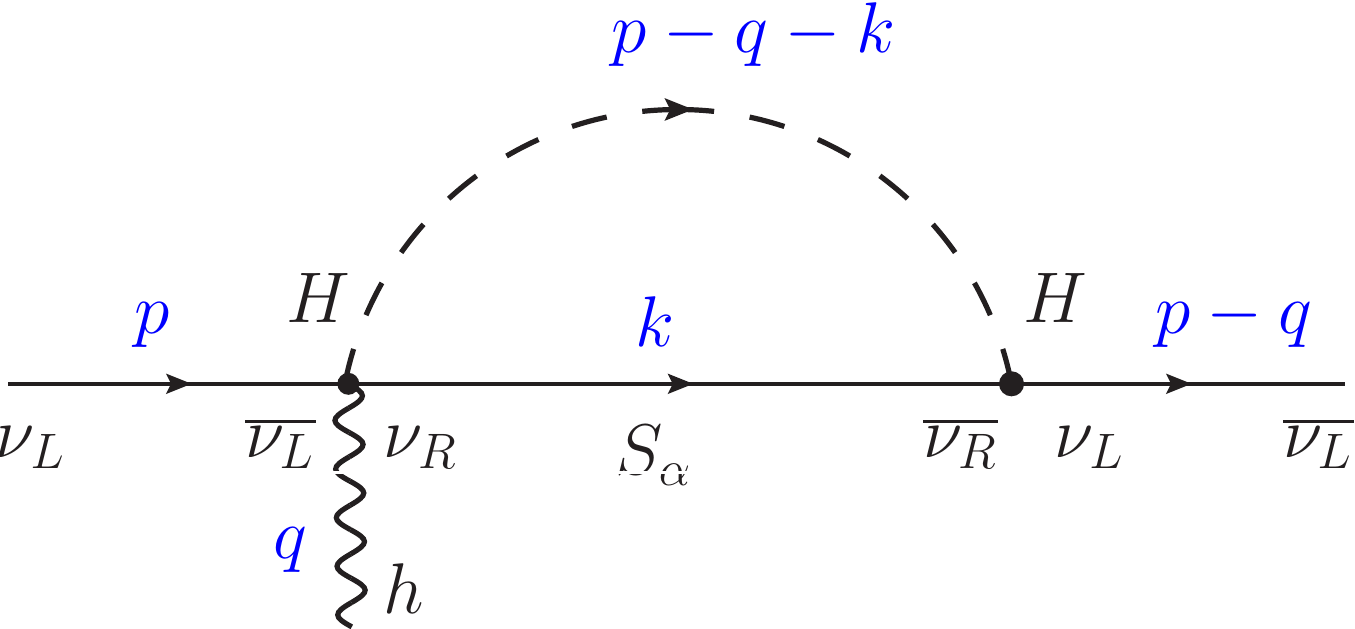}
\hskip1.5cm
\includegraphics[scale=0.50]{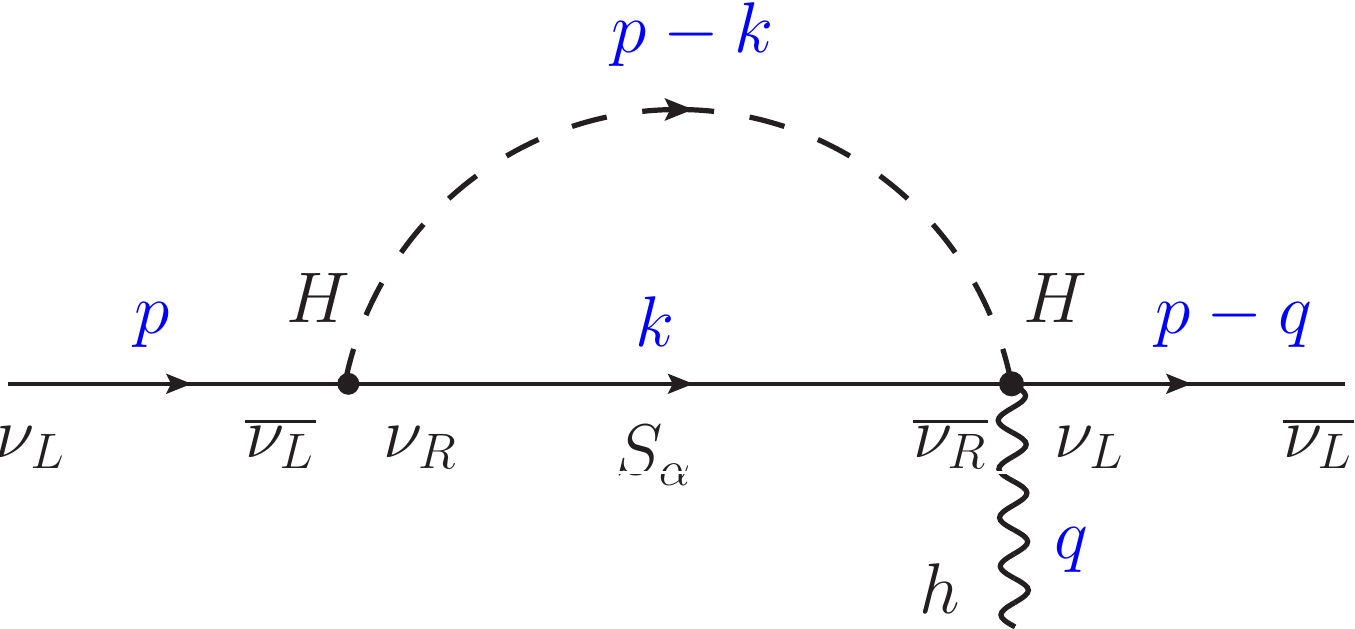}
\caption{One-loop self-energy diagrams with $h$ insertions on the Yukawa vertices.}
\label{Fig 2.6}
\end{figure}
We can then immediately read off the contribution of the diagrams with $h$ insertions on the Yukawa 
vertices, shown in Fig.~\ref{Fig 2.6}. These give\footnote{We continue to use the self-energy notation
$\S_{ij}$ here for simplicity. Clearly these Feynman diagrams are the one-loop contributions to the
three-point $h\,\overline{\n_L}^{\,i} \n_L^j$ vertex $\mathcal{V}_{ij}(p,q)$ to be matched to $L_{eff}$.}
\begin{equation}
\S_{ij}^{Yuk}(p, p-q) ~=~ \frac{i}{(4\pi)^2}\, \left(\l \l^\dagger\right)_{ij} \,(2\slashed{p} - \slashed{q}) \ .
\label{b23}
\end{equation}

\begin{figure}[h!]
\centering
\includegraphics[scale=0.48]{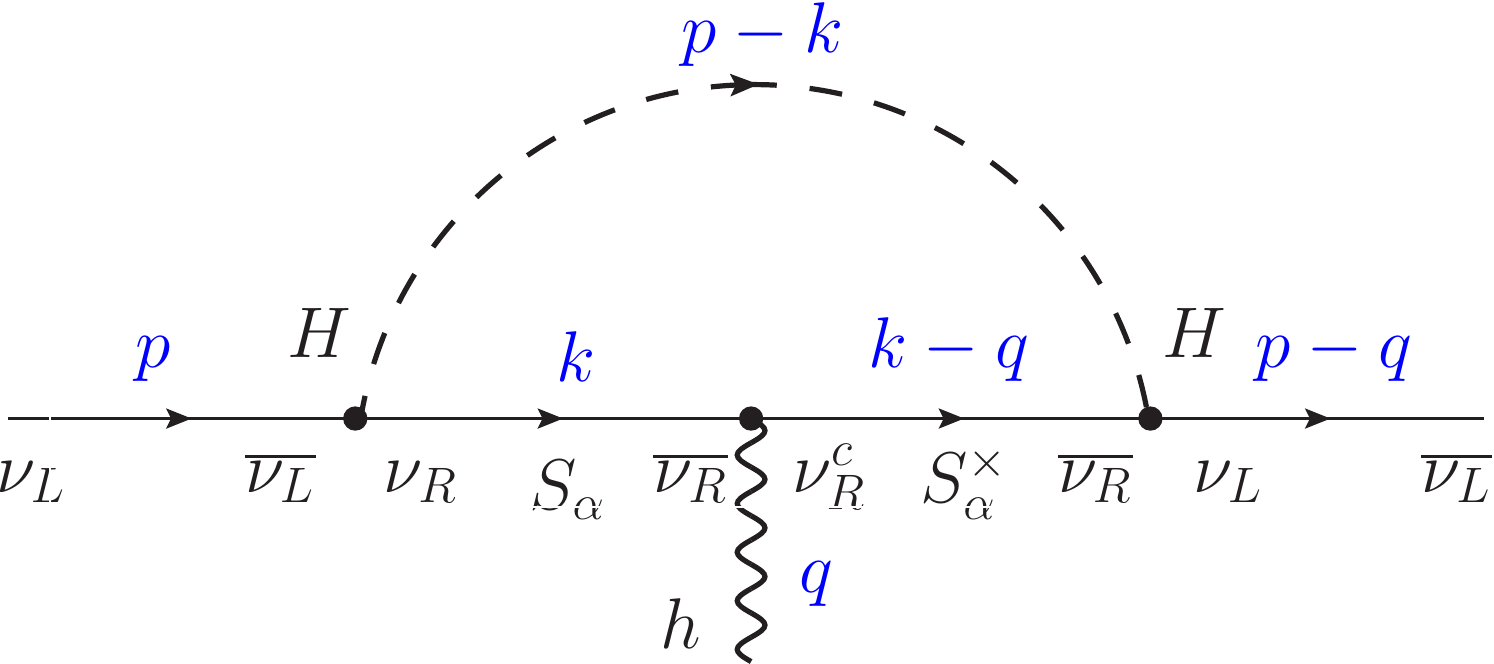}
\hskip0.8cm
\includegraphics[scale=0.48]{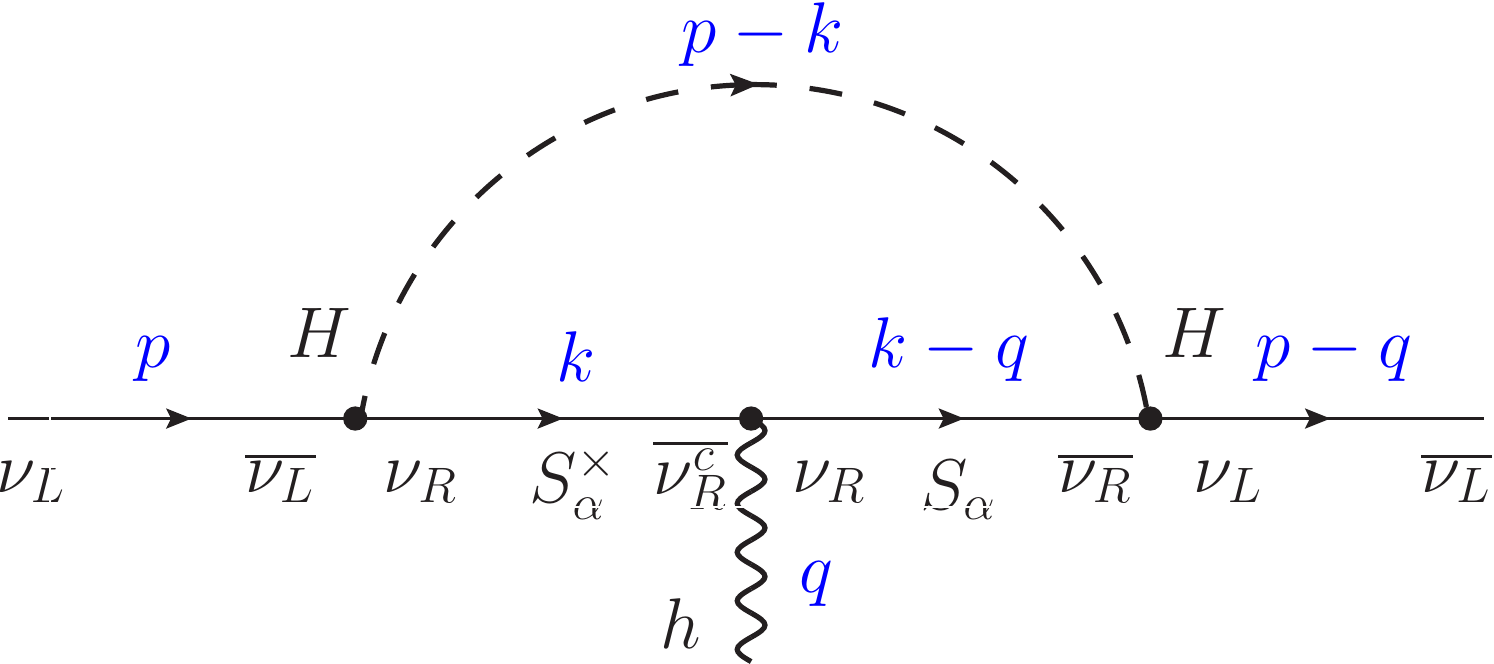}
\caption{One-loop self-energy diagrams with an $h$ insertion on the sterile neutrino propagator.}
\label{Fig 2.7}
\end{figure}
Now consider the $h$ insertion on the right-handed neutrino propagator, Fig.~\ref{Fig 2.7}.
From the sum of the two diagrams, we find
\begin{equation}
\S_{ij}^{\n_R}(p,p-q) ~=~ \frac{1}{2} \sum_\a \l_{i\a} \l_{\a j}^\dagger M_\a^2 \int \frac{d^n k}{(2\pi)^n}\,
\frac{2\slashed{k} - \slashed{q}}{\left(k^2 - M_\a^2\right)\left((k-q)^2 - M_\a^2\right)} \,
\frac{1}{(p-k)^2 - m_H^2} \ ,
\label{b24}
\end{equation}
and evaluating in the standard way introducing Feynman parameters, we find
\begin{equation}
\S_{ij}^{\n_R} ~=~ -\frac{1}{2} \frac{i}{(4\pi)^2} \sum_\a \l_{i\a} \l_{\a j}^\dagger M_\a^2 \,
\int_0^1 dz \int_0^{1-z} dx\,\left[(2x-1)\slashed{q} \,+\, 2(1-x-z)\slashed{p}\right] \,\D_\a^{-1} \ ,
\label{b25}
\end{equation}
with
\begin{equation}
\D_\a ~=~- (x+z)(1-x-z)p^2 + 2x(1-x-z)p.q -x(1-x)q^2 +(x+z) M_\a^2 + (1-x-z)m_H^2\ .
\label{b26}
\end{equation}
At the mass scales of interest for leptogenesis in this model, $m_H^2 \ll M_\a^2$, so provided 
the diagram is IR safe we can immediately set $m_H\rta 0$ at this point. This is indeed the case.
Next, in order to match coefficients with those in the effective Lagrangian, it is sufficient 
\cite{McDonald:2014yfg} to consider the leading terms in an expansion in terms of $p^2, p.q$ 
and $q^2$, neglecting terms of $O(p^4/M_\a^2)$ {\it etc}. After some calculation, we find
\begin{align}
\S_{ij}^{\n_R}(p,p-q) = &-\frac{1}{4} \frac{i}{(4\pi)^2}\, \left(\l \l^\dagger\right)_{ij} \,
(2\slashed{p} - \slashed{q}) \nonumber \\
&-\frac{1}{6} \frac{i}{(4\pi)^2} \sum_\a \l_{i\a} \l_{\a j}^\dagger \frac{1}{M_\a^2} \,
\Bigl[\bigl(2p^2 - 2p.q +\tfrac{7}{6}q^2\bigr) \slashed{p} 
+ \bigl(-p^2 + \tfrac{5}{6}p.q - \tfrac{1}{2} q^2\bigr) \slashed{q} \Bigr] \ .
\label{b27}
\end{align}
Note that the term linear in momentum in (\ref{b27}) precisely cancels the contribution (\ref{b23})
from the Yukawa insertion diagrams.

\begin{figure}[h!]
\vskip0.5cm
\centering\includegraphics[scale=0.55]{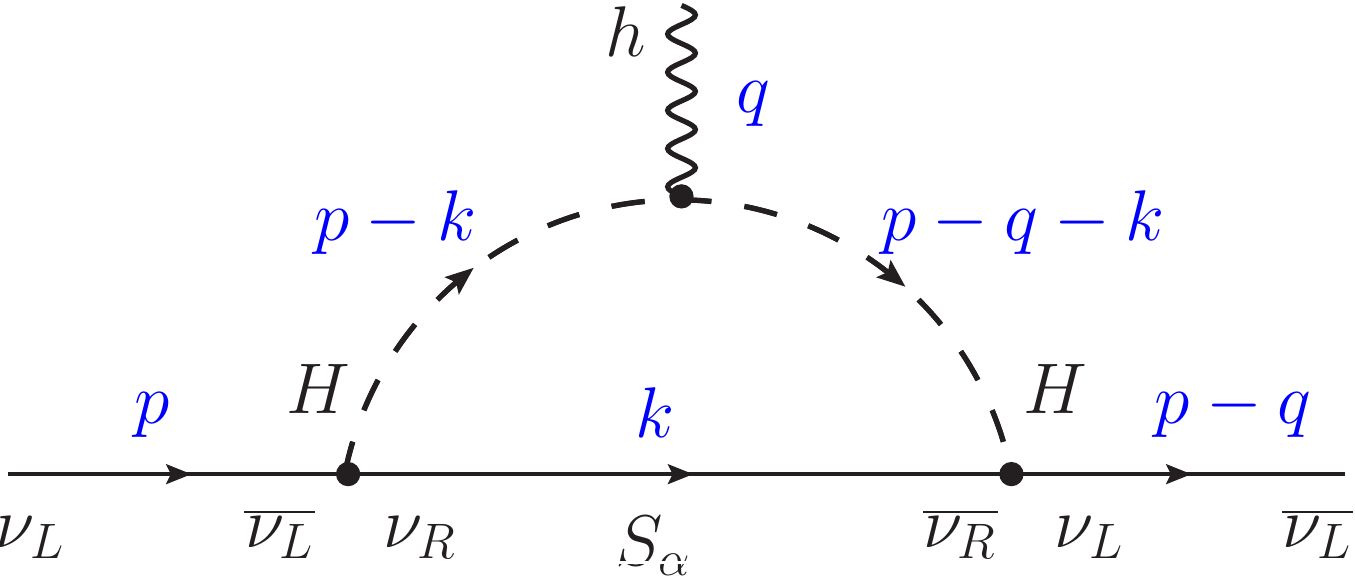}
\caption{One-loop self-energy diagram with an $h$ insertion on the Higgs propagator.}
\label{Fig 2.8}
\end{figure}
For the Higgs insertion diagram shown in Fig.~\ref{Fig 2.8}, we have
\begin{align}
\S_{ij}^H(p,p-q) ~&=~ - \sum_\a \l_{i\a}\l_{\a j}^\dagger m_H^2 \int  \frac{d^n k}{(2\pi)^n} \,
\frac{\slashed{k}}{k^2-M_\a^2} \, \frac{1}{(p-k)^2 - m_H^2} \, \frac{1}{(p-k-q)^2 -m_H^2} \nonumber \\
&= - \frac{i}{(4\pi)^2} \sum_\a \l_{i\a}\l_{\a j}^\dagger m_H^2 \int_0^1 dz \int_0^{1-z} dx \,
\Bigl[(x+z)\slashed{p} \,- \, x\slashed{q}\Bigr] \tilde{\D}_\a^{-1} \ ,
\label{b28}
\end{align}
where here
\begin{equation}
\tilde{\D}_\a = \bigl[-x(1-x) -z(1-x-z)\bigr]p^2 + 2x\bigl(1-x-\tfrac{1}{2}z\bigr)p.q -x(1-x)q^2 
 + (1-x-z)M_\a^2 +(x+z)m_H^2 \ .
\label{b29}
\end{equation}
Here, we do have to be careful since the integral is divergent as $m_H\rta 0$. 
Evaluating the divergent terms, and cancelling against the overall $m_H^2$ vertex factor, 
we find that this diagram gives a non-vanishing finite contribution in the $m_H\rta 0$ limit, {\it viz}.
\begin{equation}
\S_{ij}^{H}(p,p-q) = -\frac{1}{12} \frac{i}{(4\pi)^2} \sum_\a \l_{i\a} \l_{\a j}^\dagger \frac{1}{M_\a^2}
\Bigl[\bigl(p^2 - p.q + q^2\bigr) \bigl(2\slashed{p} -\slashed{q}\bigr)\Bigr] \ ,
\label{b30a}
\end{equation}
where again we are keeping only terms of $O(p^2/M_\a^2)$ {\it etc}.

Finally, collecting results, we find the total contribution of the $h$ insertions into the
one-loop light neutrino self-energy, at leading order in the momentum expansion, is
\begin{equation}
\S_{ij}(p,p-q) = -\frac{1}{12} \frac{i}{(4\pi)^2} \sum_\a \l_{i\a} \l_{\a j}^\dagger \frac{1}{M_\a^2}
\Bigl[\bigl(6p^2 -6 p.q + \tfrac{13}{3}q^2\bigr)\slashed{p} +
\bigl(-3p^2 + \tfrac{8}{3} p.q - 2 q^2\bigr) \slashed{q} \Bigr] \ .
\label{b30}
\end{equation}
This allows the coefficients of the three CP conserving operators in the effective Lagrangian
to be determined by matching the $h\, \overline{\n_L} \,\n_L$ vertex.

\begin{figure}[h!]
\vskip0.5cm
\centering
\includegraphics[scale=0.44]{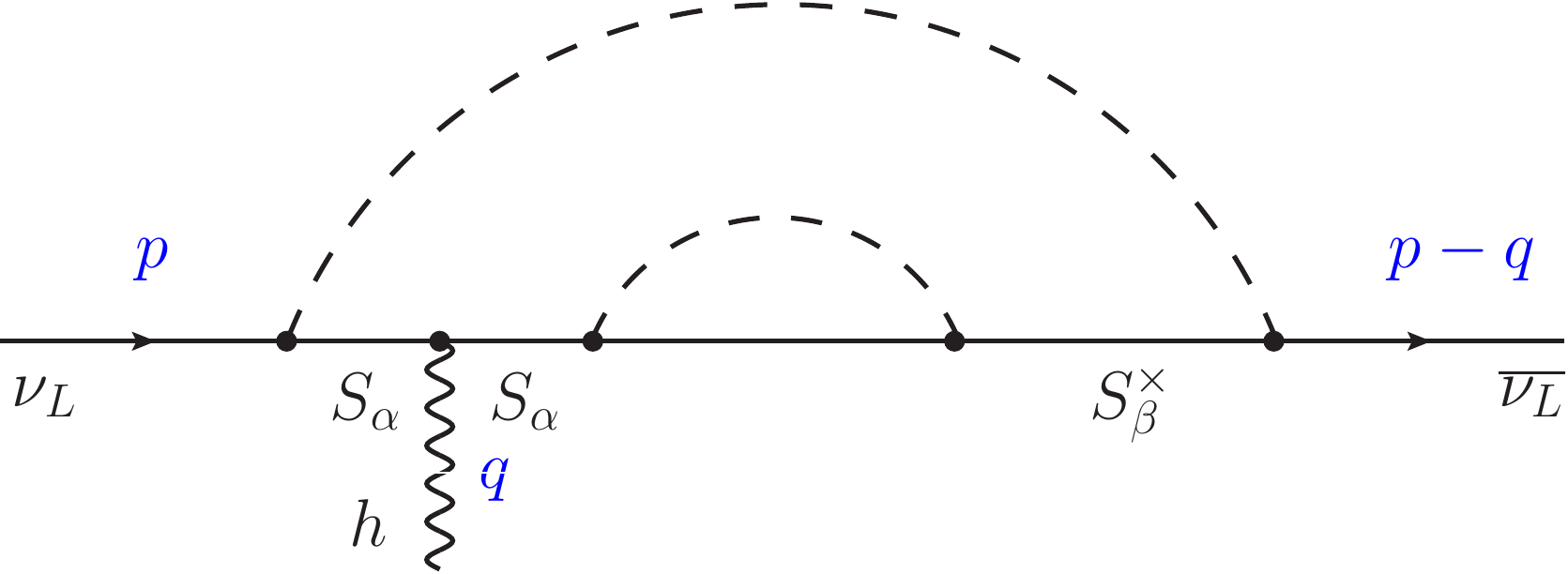}
\hskip0.4cm
\includegraphics[scale=0.44]{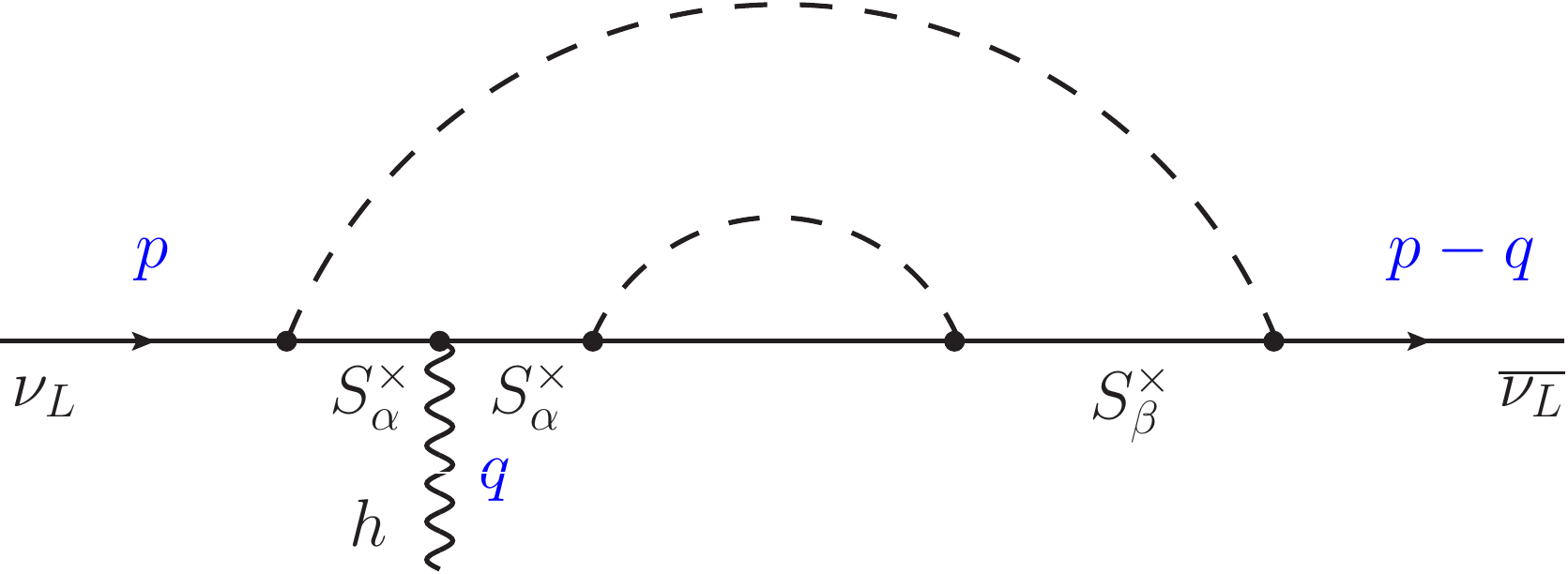}
\vskip1.2cm
\includegraphics[scale=0.44]{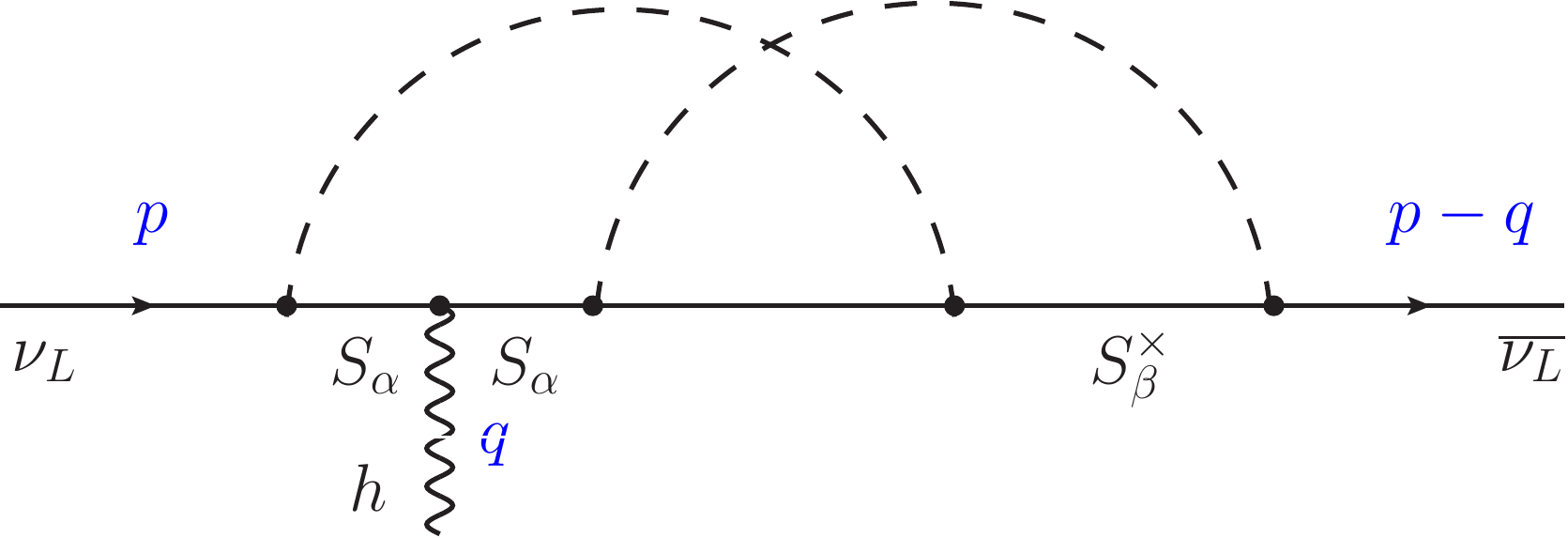}
\hskip0.4cm
\includegraphics[scale=0.44]{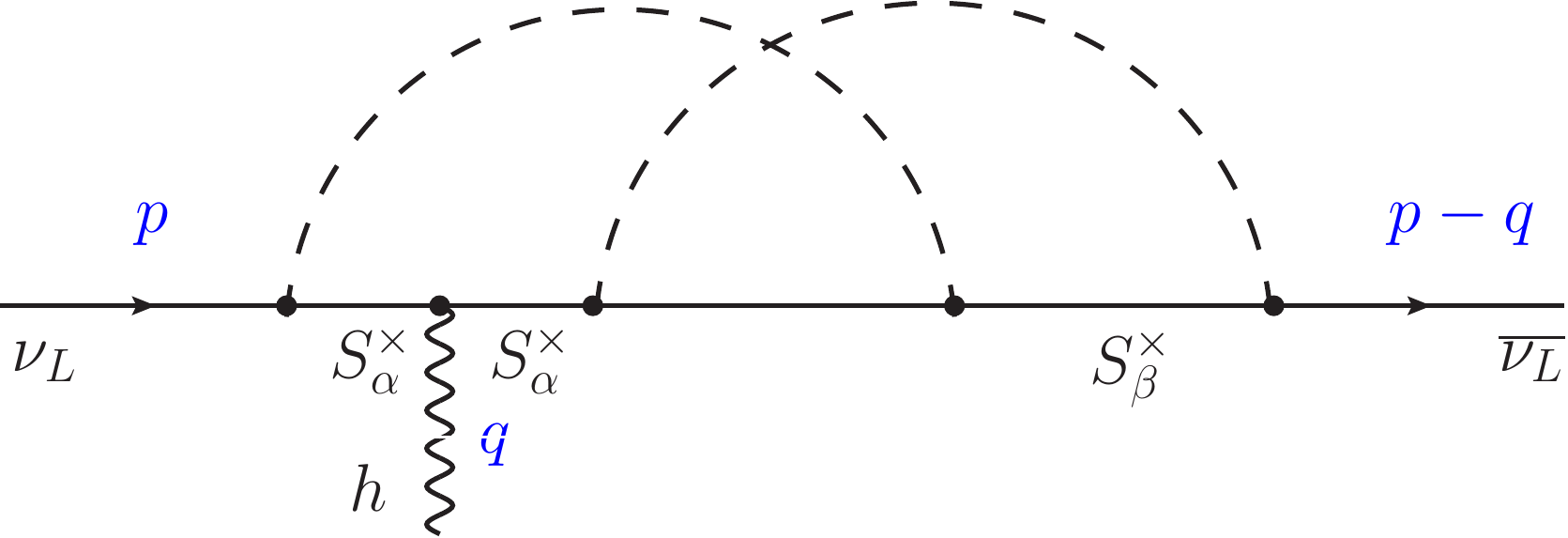}
\caption{Two-loop self-energy diagrams for the light leptons with an $h$ insertion on the 
sterile neutrino $\a$ propagator, corresponding to the two lower diagrams in Fig.~\ref{Fig 2.3}. 
Similar diagrams with the $h$ insertion on the sterile 
neutrino $\b$ propagator are also to be included, along with insertions on the Higgs propagator
and Yukawa vertices \cite{McDonald:2015iwt}. The diagrams with two $S_\alpha$ propagators
have an $h\, \overline{\nu_R}\, \nu_R^{\,c}$ vertex while those with two $S_\alpha^\times$ 
propagators have an $h\, \overline{\nu_R^{\,c}}\, \nu_R$ vertex, as seen from (\ref{b21}) and 
Fig.~\ref{Fig 2.4}.}
\label{Fig 2.9}
\end{figure}
In order to generate the vital CP violating operator in (\ref{c1}), we need to go to two loops,
where we have already demonstrated a difference in the propagation of leptons and antileptons.
The key diagrams here are shown in Fig.~\ref{Fig 2.9}, corresponding
to the two lower diagrams in Fig.~\ref{Fig 2.3}.
Note that the $h$ insertion on a sterile neutrino propagator $S_\a^\times$ gives rise to two 
contributions, one from the vertex $-i M_\a \,h\,\overline{\n_R}\,\n_R^c$ associated with two $S_\a$
propagators and another from $-i M_\a \,h\,\overline{\n_R^c}\,\n_R$ with two $S_\a^\times$ propagators.
These insertions can be made in both nested and overlapping diagrams and on the $\a$ and $\b$ 
propagators. In addition, we need to consider insertions on the Higgs propagators and on the
Yukawa vertices, which, as we have seen at one loop, can in principle contribute despite the
$h$ vertex being of $O(n-4)$ .

The resulting calculations are extensive and perhaps surprisingly stretch the limits of known
analytic methods. The hardest diagram is essentially a two-loop 3-point triangle diagram with 
arbitrary external momenta. These calculations were carried out in detail in \cite{McDonald:2015iwt},
although in that reference we were unable to complete the evaluation of this final diagram.
This leaves some uncertainty in the heavy mass hierarchy dependence of the final result.

To extract the relevant coefficient (`$b$') in the effective Lagrangian, we need only isolate the term
involving $q^2\slashed{q}$ in the momentum expansion of $\S_{ij}^{2\,loop}$. 
In \cite{McDonald:2015iwt} we found,
\begin{equation}
\S_{ij}^{2\, loop}(p,p-q)~=~ - \frac{1}{3}\frac{1}{(4\pi)^4} 
\sum_{\a,\b,k} \l_{\b j}^\dagger \l_{i\a} \l_{\b k}^\dagger \l_{k\a} \, \frac{1}{M_\a M_\b}\, I_{[\a\b]}\, 
q^2 \slashed{q} ~~+~\ldots \ ,
\label{b31}
\end{equation}
where the integral factor $I_{\a\b}$ was indeed shown to have an antisymmetric part.
The result in the large hierarchy limit $M_\b \gg M_\a$ (see \cite{McDonald:2015iwt} for the
full result) is
\begin{equation}
I_{[\a\b]} ~\sim~ \left(\frac{M_\b}{M_\a}\right)^{2p}\, \log\left(\frac{M_\b}{M_\a}\right) \ ,
\label{b32}
\end{equation}
up to an O(1) numerical factor. This shows that the gravitationally induced lepton-antilepton
asymmetry discussed above is realised explicitly at two-loop order, and consequently
generates a non-vanishing coefficient for the CP violating operator in the effective Lagrangian 
describing the low-energy dynamics in this model.

The question of whether the hierarchy parameter $p$ is 0 or 1 was left unresolved 
in \cite{McDonald:2015iwt}. The most natural result, which holds in all the diagrams
we calculated to a conclusion, is $p=0$. This would accord with the expected decoupling 
of heavy mass intermediate states in the Feynman diagram.\footnote{However, note that 
because of the external factor of $M_\b$ from the $h$ coupling, in order to find a
$p=1$ hierarchy dependence we only require the relevant diagram to tend to a constant 
or grow no faster than $\log{M_\b}$ in the large $M_\b$ limit. 
As an illustration that such a logarithmic dependence on the mass of a 
heavy neutrino propagator may arise in an individual Feynman diagram, note the asymptotic
behaviour of the UV divergent one-loop self-energy diagram in Fig.~\ref{Fig 2.5}, 
where we find a $\log{M_\a}$ dependence in the $M_\a^2 \gg p^2$ limit (see footnote \ref{fn2.4}).}
Nevertheless, in calculating 
the arbitary-momentum triangle diagram, we found contributions with $p=1$ which would
require a remarkable cancellation if they were to be absent in $I_{[\a\b]}$. For this reason, we
retain the possibility that $p$ could be 1 in what follows since, as we see in the cosmological
scenarios, the presence of a large sterile mass hierachy dependence would significantly affect
the ultimate prediction for the baryon-to-photon ratio in this particular BSM theory.

\section{Effective Lagrangian for Gravitational Leptogenesis}\label{sect 3}

In this section, we construct an effective action which encodes the tidal curvature effects
on the propagation of the light neutrinos and fix the coefficients of the induced
operators by comparing with the explicit calculations of the self-energy loop diagrams
in section \ref{sect 2}.

\subsection{Effective Lagrangian}\label{sect 3.1}

The gravitational dynamics of the light neutrinos, to leading order in the curvature,  
is encoded in the general effective action,
\begin{align}
S_{eff}= \int d^4 x\sqrt{-g} \biggl[ 
\frac{1}{2} i\,\overline{\n_L}^{\,i} \c.\overleftrightarrow{D}\, \n_L^{\, i}
&\,+ \,\tilde{a}_{ij} \, R_{\m\n} \, i \,\overline{\n_L}^{\, i} \c^\m \overleftrightarrow{D}^\n \,\n_L^{\, j}
\,+\, b_{ij} \,\partial_\m R \,\,\overline{\n_L}^{\, i} \c^\m \,\n_L^{\, j}  \nonumber \\
&~~~\,+ c_{ij} \, R \, i \,\overline{\n_L}^{\, i} \c.\overleftrightarrow{D} \,\n_L^{\,  j}
\,+\, \tilde{d}_{ij}\, i\left(D_\m \overline{\n_L}^{\, i}\right) \c.\overleftrightarrow{D} \, 
D^\m \n_L^{\,  j} \biggr] \ .
\label{c1}
\end{align}
With real coefficients $\tilde{a}, b, c, \tilde{d}$, each term in (\ref{c1}) is individually hermitian.
We have also allowed for gravitationally-induced flavour-changing effective interactions,
with the coefficients carrying flavour labels.

This effective Lagrangian arises quite generally as the low-energy limit of a fundamental UV-complete
theory, characterised by some high mass scale $M$. The coefficients $\tilde{a}, b, c, \tilde{d}$
are then of $O(1/M^2)$, so the Lagrangian is a weak gravitational field expansion in $\RR/M^2$,
where $\RR$ denotes a typical curvature.  Taking the BSM theory of section \ref{sect 2} as the 
fundamental theory, the mass $M$ is identified with the Majorana mass $M_\a$ of the heavy
sterile neutrinos.

It is also a low-energy expansion, since we have retained only leading-order terms in derivatives.
In the papers \cite{Shore:2002gn, Hollowood:2007ku},
which analysed the full energy dependence of UV-complete quantum field theories coupled to gravity, 
the relevant expansion parameter was shown to be $E\sqrt{\RR}/M^2$, for a typical particle energy $E$.
This will be an important factor in the application of our results for leptogenesis in realistic 
cosmological settings, and we will return to a careful discussion of this point later.

The presence of direct couplings to the curvature means the Lagrangian (\ref{c1}) violates
the strong equivalence principle and cannot be regarded as a fundamental theory in its own right.
In particular, in the absence of an embedding into a UV-complete theory, the theory 
described by this effective Lagrangian is not causal.
This issue has been explored extensively in the series of papers 
\cite{Hollowood:2007ku, Hollowood:2008kq, Hollowood:2009qz, Hollowood:2011yh, Hollowood:2015elj}
on the realisation of causality and unitarity in QED in curved spacetime.\footnote{See also 
\cite{deRham:2019ctd} and the discussion of causality in explicitly CPT violating theories 
in \cite{Kostelecky:2000mm}.}

The properties of the operators in (\ref{c1}) under the discrete symmetries C, P, T is important, 
and has been described carefully in \cite{McDonald:2014yfg}. Note that in curved spacetime, 
the symmetries P, T are defined only in the local Lorentz frame, or tangent space, 
at each point in spacetime. Crucially, the operators $\tilde{a}, c, \tilde{d}$ are all CP even,
while only the $b$ operator is CP odd. All are CPT even.\footnote{This is in contrast to a model
where instead of the curvatures, the bilinear neutrino operators are multiplied by constant
parameters, as in the Lorentz and CPT violating standard model extension introduced 
in \cite{Colladay:1998fq}. Importantly, the Lorentz vector and tensor operators
$i\,\overline{\n_L} \,\c^a \overleftrightarrow{D}^b \,\n_L$ and 
$\overline{\n_L}\, \c^a \,\n_L$ are CPT even and odd respectively.}

\subsection{Weak gravitational field expansion}\label{sect 3.2}

The necessary formalism to match the coefficients in the effective Lagrangian to the couplings
and masses of a fundamental theory was developed extensively in \cite{McDonald:2014yfg}
and we need only summarise the essential results here.
In \cite{McDonald:2014yfg}, we considered the low-energy limit of the standard model 
coupled to gravity; here, we are interested in the effective Lagrangian for the BSM model
of section \ref{sect 2}.

To facilitate comparison with \cite{McDonald:2014yfg}, and to simplify the application to the
eikonal expansion in section \ref{sect 5}, it is convenient to rewrite the effective action
(\ref{c1}) in the form,
\begin{equation}
S_{eff} ~=~ S_a \,+\, S_b \,+\, S_c \,+\, S_d \ ,
\label{c2}
\end{equation}
where
\begin{align}
S_a~&=~ \int d^4 x \sqrt{-g} \,\, a_{ij} \, i \,\overline{\n_L}^{\,i} \,\Bigl( 2 R_{\m\n} \c^\m D^\n \,+\, 
\frac{1}{2} \partial_\m R \,\c^\m \Bigr) \n_L^j  \ ,  \nonumber\\
S_b~&=~ \int d^4 x \sqrt{-g} \,\,  b_{ij} \, \partial_\m R \, \overline{\n_L}^{\, i} \c^\m \n_L^j   \ , \nonumber \\
S_c~&=~ \int d^4 x \sqrt{-g} \,\,  c_{ij} \, i \,\overline{\n_L}^{\, i} \, \Bigl( 2 R \c . D \,+\, 
\partial_\m R \,\c^\m \Bigr) \n_L^j  \ ,  \nonumber\\
S_d~&=~ \int d^4 x \sqrt{-g}\, \,  d_{ij} \, i \,\overline{\n_L}^{\, i} \,\Bigl( 2 D^2 \c.D \,+\, \frac{1}{4}
\partial_\m R \, \c^\m \Bigr) \n_L^j  \ .
\label{c3}
\end{align}
To relate $S_d$ and $S_{\tilde{d}}$ we use the identities,
\begin{equation}
\left[D_\m\, ,\,D_\n\right]\n_L ~=~ \frac{1}{4} R_{\m\n\r\s} \c^\r \c^\s\,\n_L \ ,
\label{c3a}
\end{equation}
and 
\begin{equation}
R_{\m\n\r\s} \c^\n \c^\r \c^\s ~=~ R_{\m\n\r\s} \left( g^{\r\s} \c^\n +  g^{\n\r} \c^\s -g^{\n\s} \c^\r
- i \e^{\l\n\r\s}\c_\l \c^5 \right) ~=~ - 2 R_{\m\n}\c^\n \ .
 \label{c3b}
\end{equation}
Then $S_{\tilde{d}} = - S_d - \tfrac{1}{2} S_a$ and we have $\,\tilde{a} = a-\tfrac{1}{2} d$, $\, \tilde{d} = -d$.

Note from (\ref{c3}) that $S_c$ and $S_d$ are equivalent up to equation of motion
operators. It follows that {\it on-shell} quantities derived from this effective action will only depend
on the combination of coefficients $d+4c$. We see below how this consistency requirement appears
in the physical results derived from (\ref{c3}), notably in the coupling $\hat{b}$ introduced 
in section \ref{sect 4.1}.

\begin{figure}[h!]
\centering\includegraphics[scale=0.55]{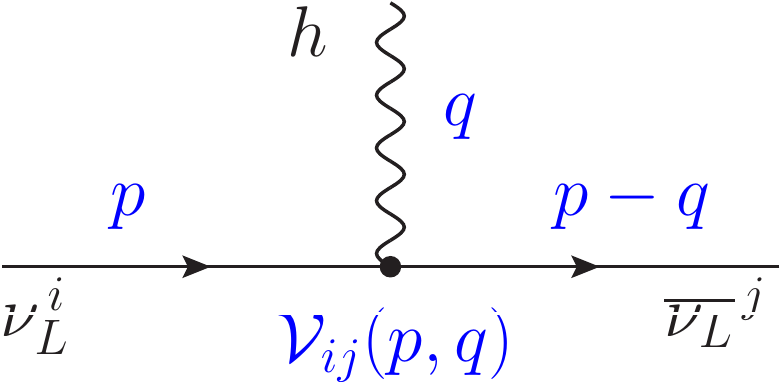}
\caption{The $h\,\overline{\n_L}^{\, i}\, \n_L^j$ vertex $\mathcal{V}_{ij}(p,q)$ in the effective Lagrangian 
(\ref{c1}), (\ref{c2}).}
\label{Fig 3.1}
\end{figure}
As discussed in section \ref{sect 2}, the simplest approach to determine the couplings in the
effective Lagrangian is to compare with the fundamental theory in the weak gravitational field
approximation. Using the weak field expansion of the curvatures in (\ref{b19a}), we can show
that the operators in (\ref{c3}) give rise to an effective $h\,\overline{\n_L}^{\, i} \, \n_L^j$ vertex
$\mathcal{V}_{ij}$ as shown in Fig.~\ref{Fig 3.1} which, after some calculation \cite{McDonald:2014yfg}, 
can be expressed in momentum space as
\begin{align}
\mathcal{V}_{ij}(p,q) ~&= i \biggl[\Bigl(2 d\, p^2 - 2 d\, p.q + (a + 6c + \tfrac{5}{2} d) \,q^2 \Bigr) \slashed{p} 
\nonumber \\
&~~~~~~~~+\, \Bigl( - d\, p^2 + 2a \,p.q + (-\tfrac{3}{2} a - 3 c - \tfrac{3}{4} d + 3 i b)\,q^2 \Bigr) \slashed{q} 
\,\biggr]_{ij} \ .
\label{c4}
\end{align}

This vertex is subject to an important constraint
from unitarity. Note first, suppressing flavour labels, that $\mathcal{V}(p,q) = i \mathcal{M}(p,q)$,
where
\begin{equation*}
\mathcal{M}(p,q) ~=~ \langle \n_L(p) \,|\, \n_L(p-q) \, h(q) \rangle \ .
\end{equation*}
This amplitude is required to satisfy the unitarity relation,
\begin{equation*}
\mathcal{M}(p,q)^* ~=~ \mathcal{M}(p-q,-q) \ .
\end{equation*} 
Writing the general expression,
\begin{equation*}
\mathcal{M}(p,q) ~=~ \Bigl[ \left(\a\,p^2 \,+\, \b\,p.q\,+\, \c\,q^2\right)\,\slashed{p} \,+\,
\left(\d\,p^2 \,+\, \epsilon\,p.q\,+\, \phi\,q^2\right)\,\slashed{q} \Bigr] \ ,
\end{equation*}
this implies the following relations (assuming all but $\phi$ are purely real),
\begin{equation*}
\b\,=\,-\a \ , ~~~~~~ \d \,=\, -\tfrac{1}{2} \,\a  \ , ~~~~~~ 
\Re \,\phi \,=\, \tfrac{1}{4} \a - \tfrac{1}{2}\c - \tfrac{1}{2}\epsilon  \ ,
\end{equation*}
which are satisfied by the expansion (\ref{c4}). 

It also satisfies the on-shell requirement following from the equivalence of $S_c$ and $S_d$ 
up to equation of motion operators. Setting $p^2 = (p-q)^2 = 0$, so that $2p.q = q^2$ in 
(\ref{c4}), we see that the remaining terms proportional to $q^2\slashed{p}$ and 
$q^2\slashed{q}$ are indeed proportional only to the combination $d+4c$ of the $S_c$ and
$S_d$ coefficients. 

\subsection{Matching conditions}\label{sect 3.3}

We can now determine the effective Lagrangian coefficients for the fundamental BSM theory 
in section \ref{sect 2} by comparing the effective vertex $\mathcal{V}_{ij}(p,q)$ in (\ref{c4})
with the explicit Feynman diagram results in (\ref{b30}) and (\ref{b31}).
With the identification,
\begin{equation}
\mathcal{V}_{ij}(p,q) ~=~ \S_{ij}(p,p-q) \ ,
\label{c5}
\end{equation}
we can read off the following values of the coefficients:
\begin{equation}
\bigl(a_{ij}, ~c_{ij},~d_{ij} \bigr) ~=~ \frac{1}{12} \frac{1}{(4\pi)^2} \,
\sum_\a \l_{\a j}^\dagger \l_{i\a} \, \frac{1}{M_\a^2} \,
\Bigl(-\frac{4}{3},~\frac{3}{4},\,-3\Bigr) \, \ .
\label{c6}
\end{equation}
It is easily checked that with these identifications, (\ref{b30}) satisfies the above unitarity conditions 
for $\mathcal{V}_{ij}$. Two features of these coefficients deserve comment. 
First, note that $d=-4c$. As we see below, this
will imply the cancellation of these coefficients from the light neutrino equation of motion
in the effective theory. Remarkably, precisely the same relation is found from the light neutrino 
self-energies in the standard model, as shown in \cite{McDonald:2014yfg}.
Second, again as in the standard model, we find the sign of the coefficient $a$ is negative.
This translates directly into the evolution of lepton number in the cosmological models.

As already noted, there is no contribution to the coefficient $b_{ij}$ of the CP violating operator 
at one loop. From (\ref{b31}) we find its leading-order contribution,
\begin{equation}
b_{ij} ~=~ \frac{1}{9} \frac{1}{(4\pi)^4} \,\sum_{\a,\b,k} \,\l_{\b j}^\dagger \l_{i\a} \l_{\b k}^\dagger \l_{k\a}\,
\frac{1}{M_\a M_\b} \, I_{[\a\b]} \ ,
\label{c7}
\end{equation}
with $I_{[\a\b]}$ in (\ref{b32}).

This completes the identification of the coefficients in (\ref{c1}) for the specific BSM theory
described in section \ref{sect 2}. In the following sections, we develop the theory of gravitational
leptogenesis based on the general effective Lagrangian.

\section{Lepton Number Evolution and the Eikonal Expansion}\label{sect 4}

We now come to the key theoretical development in this paper, the investigation of gravitational leptogenesis 
in the radiatively-induced effective Lagrangian (\ref{c1}). We do this in two complementary ways --  first,
from a detailed analysis of the eikonal expansion of the loop-corrected light neutrino equation of
motion, then, in the following section, using operator methods and non-conservation of the lepton
number current.

The four operators in $S_{eff}$ play different roles in leptogenesis. The $b$ operator 
$\partial_\m R \, \overline{\n_L} \c^\m \n_L$ acts as a chemical potential for lepton number and
allows a lepton-antilepton asymmetry, {\it i.e.}~a net lepton number, to arise in thermal (quasi-)equilibrium. 
The remaining operators modify the evolution of the lepton number in time, giving a dynamical
amplification or damping depending on the signs of the coefficients $a,\, c,\, d$.

\subsection{Eikonal expansion for the light neutrinos}\label{sect 4.1}

The propagation of the light neutrinos in a background gravitational field is governed by the equation 
of motion derived from the effective Lagrangian.\footnote{For clarity, we present the formal developments 
in sections \ref{sect 4} and \ref{sect 5} for a single light neutrino flavour. Alternatively, the formulae
given here may be viewed as matrix expressions with the flavour indices suppressed in the coefficients
$a_{ij}$ {\it etc.} and currents $J_{ij}^\m = \overline{\n_L}^i \c^\m \n_L^j$. We have also taken the light 
neutrino mass to zero in this subsection. }
This is written most easily with the parametrisation in (\ref{c3}) and we find
\begin{equation}
\left[\left(1 + 2 c R + 2d D^2\right) i \c.D \,+\, 2a R_{\m\n} i \c^\m D^\n \,+\, 
(\hat{b} - ib)\, i \c.\partial R \right] \n_L \,=\,0 \ ,
\label{d1}
\end{equation}
where $\hat{b} = \tfrac{1}{2} a + c + \tfrac{1}{4}d$. 

We are viewing $S_{eff}$ as being generated by radiative 
corrections, so we work to consistent perturbative order. Then, since $\c.D \n_L = O(\l^2)$, to this order
we must omit the pre-factor $\left(2 c R + d D^2\right)$ which gives terms of $O(\l^4)$ which we have
not computed in the other coefficients. The perturbatively consistent equation of motion therefore reduces 
to 
\begin{equation}
\left[ i \c.D \,+\, 2a R_{\m\n} i \c^\m D^\n \,+\, (\hat{b} - i b) i \c.\partial R \right] \n_L \,=\,0 \ .
\label{d2}
\end{equation}

The eikonal expansion consists of writing the field as the product of a slowly-varying amplitude $\mathcal{A}$
and a rapidly varying phase $\Theta$. In order to keep track of the relative orders in the eikonal expansion,
we temporarily introduce a formal counting parameter $\e$ and write the ansatz
\begin{equation}
\n_L \,=\, \mathcal{A}\, u_L\, e^{-i \Theta} ~=~ \left(A - i \e B + \ldots\right) \, u_L\, 
e^{-\frac{i}{\e}\left(\theta + \e \a\right)} \ ,
\label{d3}
\end{equation}
where $u_L$ is the appropriate spinor wave function. The wave vector is defined as $p_\m = \partial_\m \Theta$.

The inclusion of an $O(\e)$ correction in the phase $\Theta$ is novel and is required to accomodate the CP
violating $b$ operator. As explained in reference \cite{McDonald:2014yfg}, this operator induces a 
phase modulation of the wave solution, which in the particle interpretation produces a linear shift 
in the energy in the dispersion relation, with opposite sign for the neutrinos and antineutrinos.
This difference in the propagation of particles and antiparticles in a gravitational field is the origin
of gravitational leptogenesis.

The simplest route to a complete solution of the eikonal equation is to first act on (\ref{d2}) with
$\c.D$ to remove the explicit dependence on the gamma matrices and produce a second-order wave equation.
After some manipulation of covariant derivatives acting on spinors, using the identities 
(\ref{c3a}) and (\ref{c3b}), and again keeping terms to consistent perturbative order only, 
we eventually find\footnote{To keep the notation in this section compact, our convention here is that within 
the square parentheses, derivatives immediately to the left of a curvature (or $k,$ $\a$) act only on that term, 
while `free' derivatives act on everything outside the square parentheses.}
\begin{align}
&\c.D \left( \c.D \,+\, 2a R_{\m\n} \c^\m D^\n \,+\, (\hat{b} - i b) \c.\partial R \right) \n_L 
\nonumber \\
&=~ \biggl[ D^2 - \frac{1}{4} R + 2a \left( 2 R_{\m\n}\c^\m D^\n + \frac{1}{2} \partial_\m R D^\m 
- \frac{i}{2} D_\m R^{\m}{}_{\n\r\s} \s^{\r\s} D^\n - \frac{1}{2} R_{\m\n} R^{\m\n} \right) \nonumber \\
&~~~~~~\,+\,\left(\hat{b} - i b\right) \left(2 \partial_\m R D^\m \,+\, D^2 R \right) \,\biggr] \, \n_L ~=~ 0 
\label{d4}
\end{align}
Inserting the ansatz (\ref{d3}) and collecting terms of the same order in $\e$ then gives,
at $O(1/\e^2)$,
\begin{equation}
k^2 \,+\, 4a R_{\m\n} k^\m k^\n ~=~0 \ ,
\label{d5}
\end{equation}
and at $O(1/\e)$,
\begin{align}
&\biggl[ k.D \,+\, \frac{1}{2} D.k \,-\, i k^\m \partial_\m \a \,-\, 4 i a R_{\m\n} k^\m \partial_\m \alpha 
\,+\, (\hat{b} - i b) k.\partial R \nonumber \\
&~~~~+ 2a \Bigl( 2 R_{\m\n} \bigl(k^\m D^\n + \frac{1}{2} D^\m k^\n\bigr) + \frac{1}{4} k.\partial R 
- \frac{i}{4} k^\n D^\m R_{\m\n\r\s} \s^{\r\s} \Bigr)  \biggr]\, \mathcal{A} \,u_L ~=~0 \ ,
\label{d6}
\end{align}
where we have defined $k_\m = \partial_\m \theta$.  Higher orders in $\e$ give sub-leading corrections
to the amplitude which do not concern us here.

The first equation determines the one-loop curvature induced modification to the dispersion relation
when re-expressed in terms of the true wave vector $p_\m = k_\m + \e \partial_\m \a$.  For the second,
note that the imaginary terms cancel at $O(\l^2)$ with the choice $\a = - b R$. In fact, the entire dependence
of the wave operator on the coefficient $b$ is accounted for by this addition to the phase factor.

Now, in the absence of the loop effects, the dispersion relation from (\ref{d5}) is simply $k^2=0$,
from which we deduce $k.D\, k^\n = 0$. This is just the geodesic equation, showing that the tangent
vector $k^\m$ is parallel transported along its integral curve. The $O(1/\e)$ terms may be separated 
into equations for the amplitude and spinor wave function, {\it viz.}
\begin{align}
\Bigl( k.D + \frac{1}{2} D.k \Bigr) \, \mathcal{A} \,=\, 0 \nonumber \\
k.D\, u_L \,=\, 0 \ .
\label{d7}
\end{align}
This states that the wave function $u_L$ is parallel transported along the geodesic (see
\cite{Audretsch:1981wf} for a detailed discussion), whereas the amplitude
varies as the `expansion' $\hat{\theta} = - \tfrac{1}{2} D.k$ of the geodesic congruence, one of
the optical scalars in the Raychoudhuri equations. This has the clear interpretation of the amplitude increasing 
as the congruence focuses. In the particle interpretation of the eikonal formalism, the number density of
particles comprising the wave is proportional to the square of the amplitude, so the light neutrino
number density varies as $n_\n \sim \mathcal{A}^2$. 

Putting all this together, we can finally write the eikonal solution to the loop-corrected equation of motion
in the form,
\begin{equation}
\n_L ~=~ \mathcal{A}\, u_L\, e^{-i (\theta - bR)} \ ,
\label{d8}
\end{equation}
where the dispersion relation is given in (\ref{d5}),
while the amplitude and spinor wave function satisfy
\begin{align}
&\left[ k.D + \frac{1}{2} D.k + 
4a R_{\m\n}\Bigl(k^\m D^\n + \frac{1}{2}D^\m k^\n\Bigr) + \Bigl(\hat{b} + \frac{1}{2}a\Bigr) k.\partial R 
\right]\, \mathcal{A} ~=~0 
\label{d9}  \\
&\left[k.D + 4a R_{\m\n}k^\m D^\n - \frac{i}{2} a \,k^\n D^\m R_{\m\n\r\s} \s^{\r\s} \right]\, u_L ~=~0 \ .
\label{d10}
\end{align}
Neither $k^\m$ nor $p^\m$ satisfies the geodesic equation with the loop effects included. 
The evolution equations for both the amplitude and the wave function are also modified
as shown. The interpretation of the tidal effects on the amplitude is discussed below. 
Tidal effects on the spinor wave function, which depend on the spin generator $\s^{\r\s}$,
modify the neutrino helicity as it propagates through spacetime. These potentially important effects, 
which parallel those for polarisation and birefringence in photon propagation, do not however affect 
our discussion here of the evolution of lepton number. 

In terms of the true wave vector $p_\m= \partial_\m(\theta - bR) \,=\, k_\m - b\,\partial_\m R$, 
and including a light neutrino mass, 
the dispersion relation is
\begin{equation}
p^2 \,-\, 4a R_{\m\n} p^\m p^\n \,\pm\, 2b\, p^\m \partial_\m R \,-\, m^2 ~=~ 0 \ .
\label{d11}
\end{equation}
This is the form appropriate for an interpretation in terms of energy and momentum,
with the $\pm$ sign for the neutrino $\n_L$ and antineutrino $\n_L^{\,\,c}$ respectively.

\subsection{Interpretation in FRW spacetime}

We now show how this eikonal solution may be interpreted in terms of the evolution of lepton
number in a FRW spacetime.

The spatially flat Robertson-Walker metric,
\begin{equation}
ds^2\,=\, dt^2 - a(t)^2 \,\d_{ij}  dx^i dx^j \ ,
\label{dd1}
\end{equation}
has non-vanishing Christoffel symbols,
\begin{equation}
\Gamma^0_{ij} \,=\, - H \d_{ij} \ , ~~~~~~~~~ 
\Gamma^i_{j0} \,=\, H \d^i{}_j \ ,
\label{dd2}
\end{equation}
where $H$ is the Hubble parameter.
The FRW Ricci curvatures are
\begin{align}
&R_{00} \,=\, 4\pi G \r (1 + 3w)      \ , ~~~~~~~~ R_{0i} \,=\, 0 \ , ~~~~~~~~ 
R_{ij} \,=\, -4\pi G\r (1-w) \d_{ij}\ , \nonumber \\
&R \,=\, -8\pi G\r (1-3w) \ ,
\label{dd3}
\end{align}
where $\r$ is the energy density and $w$ is the equation of state parameter, $p=w\r$. 
The energy conservation equation,
\begin{equation}
\dot{\r} \,+\, 3(1+w) \r H \,=\, 0 \ ,
\label{dd4}
\end{equation}
then implies
\begin{equation}
\dot{R} \,=\, 8\pi G \r\,3H(1-3w)(1+w) \ .
\label{dd5}
\end{equation}
Also note the Friedmann equation gives the Hubble parameter from
\begin{equation}
3 H^2 \,=\, \r/M_p^2 \ .
\label{dd6}
\end{equation}

The geodesic equations for a massive free particle in this metric are simply solved
by $x^i \,=\, {\rm constant}$, {\it i.e.}~the particle is co-moving with the cosmological expansion.
Letting $k^\m$ be the corresponding momentum, and with dispersion relation $k^2 = m^2$,
this implies $k^0 = m$, $k^i = 0$.

The corresponding solution of the loop-corrected dispersion relation (\ref{d5}) at $O(\l^2)$ is
\begin{equation}
k^0 \,=\, \bigl(1-2aR_{00}\bigr) m \ , ~~~~~~~~k^i\,=\, 0 \ .
\label{d12}
\end{equation}

Along this trajectory\footnote{This is not in general the tangent vector to a geodesic since, 
writing $K_\m = k_\m + 2a R_{\m\n}k^\n$ so the dispersion relation is simply $K^2 = m^2$, 
we can show
\begin{equation*}
K.D K^\r \,=\, 2a\,\left[ \bigl( D_\m R^\r{}_\n - D^\r R_{\m\n} \bigr) K^\m K^\n \,-\,
\frac{1}{2} R_{\m\n} D^\r \bigl(K^\m K^\n\bigr) \right] \,+\, O(\l^4) \ .
\end{equation*}
However, in the FRW metric and taking $K^0 = m, K^i =0$ as in (\ref{d12}), this does simplify to 
\begin{equation*}
K.D \,K^\r\,=\, 2a \left[ \partial_0 R^\r{}_0 - \partial^\r R_{00} \right] \,=\, 0,    ~~~~~~(\r = 0,i)\ .
\end{equation*} 
Also note that in terms of $K^\m$, equation (\ref{d9}) for the amplitude becomes
\begin{equation*}
\left[ K.D \,+\, \frac{1}{2} D.K \,+\, 2a R_{\m\n} K^\m D^\n \,+\, a R_{\m\n} D^\m K^\n \,+\, 
\hat{b} K.\partial R \right]\,\mathcal{A} \,=\,0 \ ,
\end{equation*}
where we have made use of the Bianchi identity $D^\m R_{\m\n} = \tfrac{1}{2} \partial_\n R$.
This form is especially close to the current non-conservation equation derived in section
\ref{sect 5}. \label{fn 4.1}}
the amplitude equation (\ref{d9}) evaluates as
\begin{equation}
\left[\partial_0 \,+\, \frac{1}{2} \Gamma^i_{i0} \,+\, 2a R_{00} \,\partial_0 \,+\, a R_{ij} \Gamma^j_{i0} 
\,+\, \hat{b}\, \partial_0R \right]\,\mathcal{A} ~=~0 \ .
\label{d13}
\end{equation}
The time evolution of the amplitude is therefore
\begin{equation}
\left[\frac{d}{dt} \,+\, \frac{3}{2} H \,+\, 2a R_{00} \frac{d}{dt} \,+\, a R^i{}_i H \,+\, 
\hat{b}\, \dot{R} \right]\, \mathcal{A} ~=~ 0 \ .
\label{d14}
\end{equation}

Recalling that in the eikonal formalism, the particle number density $n \sim \mathcal{A}^2$, we 
therefore find the time evolution of the light neutrino number density $n_\n$ in a FRW spacetime 
is given by
\begin{equation}
\bigl(1 + 2a \,R_{00} \bigr) \, \frac{dn_\n}{dt} \,+\, 3H n_\n \,+\, 2a\,R^i{}_i H n_\n \,+\, 
2 \hat{b} \,\dot{R} \,n_\n ~=~ 0 \ .
\label{d15}
\end{equation}
This is one of the key equations in this paper.
It is the first step in establishing the new Boltzmann equation for gravitational leptogenesis
incorporating the radiatively-induced curvature effects. In the following section, we first show
how this equation may be found using a quite different approach using non-conservation of 
the lepton number current, then develop the full Boltzmann equation including the non-vanishing
equilibrium distribution of lepton number induced by the CP violating $b$ operator.
Note that the $b$ coefficient does not appear in the lepton number evolution equation,
entering only in the modified dispersion relation (\ref{d11}).

\section{Gravitational Leptogenesis and the Extended Boltzmann Equation}\label{sect 5}

In this section, we provide an independent derivation of the evolution equation (\ref{d15}) for the
lepton number density by showing that the lepton number current is not conserved when
the CP even curvature interactions are included in the effective action. We then formulate a new
generalised Boltzmann equation incorporating these gravitational effects.

\subsection{Current non-conservation and lepton number evolution}\label{sect 5.1}

For the free Dirac Lagrangian, the lepton number current $J^\m =\overline{\n_L} \c^\m \n_L$ is conserved,
{\it i.e.} $D_\m J^\m \sim 0$ (where $\sim 0$ indicates zero up to terms vanishing by the equation of 
motion). In the presence of the curvature terms in the effective Lagrangian, however, the current is
no longer conserved, implying a non-trivial time dependence of the lepton number. 

The current non-conservation identity is derived using standard methods. Under a variation 
$\n_L \rta e^{i\theta} \n_L$, ~$\overline{\n_L} \rta e^{-i\theta}\,\overline{\n_L}$, the effective action
transforms as 
\begin{align}
\d S_{eff} \,&\equiv\, \d \overline{\n_L} \,\frac{\d S_{eff}}{\d \overline{\n_L}} \,+\, 
\frac{\d S_{eff}}{\d \n_L} \,\d \n_L \nonumber \\
&=\, D_\m J^\m \,+\, \D ~~\sim~0 \ ,
\label{e1}
\end{align}
where the total derivative term defines the current $J^\m$, while for a non-conserved current the 
remainder $\D$ is non-zero. Taking the terms in the effective action (\ref{c3}) in turn, we find
\begin{align}
\d S_a \,&=\, 2a \, R_{\m\n}D^\m J^\n \,+\, a\,  \partial_\m R J^\m  \ , \nonumber \\
\d S_b \,&=\, 0 \ , \nonumber \\
\d S_c \,&=\, 2c\, R D_\m J^\m \,+\, 2c \,\partial_\m R J^\m \ , \nonumber \\
\d S_d \,&=\, \frac{1}{2} d\, \partial_\m R J^\m \,+\, 2d\left( \overline{\n_L} D^2 \c.D \n_L
\,+\, \overline{\n_L} \c.\overleftarrow{D} \overleftarrow{D^2} \n_L \right) \ .
\label{e2}
\end{align}
Collecting terms, the current non-conservation equation is therefore
\begin{align}
(1+2cR) D_\m J^\m \,+\, 2a R_{\m\n} D^\m J^\n \,&+\, \Bigl(a + 2c + \frac{1}{2}d\Bigr) \partial_\m R J^\m 
\nonumber \\
&+\, 2d\left( \overline{\n_L} D^2 \c.D \n_L
\,+\, \overline{\n_L} \c.\overleftarrow{D} \overleftarrow{D^2} \n_L \right)  ~~\sim ~0 \ .
\label{e3}
\end{align}

Again, since $D_\m J^\m \sim O(\l^2)$, the pre-factor $2cR$ of the $D_\m J^\m$ term must be omitted 
to consistent perturbative order. So, up to terms vanishing by the equation of motion, we find
simply,
\begin{align}
D_\m J^\m \,+\, 2a\,R_{\m\n} D^\m J^\n \,+\, 2 \hat{b} \, \partial_\m R J^\m ~\sim~ 0 \ ,
\label{e4}
\end{align}
recalling $\hat{b} = \tfrac{1}{2} a + c + \tfrac{1}{4}d$. Note here the similarity with the eikonal 
amplitude equation, especially the form in footnote {\ref{fn 4.1}}.
Note also that (\ref{e3}), (\ref{e4}) may be derived directly using the equations of motion (\ref{d1}),
(\ref{d2}).

Now in the FRW spacetime, using the formulae (\ref{dd2}), (\ref{dd3}), this current non-conservation
equation becomes
\begin{equation}
\partial_0 J^0 \,+\, \partial_i J^i \,+\, \Gamma^i_{i0} J^0 \,+\,
2a \left( R^0{}_0 \,\partial_0 J^0 \,+\, R^i{}_j \partial_i J^j \,+\, R^i{}_j \Gamma^j_{i0} J^0 \right)
\,+\, 2 \hat{b} \, \dot{R} J^0 ~\sim~0 \ .
\label{e5}
\end{equation}
The charge density corresponding to the current is $J^0$, which we identify as the lepton number 
density $n_L$. In an isotropic universe, the spatial gradients in (\ref{e5}) vanish. 
We therefore find the time evolution equation for the lepton number density,
\begin{equation}
(1 + 2a R_{00}) \frac{dn_L}{dt} \,+\, 3H n_L \,+\, 2a R^i{}_i H n_L \,+\, 2 \hat{b}\,\dot{R}\, n_L \,=\, 0 \ .
\label{e6}
\end{equation}
This reproduces (\ref{d15}), derived using the eikonal formalism, with $n_L = n_\n - n_{\n^{c}}$.

To develop this, noting that $dn_L/dt = - 3 H n_L + O(\l^2)$, we can rewrite the pre-factor term in 
(\ref{e6}) so that to $O(\l^2)$,
\begin{equation}
\frac{dn_L}{dt} \,+\, 3 H n_L \,+\, 2a \bigl(-3 R_{00} + R^i{}_i\bigr) H n_L \,+
\, 2\hat{b} \,\dot{R}\, n_L \,=\, 0 \ .
\label{e7}
\end{equation}
Substituting for the curvatures finally gives,
\begin{equation}
\frac{dn_L}{dt} \,+\, 3 H n_L \left( 1 \,-\, 8\pi G \r (1+w) 
\big[2a - 2 \hat{b} (1-3w)\bigr] \right) \,=\, 0 \ .
\label{e8}
\end{equation}

Whether the radiative curvature corrections amplify or reduce the lepton number density as the 
universe evolves therefore depends on the sign of the combination of coefficients
$\big[2a - 2 \hat{b} (1-3w)\bigr]$.

For a radiation dominated FRW universe, $w \simeq 1/3$, with a deviation arising purely from the beta
functions characterising the trace anomaly in the energy-momentum tensor, $T^\m{}_\m \neq 0$. 
With standard model fields, this gives $(1-3w)\simeq 0.1$.
Since this is small, the dominant factor in (\ref{e8}) is the coefficient $2a$. A negative $a$ corresponds
to damping of the lepton number with time, while a positive $a$ implies an amplification.
Recall that in the particular BSM model described in sections \ref{sect 2} and \ref{sect 3},
we found in (\ref{c6}) that $a$ is negative.

Other cosmological scenarios with $(1-3w) \neq 0$ are considered in section \ref{sect 6}.
At this point, however, note that in the BSM model we have $d = -4c$, so the coefficient 
$\hat{b} = a/2$. So in this particular model, the combination $\big[2a - 2 \hat{b} (1-3w)\bigr] 
= a (1+3w)$.  Curiously, this is negative unless $(1+3w) < 0$, which is precisely the condition
for an accelerating, or inflationary, universe.

We return to (\ref{e8}) in section \ref{sect 5.3} below, where we incorporate it into the new Boltzmann
equation for lepton number. First, we consider the CP violating $b$ operator and the
mechanism for leptogenesis.

\subsection{Gravitational leptogenesis}\label{sect 5.2}

The CP violating operator in the effective Lagrangian (\ref{c1}) may be written in FRW spacetime
as
\begin{align}
S_b \,&=\, b \int d^4 x \sqrt{-g} \, \partial_\m R\, \overline{\n_L} \c^\m \n_L  \nonumber \\
&=\, \int dt \, b \dot{R} \int d^3 x \sqrt{-g^{(3)}}\, J^0 \ ,
\label{e9}
\end{align}
since $R = R(t)$. In this form, given that $J^0$ is the lepton number density $n_\n$, 
we see that this may be interpreted as introducing a chemical potential $\m = b \dot{R}$ 
for lepton number.

Since the neutrinos interact via the Higgs field with the finite temperature medium in the early
universe, this chemical potential biases the net lepton number $n_L = n_\n - n_{\n^c}$ to produce a 
non-vanishing value $n_L^{eq}$ in thermal equilibrium, with $n_L^{eq} \sim \m T^2$.

The equilibrium is maintained by the lepton number violating reactions described in section 
\ref{sect 2.1}. As well as the $\D L = 2$ reactions $\n_L H \leftrightarrow \n_L^{\,c} H$ and 
$\n_L \n_L\leftrightarrow HH$ considered there, there are further $\D L = 1$ reactions 
involving other standard model fields which are described in detail elsewhere 
(see for example \cite{Buchmuller:2004nz}.

Of course, since we need the Ricci scalar to be varying in time to produce the chemical potential,
we are not in true thermal equilibrium. However, provided the $\D L \neq 0$ reactions are faster
than the rate of change $\dot{R}$, the neutrinos and antineutrinos are in quasi-equilibrium
and the formalism here is an accurate description. As discussed in section 2, the time-dependence 
inherent in $R(t)$ is necessary to satisfy the third Sakharov condition and allow leptogenesis 
to occur.

To see this in detail, recall the dispersion relation (\ref{d11}). Rewriting in components, 
this reads
\begin{equation}
(1 - 4a R_{00})\, (p^0)^2 \,-\, \bigl(1 + \tfrac{4}{3} a R^i{}_i\bigr) |{\bf p}|^2 \, \pm\,
2 b\, p^0 \dot{R} \,-\, m^2 \,=\,0 \ ,
\label{e10}
\end{equation}
and identifying energy as $E = p^0$, we find
\begin{equation}
E\,=\, \mathcal{E}(|{\bf p}|) \,\mp\, \m \ ,
\label{ee10}
\end{equation}
with $\m = b \dot{R}$ and, at $O(\l^2)$,
\begin{equation}
\mathcal{E}(|{\bf p}|) \,=\, \left[\left(1 + 4a\left(R_{00} + \tfrac{1}{3} R^i{}_i\right)\right) |{\bf p}|^2
\,+\, \bigl(1 + 4a R_{00}\bigr) m^2 \right]^{1/2} \ .
\label{e11}
\end{equation}

From standard statistical mechanics, the equilibrium lepton number density is then given,
for small $\m/T$, by 
\begin{align}
n_L^{eq} \,=\, n_\n\,-\,n_{\n^c} 
\,&=\, \int \frac{d^3|{\bf p}|}{(2\pi)^3}\,\left(\frac{1}{e^{(\mathcal{E} -\m)/T} + 1} \,-\, 
\frac{1}{e^{(\mathcal{E}+\m)/T} + 1} \right) \nonumber \\
&{} \nonumber  \\
&\simeq \, \frac{2\m}{T}\, \frac{1}{2\pi^2} \, \int_0^\infty d|{\bf p}| \,
\frac{|{\bf p}|^2\, e^{\mathcal{E}}}{\left(e^{\mathcal{E}} + 1\right)^2} ~+~ O\left(\frac{\m}{T}\right)^2 \ .
\label{e12}
\end{align}
Since this is already $O(\l^4)$, we may neglect the $O(a)$ corrections to $\mathcal{E}(|{\bf p}|)$
in the integral in (\ref{e12}), and find to a good approximation,
\begin{equation}
n_L^{eq} \,\simeq\,  \frac{2\m}{T}\,\frac{1}{2\pi^2}\, T^3 \int_0^\infty dx\, 
\frac{x^2\, e^x}{\left(e^x + 1\right)^2} ~~=~~ \frac{1}{3}\,\m\, T^2 \ .
\label{e13}
\end{equation}

This may be compared to the photon number density $n_\c = \frac{2\zeta(3)}{\pi^2} T^3$ or 
the entropy density $s = \frac{2\pi^2}{45} g_{*s}T^3$,
where $g_{*s}$ is the effective number of degrees of freedom at the energy scale $T$,
both of which are commonly used to normalise the lepton asymmetry.

To summarise, we have shown that radiative corrections in the BSM model of section 2 induce 
a non-vanishing lepton number density asymmetry in thermal equilibrium, given by
\begin{equation}
n_L^{eq} \,=\, \frac{1}{3}\, b\, \dot{R}\, T^2 \ ,
\label{e14}
\end{equation}
with the coefficient $b = {\rm tr}\, b_{ij} = O(\l^4)$ given by (\ref{c7}). 
This is the mechanism of radiatively-induced gravitational leptogenesis, first proposed 
in \cite{McDonald:2015ooa}.

\subsection{Boltzmann equation}\label{sect 5.3}

The final step is to incorporate these two new effects into the Boltzmann equation describing 
the time, or temperature, dependence of the lepton number density.

We begin with the modification to the evolution term (\ref{e8}). It is traditional in discussions 
of leptogenesis to give results in terms of the ratio of lepton to photon number, 
$N_L = n_L/n_\c$. Since $n_\c \sim T^3$ and $T\sim 1/a$, where $a(t)$ is the FRW scale
parameter, we have
\begin{equation}
\frac{dN_L}{dt}\,=\, \frac{1}{n_\c} \left( \frac{dn_L}{dt} \,+\, 3 H n_L\right) \ .
\label{e15}
\end{equation}

We also usually express the evolution in terms of temperature, or more specifically 
$z = M_1/T$, where in our BSM model we can take the mass scale $M_1$ to be the 
mass of the lightest sterile neutrino.  Then, $d/dz = (1/Hz) d/dt$. Recalling that
$8\pi G = 1/M_p^2$ defines the reduced Planck mass, we can finally re-express 
(\ref{e8}) in the form,
\begin{equation}
\frac{dN_L}{dz} \,= 3(1+w) \, \left[2a - 2 \hat{b} (1-3w)\right]\, 
\frac{\r}{M_p^2} \, \frac{1}{z} \,  N_L \ .
\label{e16}
\end{equation}

\vskip0.2cm
Next, consider the modification due to the non-vanishing equilibrium number density $n_L^{eq}$.
For this, we need some kinetic theory. We only sketch the key results we need, referring
to standard texts \cite{Dodelson:2003ft} 
for the general theory. We also need to step back from only using the effective Lagrangian here,
since we need to consider the specific lepton number violating reactions described in section
\ref{sect 2} in the BSM model itself.

For a $\D L = 2$ reaction such as $\n_L H \leftrightarrow \n_L^{\,c} H$, the number density of
neutrinos is given by the Boltzmann equation \cite{Baumann} in FRW spacetime,
\begin{equation}
\frac{dn_\n}{dt} \,+\, 3H n_\n \,=\, \langle \s |v|\rangle \left( - n_\n \,n_H \,+\, 
\frac{n_\n^{eq} \, n_H^{eq}}{n_{\n^c}^{eq}\, n_H^{eq}}\, \,n_{\n^c} \,n_H \right) \ ,
\label{e17}
\end{equation}
with the corresponding result for $dn_{\n^c}/dt$.

The cross-section term above is re-expressed in terms of the thermally-averaged reaction
rate for particle $\n$, {\it viz}. $\Gamma_{\D L=2}^\n$, given by
\begin{align}
\Gamma_{\D L=2}^\n \,&=\, \frac{1}{n_\n^{eq}} \, \c_{\D L=2} \,=\, 
\frac{1}{n_\n^{eq}} \bigl( n_\n^{eq}\, n_H^{eq}\, \langle \s |v|\rangle \bigr) \nonumber \\
&=\, \frac{1}{64} \frac{1}{(2\pi)^3} \frac{1}{T^2} 
\int ds \,\sqrt{s} \, K_1\left(\frac{\sqrt{s}}{T}\right) \, \frac{1}{s} \int_{-s}^0 du\, 
|\mathcal{M}(u,s)|^2 \ ,
\label{e18}
\end{align}
in terms of the $u$-average of the amplitude $\mathcal{M}(u,s)$ for the $\D L=2$ reaction.
(See for example \cite{Buchmuller:2004nz, McDonald:2016ehm} and references therein for 
details and notation.)

We may take $n_H \simeq n_H^{eq}$ and, using Maxwell-Boltzmann statistics for the neutrinos
({\it i.e.} neglecting the `Fermi blocking' effect), approximate
\begin{equation}
\frac{n_\n^{eq}}{n_{\n^c}^{eq}} \, \simeq\, e^{2\m/T} \ ,
\label{e19}
\end{equation}
from (\ref{e12}) due to the non-vanishing chemical potential. It then follows that,
with $n_L = n_\n - n_{\n^c}$,
\begin{align}
\frac{dn_L}{dt} \,+\, 3Hn_L  \,&=\, \Gamma_{\D L=2}^\n \, \Bigl( - \left(1 + e^{-2\m/T}\right) n_\n \,+\, 
\left(1 + e^{2\m/T}\right) n_{\n^c} \Bigr) \nonumber \\
&=\, -2 \Gamma_{\D L=2}^\n \, \left( n_\n - n_{\n^c} \,-\, \frac{\m}{T} \left(n_\n +
n_{\n^c} \right)\right) \ .
\label{e20}
\end{align}

Next, since in the free theory, $n_\n$ and $n_{\n^c}$ take their common equilibrium values,
and referring to (\ref{e12}) for $n_L^{eq}$, we see that to $O(\l^4)$,
\begin{equation}
\frac{\m}{T} \bigl(n_\n + n_{\n^c}\bigr) \,\simeq\, n_L^{eq} \ .
\label{e21}
\end{equation}
We therefore find,
\begin{equation}
\frac{dn_L}{dt} \,+\, 3Hn_L \,=\, -2\, \Gamma_{\D L=2}^\n \,\bigl(n_L \,-\, n_L^{eq}\bigr) \ .
\label{e22}
\end{equation}
Despite the technicalities of its derivation, (\ref{e22}) has a very straightforward physical 
interpretation, {\it viz}.~that provided the $\D L=2$ reactions are active, the lepton number
density $n_L$ is driven towards its equilibrium value $n_L^{eq}$.

Finally, re-expressing in terms of $dN_L/dz$ as before, we find
\begin{equation}
\frac{dN_L}{dz} \,=\, -2\, W_{\D L=2}\, \bigl(N_L\,-\, N_L^{eq}\bigr) \ ,
\label{e23}
\end{equation}
where $W = \Gamma/zH$ is essentially the ratio of the reaction rate to the expansion rate
of the universe. In the full theory, $W_{\D L=2}$ is replaced by the complete rate term 
$W$ incorporating the inverse sterile neutrino decay and $\D L=1$ reaction rates as well
as the $\D L=2$ reactions considered in detail here.

\vskip0.2cm
The full Boltzmann equation is therefore given by combining (\ref{e16}) and (\ref{e23})
and we find
\begin{equation}
\frac{dN_L}{dz} \,=\, - W \, \bigl(N_L\,-\, N_L^{eq}\bigr) \,-\, \mathcal{W}\, N_L \ ,
\label{e24}
\end{equation}
with $\mathcal{W}$ given by
\begin{equation}
\mathcal{W} \,=\, -3(1+w) \, \left[2a - 2 \hat{b} (1-3w)\right]\,
\frac{\r}{M_p^2}\, \frac{1}{z}  \ ,
\label{e25}
\end{equation}
and, using the expression  (\ref{e14}) for $n_L^{eq}$ together with (\ref{dd5}) for the curvature
$\dot{R}$,
\begin{equation}
N_L^{eq} \,=\, \frac{\sqrt{3}\, \pi^2}{2\,\zeta(3)}\, (1-3w)(1+w) \,b\,\,
\frac{\r^{3/2}}{M_p^3}\, \frac{1}{M_1}\, z \ .
\label{e25b}
\end{equation}

This is the key equation of the paper. The new features compared to conventional 
leptogenesis models are the radiatively-induced curvature dependent terms --
the non-vanishing equilibrium asymmetry $N_L^{eq}$ already presented in
\cite{McDonald:2015ooa, McDonald:2015iwt, McDonald:2016ehm},
and the new evolution term $\mathcal{W} N_L$ derived here.

In the next section, we study this equation in detail in different cosmological settings
using parametrisations of the reaction rate $W$ and other couplings expressed 
in terms of neutrino mass parameters in the specific BSM model in section \ref{sect 2}.

Equation (\ref{e24}) is, however, far more general than this particular model
and it is interesting at this point to give a general estimate of the temperature dependence 
of the various terms in a radiation-dominated cosmology, with $w\simeq 1/3$.
The energy density $\r$ in this case is given by $\r = \s T^4$, where
$\s = \pi^2 g_*/30$ with $g_*$ the effective number of degrees of freedom.
Then, recalling that the loop coefficients in (\ref{e25}) are $O(\l^2/M_1^2)$,
while $b$ in (\ref{e25b}) is $O(\l^4/M_1^2)$, 
we find
\begin{equation}
\mathcal{W} \,\sim\, \l^2\, \left(\frac{M_1}{M_p}\right)^2\, \frac{1}{z^5} \ , ~~~~~~~~~~
N_L^{eq} \,\sim\, \l^4\, \left(\frac{M_1}{M_p}\right)^3 \, \frac{1}{z^5} \ .
\label{e27}
\end{equation}
In (\ref{e18}), the $u$-averaged cross-section $\sim s^2/M_1^2$ so, rescaling 
within the integral over $s$, we deduce $\Gamma \sim T^3/M_1^2$. The Friedmann equation
gives the Hubble constant as $H^2 = \r/3M_p^2$, so  here $H \sim T^2/M_p$ and we have
\begin{equation}
W \,=\, \frac{\Gamma}{zH} \,\sim\, \l^4\, \left(\frac{M_1}{M_p}\right)^{-1} \frac{1}{z^2}\ .
\label{e26}
\end{equation}

Note that in this case, both the curvature-dependent
terms $N_L^{eq}$ and $\mathcal{W}$ fall off rapidly as $T^5$ as the temperature falls and
the universe expands. Conversely, they become increasingly important in the 
high-temperature, strong-curvature regime of the early universe.

The lepton number violating rate term $W$ is conventionally known as the `washout' factor,
since in the absence of a non-vanishing $N_L^{eq}$ this would dampen out any 
pre-existing lepton asymmetry. In the gravitational leptogenesis theory described here, 
however, it plays a quite different role, driving the asymmetry towards its equilibrium value 
$N_L^{eq}$. The new evolution term $\mathcal{W}$, on the other hand, can act at early times
either to slow the drive to $N_L^{eq}$ or as a source of amplification of any existing lepton 
asymmetry, depending on the coefficients in the effective Lagrangian. When these are generated 
by loop corrections, the sign of these contributions is not arbitrary but is determined by the
dynamics of the fundamental BSM model. In our particular model, we found $\mathcal{W} > 0$,
but it is not clear whether there is some general principle enforcing this sign.

\begin{figure}[h!]
\vskip0.5cm
\centering{\includegraphics[scale=1.1]{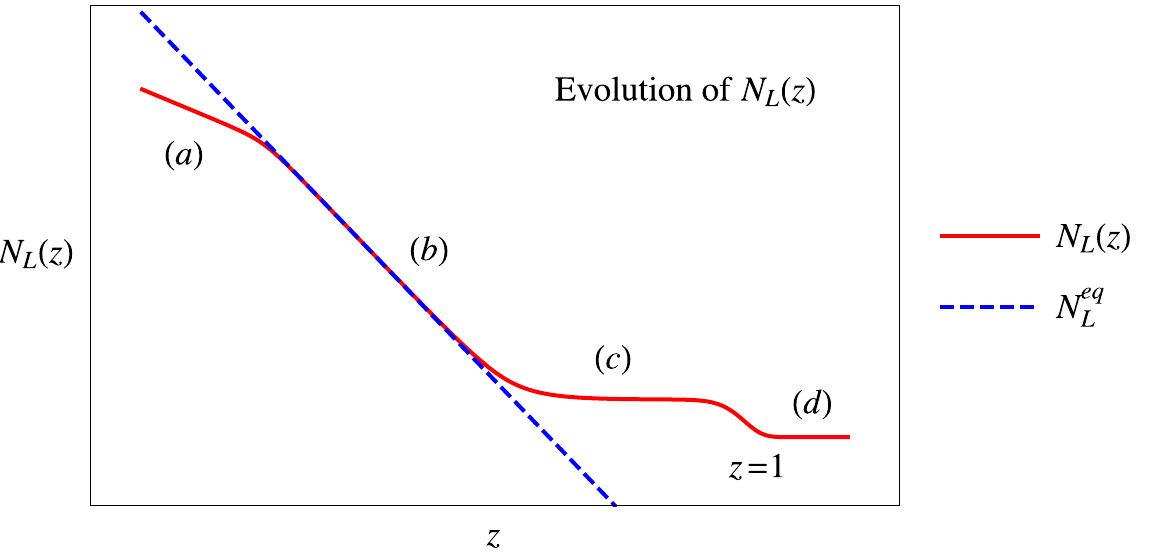}}
\caption{Illustration of the key stages in the evolution of the lepton asymmetry $N_L(z)$
with temperature, given by the Boltzmann equation (\ref{e24}).}
\label{Fig 5.1}
\end{figure}

The full evolution of the lepton asymmetry $N_L$ derived from the Boltzmann equation (\ref{e24})
is illustrated in Fig.~\ref{Fig 5.1}, which extends the picture already found in \cite{McDonald:2016ehm}
to higher temperatures. In the temperature region (b), with $z<1$, the scattering 
factor $W(z)$ dominates and drives $N_L$ to its gravitationally-induced equilibrium value 
$N_L^{eq}$. At still higher temperatures, the $1/z^5$ dependence of $\mathcal{W}(z)$ means
we reach a regime with $|\mathcal{W}(z)| > W(z)$, labelled as (a) in the Figure, where 
$N_L$ is forced away from the equilibrium value, either reducing from $N_L^{eq}$ 
if $\mathcal{W}(z) > 0$ as shown, or increasing very sharply if $\mathcal{W}(z) < 0$. 
For lower temperatures (region (c)), as the universe expands and the temperature falls, 
the lepton number violating rate $W(z)$ eventually becomes too
slow to maintain equilibrium and $N_L$ decouples, becoming essentially constant.
Finally, around $z \simeq 1$, that is $T \simeq M_1$, the rate $W(z)$ has a resonance peak
due to the sterile neutrino propagator in the diagrams of Fig.~\ref{Fig 2.1} and is sufficiently
strong to force $N_L$ back towards $N_L^{eq}$ (region (d)), before once more becoming 
negligibly small leaving $N_L$ to converge to its final, low-temperature, constant
value as shown. 

These features can be quantified using an exact analytic solution for the Boltzmann
equation in the regions (a), (b), (c) for temperatures above the resonance regime 
$T\simeq M_1$. This is given in Appendix \ref{Appendix A}, where we show that in
the ultra-high temperature region (a), the new evolution term induces a dependence
$N_L \sim 1/z^2$ in the lepton asymmetry, moderating the sharp rise $N_L^{eq} \sim 1/z^5$
in the equilibrium value.

The balance of these competing terms in the extended Boltzmann equation (\ref{e25})
depends on the detailed choice of masses and couplings in the original BSM model.
In our case, these are constrained by the light lepton mass spectrum, since the theory
also plays the role of generating neutrino masses via the see-saw mechanism with 
the heavy sterile neutrinos. In the following section, we study in detail how this works out 
in particular cosmological scenarios.

\section{Evolution of Gravitational Leptogenesis in Cosmology}\label{sect 6}

In this final section, we explore the consequences of the generalised Boltzmann equation
(\ref{e24}) for the creation and evolution of the lepton number asymmetry in the early
universe. As in \cite{McDonald:2016ehm}, we consider two scenarios in detail - first, 
when leptogenesis occurs in a
conventional radiation-dominated FRW spacetime and second, in a post-inflationary
era preceding radiation dominance when the expansion is driven by a source with 
effective equation of state $w>1/3$, including the extreme `kination' scenario with $w=1$.

As usual in leptogenesis models, we assume that the lepton asymmetry generated in the 
early, post-inflationary universe is subsequently converted to a baryon asymmetry by
sphaleron processes at the electroweak scale \cite{Klinkhamer:1984di, Khlebnikov:1988sr}.
The required relation for the baryon-to-photon ratio $\eta$ is (see for example
\cite{Buchmuller:2002zs})
\begin{equation}
\eta \,=\, \frac{1}{f}\,\frac{C_{sph}}{C_{sph} -1}\, N_L \ ,
\label{f1}
\end{equation}
where $C_{sph}$ is the fraction of the lepton asymmetry converted into a baryon asymmetry 
by sphaleron processes, and $f$ is a dilution factor accounting for photon production 
between leptogenesis and recombination. In this model, $C_{sph} = (8n+4)/(22n+13)$
with $n=3$ fermion generations and $f=2387/86$, leaving $\eta \simeq 0.02 |N_L|$.
To achieve the observed value of $\eta = 6 \times 10^{-10}$, we therefore require
a leptogenesis mechanism to yield a lepton asymmetry $|N_L| \simeq 10^{-8}$.

\subsection{Radiation-dominated FRW cosmology}\label{sect 6.1}

Our focus in this paper has been to derive and study the two radiatively-induced gravitational
terms in the generalised Boltzmann equation (\ref{e24}). We have developed this in terms of
the effective Lagrangian (\ref{c1}) which describes the dynamics of the light neutrinos in
curved spacetime, including the contributions from loop diagrams with virtual sterile
neutrinos $\n_R^{\,\a}$. To describe the complete dynamics of leptogenesis in the BSM
model (\ref{b1}), however, we also need to take into account the lepton number violating
decays of the sterile neutrinos themselves.

The Boltzmann equations describing leptogenesis through out-of-equilibrium decays of the 
sterile neutrinos are well-known (see \cite{Buchmuller:2004nz} and references therein for reviews), 
this being the original Fukugita-Yanagida model \cite{Fukugita:1986hr}. We can therefore 
easily include these along with our gravitational terms, giving the complete Boltzmann equations,
\begin{align}
\frac{d N_{\n_R}}{d z} \,&=\, - D \left(N_{\n_R} \,-\, N_{\n_R}^{eq} \right) \nonumber \\
\frac{d N_L}{d z} \,&=\, - D \,\varepsilon_1 \left(N_{\n_R} \,-\, N_{\n_R}^{eq} \right)
\,-\, W \left(N_L \,-\, N_L^{eq}\right)  \, - \, \mathcal{W} \, N_L \ .
\label{f2}
\end{align}
Here, $N_{\n_R}$ is the ratio of the number density of the sterile neutrino $\n_R^{\,1}$
to photons (since it is the lightest of the $\n_R^{\,\a}$ that gives the biggest contribution 
to the Boltzmann equation for $N_L$) and $N_{\n_R}^{eq}(z)$ is its equilibrium value
at temperature $T = M_1/z$. The decay parameter is given as
\begin{equation}
D(z) \,=\, \Gamma(\n_R \rightarrow \n_L\,H)\, / z H \ ,
\label{f3}
\end{equation}
and $\varepsilon_1$ is a CP-violating parameter of $O(\l^4)$ characterising the asymmetric
contributions of the $\D L = 1$ decays of the sterile neutrino $\n_R^{\,1}$
to $\n_L\, H$ and $\n_L^{\,c}\, H$. Together with $W$, and unlike $N_L^{eq}$ and 
$\mathcal{W}$, these terms are non-zero in flat spacetime and we can neglect any gravitational 
contributions to them in (\ref{f2}). 

A natural question at this point is why it is consistent to use expressions derived from the 
`low-energy' effective Lagrangian, incorporating the effects of integrating out the virtual
heavy sterile neutrinos in loop diagrams, at an energy scale where the sterile neutrinos
are themselves dynamical. The resolution depends on the extra (curvature) scale in the
gravitational theory.

Conventionally, effective Lagrangians in flat spacetime are valid for momenta such that 
$p^2< M^2$, where $M$ is the characteristic mass scale of the fundamental theory, 
and the low-energy expansion is in the parameter $p^2/M^2$.  As noted in section \ref{sect 3},
however, it has been shown \cite{Shore:2002gn, Hollowood:2007ku}
that the relevant expansion parameter in the curved spacetime theory is instead $E\sqrt{\RR}/M^2$. 
Equivalently, using the typical energy $E\sim T$, the low-energy expansion here is valid 
for $z > \sqrt{\RR}/M_1$. This means that for $\RR/M_1^2 \ll 1$, as required for the validity of the 
weak gravitational field approximation, we may legitimately use the effective Lagrangian 
at scales $T$ significantly above the sterile neutrino mass $M_1$. Recalling that for a 
radiation-dominated FRW universe, $\RR \sim T^4/M_p^2$, a reasonable estimate
for the validity of the effective Lagrangian in this case is therefore\footnote{In fact,
in previous work we found that the effective Lagrangian generally gives a good description
up to energy scales $\log E\RR/M^2 \sim O(4)$, weakening the constraint by 
the factor of $10^{-2}$ shown.} $z^3 \gtrsim 10^{-2} M_1/M_p$.
The corresponding weak field condition is a lesser constraint, $z^2 \gtrsim M_1/M_p$.

\subsubsection{Neutrino parameters and the Boltzmann equations}

Since the BSM model incorporates the see-saw mechanism for generating the light neutrino masses,
the parameters are constrained and can be re-expressed in terms of the experimentally known
neutrino masses. Here, we follow \cite{Buchmuller:2004nz} and references therein, summarised for our
purposes in \cite{McDonald:2016ehm}, so we only quote a few essential relations without further 
motivation.

First, it is convenient to introduce the notation $K_{\a\b} = \sum_i \l^\dagger_{\a i} \l_{i \b}$ for the
Yukawa couplings. The sum of the light neutrino masses is $\bar{m}^2 = m_1^2 + m_2^2 + m_3^2$
and we neglect the lightest, $m_1\sim 0$, and write $\bar{m}^2 \simeq \D m_{31}^3 + \D m_{21}^2
= \D m_{sol}^2 + \D m_{atm}^2$. Using the values $\D m_{sol}^2 = 7.53 \times 10^{-5} \,{\rm eV}^2$
and $\D m_{atm}^2 = 2.44 \times 10^{-3}\,{\rm eV}^2$ for the mass differences measured in 
solar and atmospheric neutrino experiments, we have $\bar{m} \simeq 0.05 \,{\rm eV}$.  
This is related to the Yukawa couplings through the see-saw relation,
\begin{equation}
\bar{m}^2 \,=\, v^4\, \sum_{\a,\b} \frac{{\rm Re}(K_{\a\b}^2)}{M_\a M_\b} \ .
\label{f4}
\end{equation}
A convenient mass scale in this context is set by $m_* = 8\pi (\s/3)^{1/2} \,v^2/M_p = 
1.08 \times 10^{-3}\, {\rm eV}$, where $v = 174\,{\rm GeV}$ is the electroweak scale and 
$M_p = 2.4 \times 10^{18}\,{\rm GeV}$ is the reduced Planck mass.
 
Next, we introduce the key parameter $K$ characterising the strength of the Yukawa 
interactions\footnote{$K$ is equivalently defined as the ratio of the zero-temperature decay
rate to the Hubble parameter at $T = M_1$,
\begin{equation*}
K\,=\, \Gamma_D(z=\infty) / H(z=1) \ ,
\end{equation*}
which controls whether the $\n_R^1$ decays are in equilibrium.},
\begin{equation}
K \,=\, \frac{v^2}{M_1 m_*} \, K_{11} ~~~~~~~~~\Leftrightarrow ~~~~~~~~~ 
K_{11} \,=\, 8\pi \sqrt{\frac{\s}{3}}\, K\, \frac{M_1}{M_p} \ .
\label{f5}
\end{equation}
To accommodate the conventional physics of see-saw neutrino masses, we choose 
$M_1 \sim 10^{10}\,{\rm GeV}$ here, with low values of $K$ between around 1 and 5.

The sterile neutrino decay parameter is readily expressed in terms of $K$ and Bessel functions as
\begin{equation}
D(z) \,=\, K z \,\frac{K_1(z)}{K_2(z)} \ ,
\label{f6}
\end{equation}
while the equilibrium value $N_{\n_R}^{eq}$ is 
\begin{equation}
N_{\n_R}^{eq}(z) \,=\, \frac{3}{8}\, z^2\, K_2(z) \ .
\label{f6b}
\end{equation}

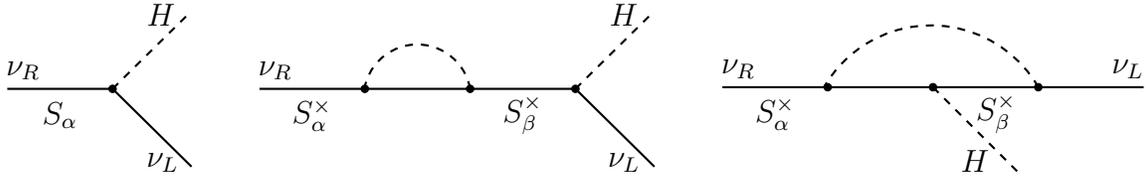
\begin{figure}[h!]
\begin{center}
\begin{tikzpicture}[scale=0.7]
\draw[thick] (0,0) -- (2,0);
\draw[thick,dashed] (2,0) -- (3.5,1.48);
\draw[thick] (2,0) -- (3.5,-1.48);
\filldraw[black] (2,0) circle (2pt);

\node at (0.3,0.35) {$\nu_R$}; 
\node at (2.92,1.4) {$H$};
\node at (2.95,-1.4) {$\nu_L$};
\node at (1.0,-0.5) {$S_\alpha$};
\end{tikzpicture}  
\hskip0.6cm
\begin{tikzpicture}[scale=0.7]
\draw[thick] (0,0) -- (2,0);
\draw[thick] (2,0) -- (4,0);
\draw[thick] (4,0) -- (6,0);
\draw[thick,dashed] (6,0) -- (7.5,1.48);
\draw[thick] (6,0) -- (7.5,-1.48);
\draw[dashed,thick] (4,0) arc (10:170:1.02); 
\filldraw[black] (2,0) circle (2pt);
\filldraw[black] (4,0) circle (2pt);
\filldraw[black] (6,0) circle (2pt);
\node at (0.3,0.35) {$\nu_R$}; 
\node at (6.92,1.4) {$H$};
\node at (6.92,-1.4) {$\nu_L$};
\node at (1.0,-0.5) {$S_\alpha^\times$};
\node at (5.0,-0.5){$S_\beta^\times$};
\end{tikzpicture}  
\hskip0.6cm
\begin{tikzpicture}[scale=0.7]
\draw[thick] (0,0) -- (2,0);
\draw[thick] (2,0) -- (4,0);
\draw[thick] (4,0) -- (6,0);
\draw[thick,dashed] (4,0) -- (5.6,-1.6);
\draw[thick] (6,0) -- (8,0);
\draw[dashed,thick] (6,0) arc (30:150:2.35); 
\filldraw[black] (2,0) circle (2pt);
\filldraw[black] (4,0) circle (2pt);
\filldraw[black] (6,0) circle (2pt);
\node at (0.3,0.35) {$\nu_R$}; 
\node at (4.8,-1.4) {$H$};
\node at (7.7,0.35) {$\nu_L$};
\node at (1.0,-0.5) {$S_\alpha^\times$};
\node at (5.2,-0.5){$S_\beta^\times$};
\end{tikzpicture}  
\end{center}
 \caption{Diagrams for the decay $\n_R^{\,\a} \rightarrow \n_L^{\, i} \, H$ which contribute to
the decay rate asymmetry factor $\varepsilon_\a$.  At $O(\l^4)$ the relevant contribution to the 
$\Gamma(\n_R\rightarrow\n_L\,H)$ decay rate arises from the interference of the tree and
one-loop diagrams shown. (Two further loop diagrams, with a self-energy insertion on the
incoming $\n_R$ line or on the outgoing $\n_L$ line, do not contribute to the asymmetry.)
Similar diagrams, where the $S_\a$ and $S_\a^\times$ type $\n_R$ propagators are switched 
compared to the figure,  give the decay rate for $\n_R^{\,\a} \rightarrow \n_L^{\, i\, c}\, H$.
Reading off the vertices from the action (\ref{b2}) shows that the asymmetry depends on the 
combination $\sum_\b\,{\rm Im}\left[ (\l^\dagger \l)_{\a\b} (\l^\dagger \l)_{\a\b}\right] f(M_\a,M_\b)$, where 
$f(M_\a,M_\b)$ is a kinematical factor from evaluating the diagrams. }
\label{Fig 6D}
\end{figure}

The CP violating decay parameter controlling the contribution to the lepton asymmetry from the
out-of-equilibrium $\n_R^{\,\a}$ decays is given by \cite{Buchmuller:2003gz}
\begin{equation}
\varepsilon_\a \,\simeq\, \frac{3}{16\pi} \sum_{\b \neq \a} \frac{{\rm Im}(K_{\a\b}^2)}{K_{\a\a}} \, 
\frac{M_\a}{M_\b} \ .
\label{f7}
\end{equation}
This arises from the interference of the tree and one-loop diagrams for the decays
$\n_R^{\,\a} \rightarrow \n_L^{\, i} \, H$ and $\n_R^{\,\a} \rightarrow \n_L^{\,i\,c}\, H$ 
shown in Fig.~\ref{Fig 6D}.
Here, we only require the parameter $\varepsilon_1$ for the dominant decays of $\n_R^{\,1}$.
Note that (\ref{f7}) then involves a linear combination of both ${\rm Im} (K_{12}^2)$ and 
${\rm Im} (K_{13}^2)$. 
A reasonable range of values for $\varepsilon_1$, consistent with the constraints from
neutrino masses, is $\varepsilon_1 \sim 10^{-5}  -  10^{-10}$.

The final element of the conventional Boltzmann equation is the factor $W(z)$ including
the $\D L=2$ scatterings, $\D L = 1$ reactions and inverse decays $\n_L H \rta \n_R$. 
In the usual theory this is known as the `washout' term, since it removes any pre-existing
lepton asymmetry. However, as we have seen, once we have a non-vanishing $N_L^{eq}(z)$
it plays a quite different role, driving $N_L(z)$ towards this equilibrium value. 

The full derivation of $W(z)$ is quite lengthy and involves a number of subtleties which need not 
concern us here. A summary of the necessary results is given in \cite{McDonald:2016ehm}.
The form of $W(z)$ is especially simple for large and small $z$. In terms of the parameters defined
above, we have
\begin{equation}
W(z\ll 1) \, \simeq \, \frac{12}{\pi^2} \, \frac{m_* M_1}{v^2}\, 
\left(\frac{\bar{m}^2}{m_*^2} \,+\, K^2\right) \, \frac{1}{z^2}  ~~=~~ 
17.6  \,\s^{1/2}\, \frac{M_1}{M_p}\,
\left(\frac{\bar{m}^2}{m_*^2} \,+\, K^2\right) \,\frac{1}{z^2} \ ,
\label{f8}
\end{equation}
with $W(z\gg 1)$ satisfying the same formula with the $K^2$ term omitted.
Since $\bar{m}^2/m_*^2 \simeq 2150$, the effect of the $K^2$ term is negligible for the
values considered here, and the coefficient of the asymptotic $1/z^2$ dependence of $W(z)$ 
is essentially the same for small and large $z$. In the region $z\simeq 1$, $W(z)$ displays a
resonance arising from the $\n_R^{\,1}$ intermediate state in the scattering diagrams (see
Fig.~\ref{Fig 2.1}). 

\begin{figure}[h!]
\vskip0.5cm
\centering{
\includegraphics[scale=1.0]{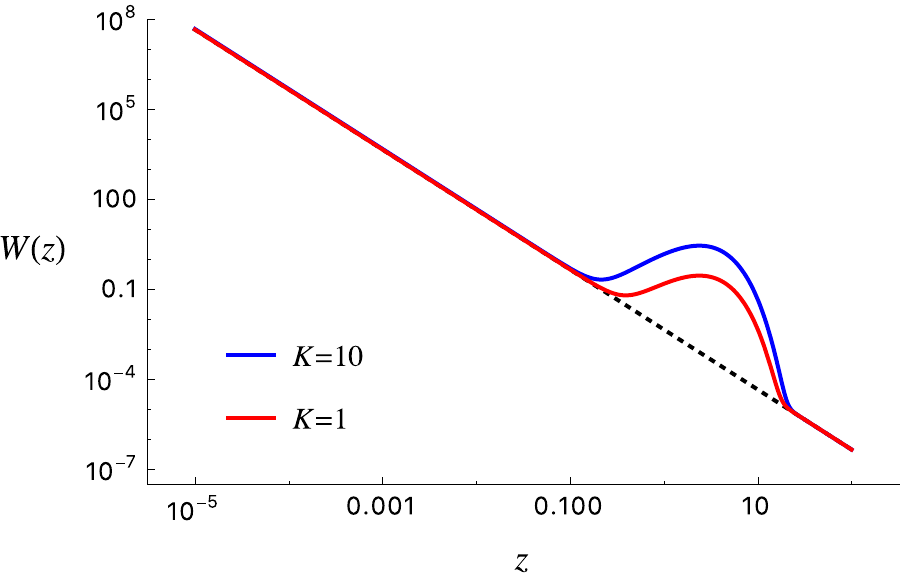}
}
\caption{This shows the dependence on $z = M_1/T$ of the coefficient $W(z)$ in the Boltzmann 
equation for $K=1$ and $K=10$, with $M_1=5\times 10^{10}\,{\rm GeV}$ 
and the neutrino parameters in the text. The resonance peak around $T\simeq M_1$ increases
with $K$.} 
\label{Fig 6.1}
\end{figure}

The complete temperature dependence of $W(z)$ for low values of $K$ is shown in 
Fig.~\ref{Fig 6.1}, illustrating the characteristic $1/z^2$ behaviour anticipated in 
section \ref{sect 5} together with the resonance enhancement at $T\simeq M_1$.

This brings us to the new gravitationally-induced terms in the Boltzmann equation (\ref{f2}).
The equilibrium lepton number asymmetry $N_L^{eq}$ has been derived and discussed in detail
in our earlier work \cite{McDonald:2015ooa, McDonald:2015iwt, McDonald:2016ehm}. 
Here, from (\ref{e14}) and (\ref{c7}) we have
\begin{align}
N_L^{eq}(z) \,&=\, \frac{1}{3}\,\frac{1}{n_\c} \, b \,\dot{R} \,T^2 \nonumber \\
&= \, \frac{1}{27} \, \frac{1}{n_\c}\, \dot{R}\, T^2 \, \sum_{\a,\b} \frac{{\rm Im} (K_{\a\b}^2)}{(4\pi)^4} \,
\frac{1}{M_\a M_\b} \, I_{[\a\b]} \ .
\label{f9}
\end{align}
With a sterile neutrino mass hierarchy $M_1 < M_2 < M_3$, the dominant term is $\a = 1$.
Then, substituting for $\dot{R}$ from (\ref{dd5}) and with the equation of state $w \simeq 0.3$
for a radiation-dominated background (including the trace anomaly), we find
\begin{equation}
N_L^{eq}(z) \,=\,  0.034\, \s^{3/2} \,\left(\frac{M_1}{M_p}\right)^3\, \sum_{\b\neq1} 
\frac{{\rm Im} (K_{1\b}^2)}{(4\pi)^4} \,\left(\frac{M_\b}{M_1}\right)^{2p-1}\, 
\log\left(\frac{M_\b}{M_1}\right) \, \frac{1}{z^5} \ ,
\label{f10}
\end{equation}
where we have used (\ref{b32}) for $I_{[\a\b]}$ and collected the numerical terms into a single
pre-factor.

In writing (\ref{f10}), we have retained the possibility of a hierarchy enhancement with $p=1$
or the, perhaps more likely, form of $I_{[\a\b]}$ with $p=0$. With $p=1$, the dominant contribution
comes from the heaviest sterile neutrino $\n_R^{\,3}$, whereas for $p=0$ it is $\n_R^{\,2}$ that 
gives the largest contribution. We consider both possibilities in the solutions of the Boltzmann
equations described below.

A key observation here is that the dependence on the Yukawa couplings through combinations of
${\rm Im} (K_{\a\b}^2)$ is {\it different} in $N_L^{eq}$ and $\varepsilon_1$, so these two CP-violating 
terms in the Boltzmann equation may be chosen independently.\footnote{In more detail,
see \cite{McDonald:2016ehm}, if we impose the relation
\begin{equation*}
M_3\,{\rm Im}(K_{12}^2) \,+\, M_2\, {\rm Im}(K_{13}^2) \,\simeq\, 0 \ ,
\end{equation*}
then $\varepsilon_1 \simeq 0$ and the sterile neutrino decays play no role in the generation of the
lepton asymmetry. Leptogenesis is then due entirely to the radiatively-induced gravitational 
mechanism. Imposing this condition, the hierarchy-sensitive terms in (\ref{f10}) become
approximately, for $p=1$ and $p=0$ respectively,
\begin{equation*}
{\rm Im}(K_{13}^2) \,\frac{M_3}{M_1}\, \log\left(\frac{M_3}{M_1}\right) \ ,   ~~~~~~~~~~ (p=1) \ ,
\end{equation*}
and 
\begin{equation*}
{\rm Im}(K_{12}^2) \,\frac{M_1}{M_2}\, \log\left(\frac{M_2}{M_3}\right) \ ,   ~~~~~~~~~~ (p=0) \ .
\end{equation*}
}

The final element is the evolution rate $\mathcal{W}(z)$, derived for the first time in this paper.
Writing (\ref{e25}) in terms of the neutrino parameters introduced above,
and with the particular perturbative coefficients (\ref{c6}) in this BSM model,
we have
\begin{equation}
\mathcal{W}(z) \,=\,  \frac{1}{3}\, (1+w)(1+3w)\big|_{w=0.3} \,\, \s \,\left(\frac{M_1}{M_p}\right)^2\,
\frac{K_{11}}{(4\pi)^2} \, \frac{1}{z^5} \ ,
\label{f11}
\end{equation}
where we keep only the dominant contribution from the lightest sterile neutrino.  Re-expressing in terms
of the parameter $K$ from (\ref{f5}) then gives,
\begin{equation}
\mathcal{W}(z) \,=\,  0.075\,  \s^{3/2} \, K \left(\frac{M_1}{M_p}\right)^3\frac{1}{z^5} \ .
\label{f12}
\end{equation}
Notice that in this parametrisation, the small value of the Yukawa coupling constants are exchanged
for an extra power of $M_1/M_p$ compared to the estimates in (\ref{e27}), assuming $K$ is chosen
of $O(1)$ as we do here.

As already observed, $\mathcal{W}(z)$ has a $1/z^5$ dependence, characteristic of the gravitational
terms. It therefore rises sharply for temperatures $T > M_1$. Comparing 
$\mathcal{W}(z)$ with the coefficient $W(z)$, which has a milder $1/z^2$ temperature dependence,
we can determine the value of $z$ at which $\mathcal{W}(z)$ begins to dominate and force the 
lepton asymmetry away from its equilibrium value. 
With parameters chosen conventionally to reproduce light neutrino phenomenology, and with
$M_1 \simeq 10^{10}\,{\rm GeV}$ and $K\simeq 1-10$, this crossover value is around
$z\simeq 10^{-6}$, which unfortunately lies 
outside the range of $z$ where we can unambiguously rely on the low-energy expansion of
the effective Lagrangian. It is, however, very model dependent and for that reason, and with this 
caveat, we include its effects in the Figures below where we illustrate the evolution of $N_L(z)$
with temperature. 

Note especially that this crossover value of $z$ is particularly sensitive to the
observed light neutrino masses, which control the scale of $W(z)$ in (\ref{f8}). This emphasises
again the strong constraints on the gravitational leptogenesis mechanism when implemented in this
particular BSM model, which is playing the dual role of generating the light neutrino masses 
through the see-saw mechanism.
In section \ref{sect 6.2}, we show how this value also depends on the equation of state
parameter in generalised cosmological scenarios, and occurs for significantly lower temperatures
for $w > 1/3$.

\subsubsection{Evolution of the lepton number asymmetry}

We now present a variety of plots of the evolution of the lepton number asymmetry $N_L(z)$
with temperature, found by numerically solving the coupled Boltzmann equations (\ref{f2}) 
with different assumptions and parameters in the BSM model.

\begin{figure}[h!]
\vskip0.5cm
\centering{
\includegraphics[scale=0.82]{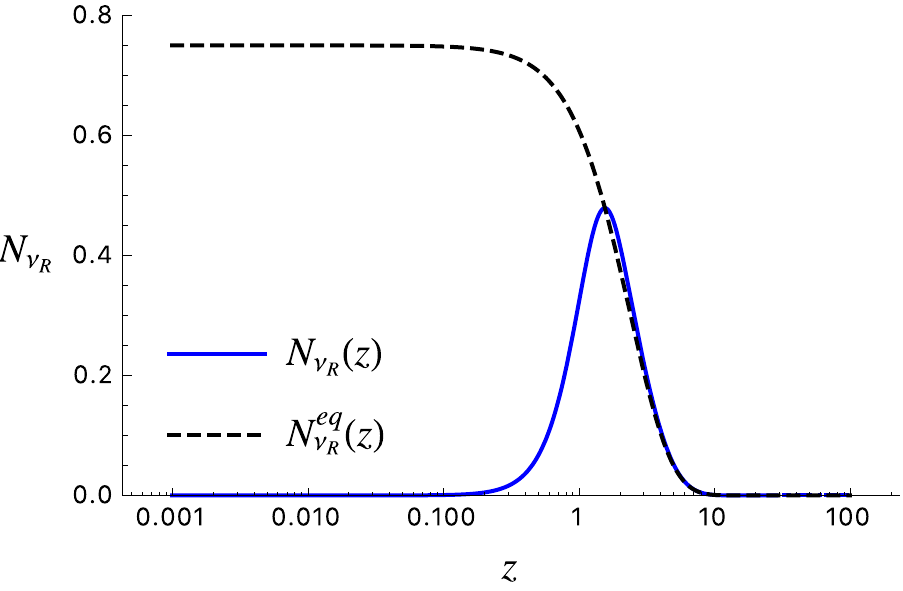}
\hskip0.25cm
\includegraphics[scale=0.82]{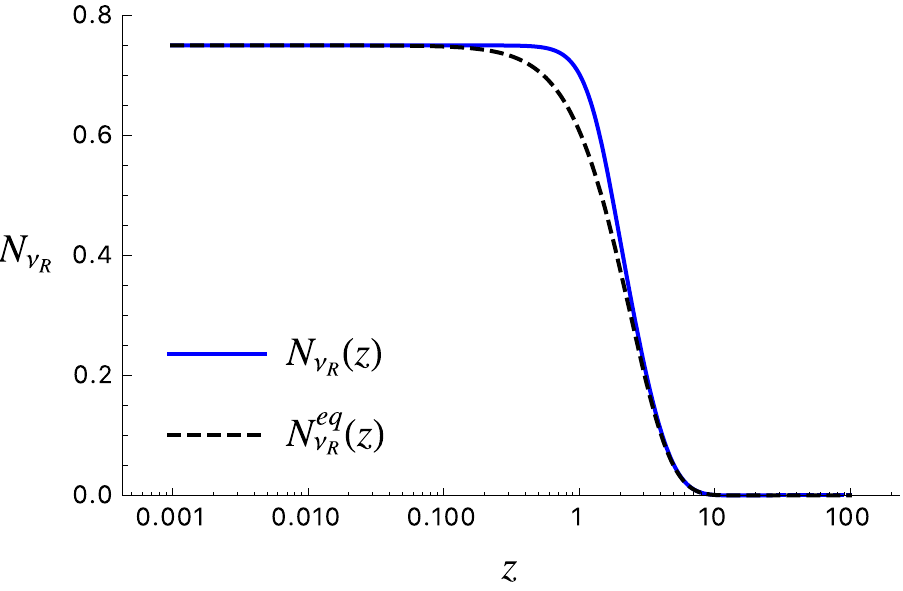}
\vskip0.5cm
\includegraphics[scale=0.6]{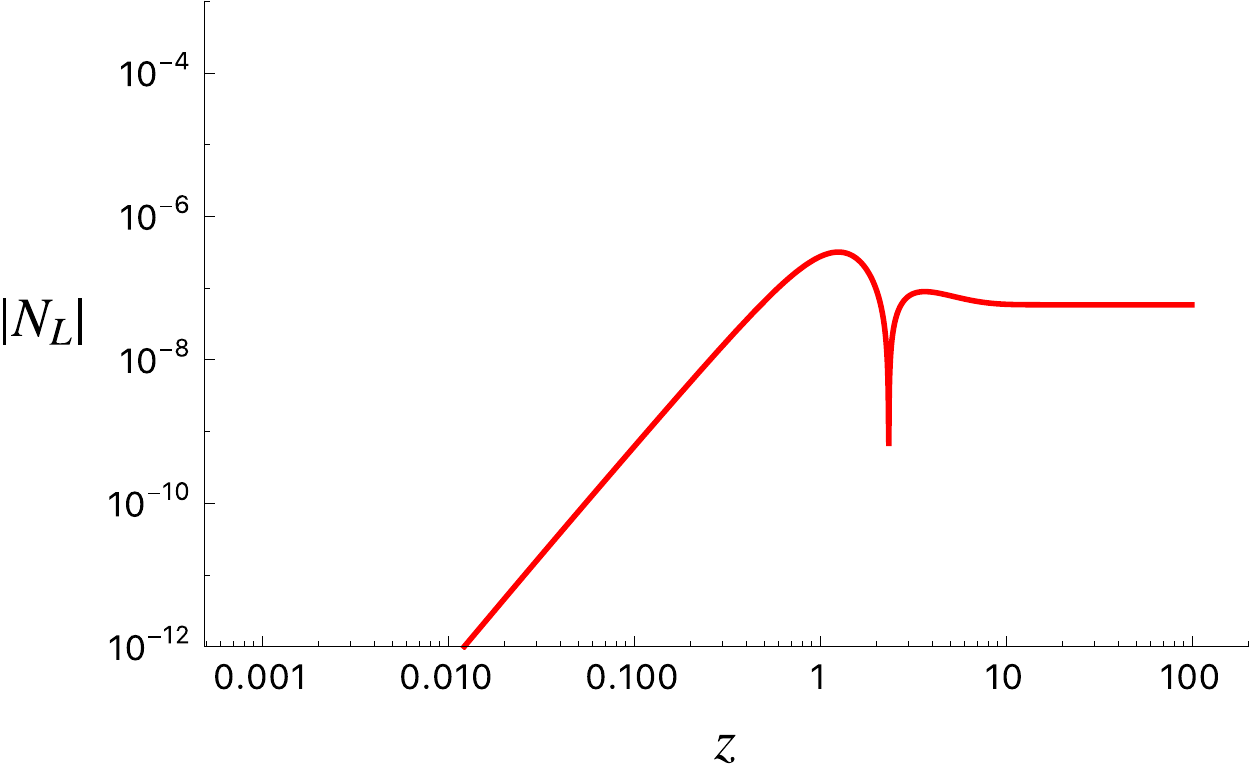}
\hskip0.12cm
\includegraphics[scale=0.6]{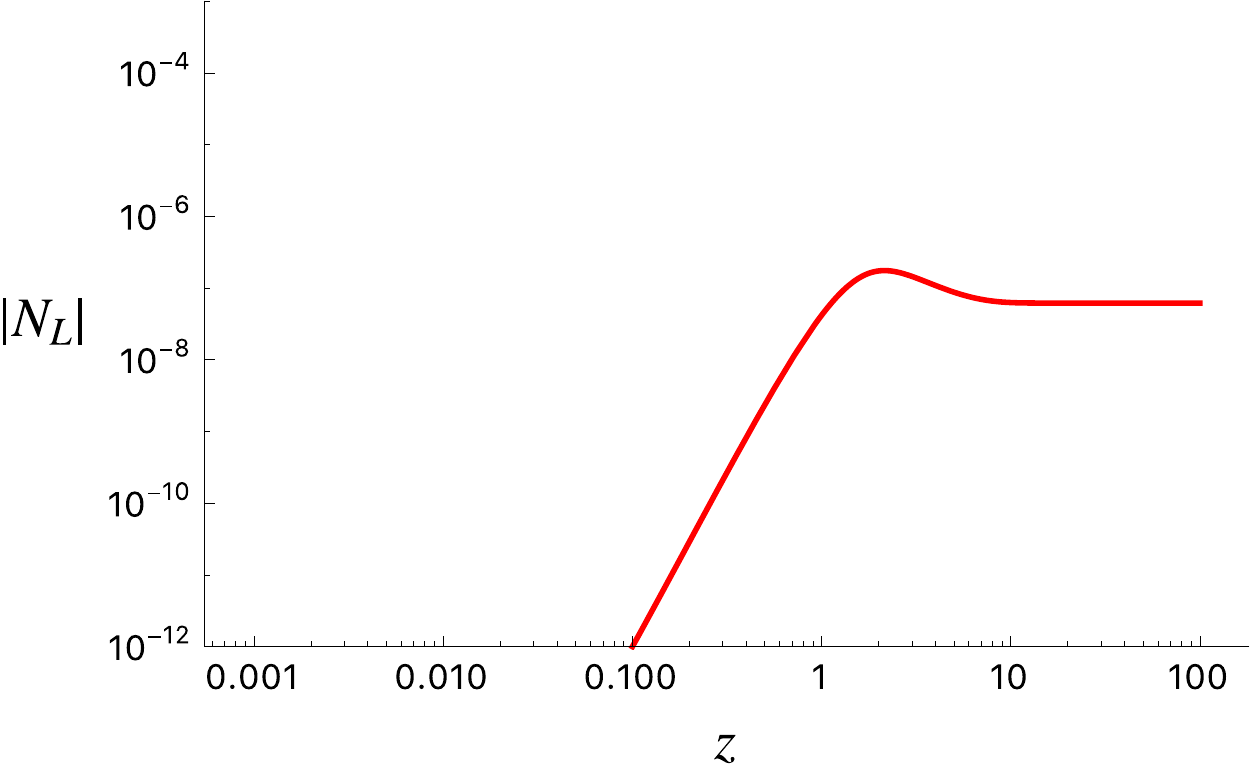}
}
\caption{The upper figures show the development of the sterile neutrino densities $N_{\n_R}(z)$.
The left-hand diagram corresponds to an initial condition with $N_{\n_R}(z)=0$ while the 
right-hand plot starts with $N_{\n_R}(z)$ already at its equilibrium value. 
The lower figures show the corresponding absolute value of the lepton asymmetry induced by the 
out-of-equilibrium decays of $\n_R^{\,1}$.  Notice the cusp in the left-hand plot, indicating 
that the asymmetry $N_L(z)$ changes sign as the sterile neutrino density $N_{\n_R}(z)$ switches
from under to just over its equilibrium value. 
The parameters here are $M_1 = 5 \times 10^{10}\,{\rm GeV}$ and $K=5$, 
with $\varepsilon_1 = 10^{-6}$ .}
\label{Fig 6.2}
\end{figure}

To begin, in Fig.~\ref{Fig 6.2}, we show the sterile neutrino density $N_{\n_R}(z)$
and the lepton asymmetry $N_L(z)$ in the conventional model neglecting the gravitational 
contributions. The two options of starting at high temperature with $N_{\n_R}(z)$ taking its
equilibrium value, or with $N_{\n_R}(z\ll 1) \simeq 0$, are shown separately.
The effect of the decay term proportional to $\left(N_{\n_R} - N_{\n_R}^{eq}\right)$ in the 
Boltzmann equation for $dN_L(z)/dz$ is readily understood, becoming appreciable only when 
the decay rate $D(z)$, which grows with $z$, becomes sufficiently large. The above equilibrium
abundance of $\n_R^{\,1}$ in the upper right-hand plot in Fig.~\ref{Fig 6.2} drives $N_L(z)$ negative, 
as shown in the corresponding lower plot for $|N_L(z)|$,
while the initial under abundance in the upper left-hand plot in Fig.~\ref{Fig 6.2} drives a positive value 
for $N_L(z)$ before switching, at the cusp in $|N_L|$, to a negative value as $N_{\n_R}(z)$ 
becomes temporarily over abundant.
This describes the well-known mechanism of generating the lepton asymmetry from the 
out-of-equilibrium decays of the sterile neutrinos.

\begin{figure}[h!]
\vskip0.5cm
\centering{
\includegraphics[scale=1.1]{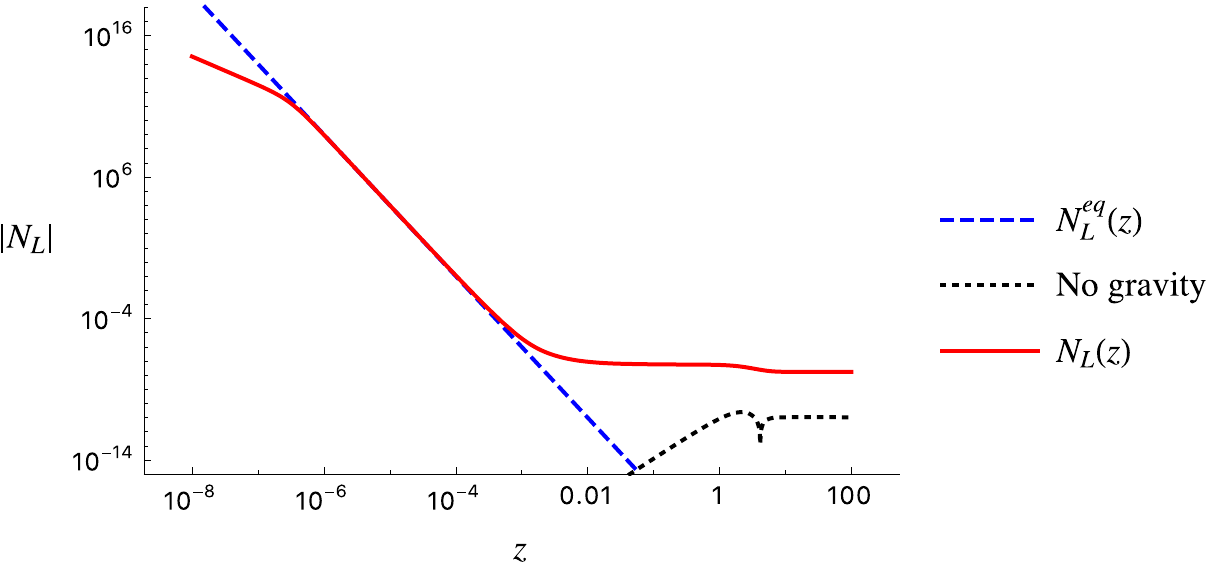}
}
\caption{Dynamical evolution of the lepton number asymmetry $N_L(z)$
in the case of a hierarchy enhancement, $p=1$, with parameters chosen so that the gravitational
leptogenesis mechanism dominates over the out-of-equilibrium $\n_R^{\,1}$ decays.
Here, $M_1=5\times 10^{10}\,{\rm GeV}$, $K=1$, with $M_3= 10^{16}\,{\rm GeV}$,
${\rm Im}K_{13}/(4\pi)^2 = 5\times 10^{-4}$ and $\varepsilon_1 = 10^{-10}$. }
\label{Fig 6.4}
\end{figure}

Our main interest here is of course the gravitational contributions, $N_L^{eq}$ and $\mathcal{W}$.
Consider first the scenario in which $N_L^{eq}$ has a hierarchy enhancement, {\it i.e.} $p=1$ in
(\ref{f10}). In this case, the dominant contribution to $N_L^{eq}$ comes from the heaviest sterile
neutrino, so $\b=3$ in (\ref{f10}). A reasonable choice of BSM parameters then gives the evolution 
shown in Fig.~\ref{Fig 6.4}, resulting in a final value of $|N_L| \simeq 10^{-8}$ as required to give the
observed $\eta$. 

Here, we have chosen the CP violating decay parameter $\varepsilon \simeq 10^{-10}$,
sufficiently small that the effect of the out-of-equilibrium $\n_R$ decays makes a negligible
effect on the final lepton asymmetry. $N_L(z)$ therefore evolves just as described at the end
of section \ref{sect 5}. At high temperatures $z\ll 1$, $N_L(z)$ is driven to its equilibrium
value $N_L^{eq}\sim 1/z^5$, moderated to $\sim 1/z^2$ at ultra-high temperatures 
(with these parameters only for $z \ll 10^{-6}$, which lies outside the range of validity 
of the effective Lagrangian) by the damping effect of the new evolution term $\mathcal{W}(z)$. 
After a period following its gravitationally-induced equilibrium value, 
$N_L(z)$ decouples as the scattering rate maintaining equilibrium falls below the Hubble 
expansion rate, {\it i.e.} $W(z) < 1$, and becomes essentially constant. A final dip occurs 
for $z\simeq 1$ when the resonant increase in $W(z)$ gives a further temporary push towards
$N_L^{eq}$, which is falling away very rapidly as $1/z^5$. These features are quantified in the
analytic solution in Appendix \ref{Appendix A}.

We therefore see that given the hierarchy enhancement $p=1$, the radiatively-induced gravitational
leptogenesis mechanism can reproduce the observed lepton asymmetry for conventional choices
of the BSM parameters.

\begin{figure}[h!]
\vskip0.5cm
\centering{
\includegraphics[scale=1.1]{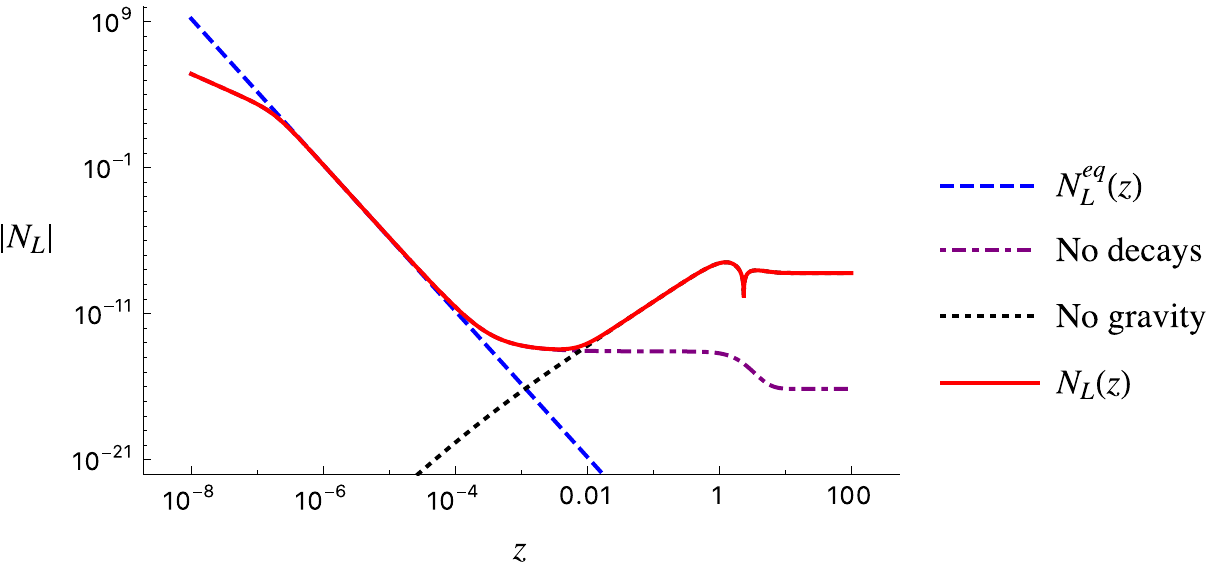}
}
\caption{Dynamical evolution of the lepton number asymmetry $N_L(z)$
in the case of no hierarchy enhancement, $p=0$. Here, the final lepton asymmetry is determined
by the out-of-equilibrium $\n_R^{\,1}$ decays, with the gravitational mechanism producing a steep
rise in the asymmetry at earlier times for temperatures $z \lesssim 0.01$.
Here, $M_1=10^{10}\,{\rm GeV}$, $K=5$, with $M_2= 10^{12}\,{\rm GeV}$,
${\rm Im}K_{12}/(4\pi)^2 = 5 \times  10^{-4}$ and $\varepsilon_1 = 10^{-7}$.}
\label{Fig 6.5}
\end{figure}

Without a hierarchy enhancement, $p=0$, the largest contribution to $N_L^{eq}$ arises from the
lighter sterile neutrino $\n_R^{\,2}$ (that is $\b=2$ in \ref{f11}), but is still significantly suppressed
relative to $p=1$. In this case, however, the gravitational terms alone do not give a big enough 
final value for $N_L$ to reproduce the observed $\eta$, given the parameter constraints 
in this particular BSM model from the light neutrino masses in setting the magnitude of $W(z)$.
So here we rely on the conventional decay mechanism to produce the final value of $N_L$ at 
low temperatures. Nevertheless, the gravitational effects still produce a radical change in the
high temperature evolution of $N_L(z)$, illustrated in Fig.~\ref{Fig 6.5}.

Following this evolution back in time towards the early universe, the post-leptogenesis value
$N_L \simeq 10^{-8}$ arises  around $z\simeq 1$ due to out-of-equilibrium decays of sterile
neutrinos. As the temperature rises, $N_L(z)$ initially falls as in the conventional model.
However, this drop is then overtaken by the gravitational effects (from around $z\sim 10^{-3}$
in Fig.~\ref{Fig 6.5}) and $N_L(z)$ rises sharply, driven to its equilibrium value $N_L^{eq}$. 
Eventually, though outside the range of validity of $L_{eff}$ in this model, $N_L(z)$ evolves away 
from $N_L^{eq}$ due to the new gravitational term $\mathcal{W}(z)$ in the Boltzmann equation.
Far from reducing to zero for high temperatures above the sterile neutrino mass scale $M_1$, 
the lepton asymmetry drops then rises steeply, its dynamical evolution being driven by the 
radiatively-induced gravitational tidal effects.

The physics applications of a substantial lepton number asymmetry in the early universe above
the conventional leptogenesis scale remain to be explored. In the following section,
we generalise the discussion to allow for a post-inflationary era preceding radiation dominance,
where the universe expansion is driven by a source with an effective equation of state $w\neq 1/3$.
As we shall see, this has a very significant effect on the dynamical evolution described here.

\subsection{Gravitational leptogenesis in the pre-radiation era}\label{sect 6.2}

This study of the effects of gravity on the dynamical evolution of lepton asymmetry in the 
radiation-dominated era motivates an exploration of how our gravitational leptogenesis
mechanism would be modified in an earlier, post-inflationary era where the universe expansion 
is controlled by a source with $w \neq 1/3$. 

An economical way to illustrate how such an equation of state can arise is to consider a single
(inflaton) scalar field with different potentials. In an isotropic spacetime, the energy density
and pressure are given by
\begin{align}
\r \,&=\, \frac{1}{2} \dot{\phi}^2 \,+\, V(\phi)  \nonumber \\
p \,&=\, \frac{1}{2}\dot{\phi}^2 \,-\, V(\phi) \ .
\label{f13}
\end{align}
For a potential $V\sim \phi^{2n}$, it is easy to show \cite{Turner:1983he} that in a (reheating) phase 
where the scalar field is oscillating around its minimum, the average value of $\r$ and $p$ over 
a cycle are related by $\langle p\rangle = w \langle \rho\rangle$ with $w = (n-1)/(n+1)$.
For example, a non-interacting scalar field, with a $\phi^2$ potential, has $w=0$.
The conformal case $V(\phi)\sim \phi^4$ gives $w = 1/3$, just as for radiation.
Steeper potentials give higher values of $w$, while in the limiting case of `kination' scenarios,
where the kinetic term dominates completely over the potential, the equation of state is 
$w=1$. 

In the conventional picture of the early universe, radiation dominance is preceded by a 
post-inflationary reheating era during which the inflaton field oscillates around the minimum
of a shallow potential (such that $V \sim \tfrac{1}{2}m^2 \phi^2$) while decaying into 
relativistic standard model particles at temperature $T$. This requires a non-vanishing but
very weak coupling of the inflaton to the standard model fields. The process ends when the inflaton
decay rate falls below the Hubble parameter and the universe enters the radiation-dominated
era at the reheating temperature $T_{rh}$. During reheating, the effective equation of state
parameter therefore evolves in a model-dependent way from an initial value $w\simeq 0$ 
towards $1/3$ as a successively greater fraction of the initial energy density is in the
form of radiation. If $T_{rh} \lesssim M_1$, then the gravitational effects described here
would be active during reheating.

Gravitational leptogenesis in this era would therefore be characterised by 
$0 < w < 1/3$, a lower value than for the radiation dominance scenario
just described. Now, as can be seen from the analysis below, for an equation of state $w < 1/3$,
the Hubble parameter is decreased relative to $w=1/3$ resulting in an increase in the
factor $W(z) = \Gamma/zH$ in the Boltzmann equation. Compared to the evolution
shown in Fig.~\ref{Fig 6.5}, this means the lepton asymmetry $N_L(z)$ is driven more strongly 
to the equilibrium value $N_L^{eq}$, reaching it at an earlier time and decoupling later at a 
higher value of $z$. This results in a lower final asymmetry for $N_L$ due to the gravitational 
effects prior to the $\n_R$ decays. Moreover, this value of $N_L$ is further diluted by entropy 
production during the reheating phase. Overall, therefore, the gravitational leptogenesis
mechanism is less effective at generating the required final lepton asymmetry during the 
reheating phase than it is during radiation dominance. The sharp rise in the asymmetry 
to $N_L^{eq}$ for temperatures above $M_1$ shown Fig.~\ref{Fig 6.5} would however still
occur.

Instead, we consider here a scenario in which at early times the universe comprises a 
non-thermalised `exotic' component with $w>1/3$ which drives the expansion, together
with an initially sub-dominant radiation component at temperature $T$ 
\cite{Davoudiasl:2004gf, McDonald:2016ehm}.
As the universe expands, its energy density naturally decreases faster than the energy density 
of the radiation, becoming equal at some critical temperature $T_* = M_1/z_*$. 
Below this temperature, the universe becomes radiation dominated as before.

There are several ways in which such a scenario could arise \cite{Davoudiasl:2004gf}. 
An attractive picture \cite{Ford:1986sy,Peebles:1998qn} is that instead of conventional reheating, 
the relativistic particles giving rise to the entropy of the universe arise through gravitational
particle creation at the transition from the de Sitter inflationary vacuum to the FRW vacuum 
at the end of inflation. 
This obviates the need for a direct coupling of the inflaton to the standard model fields.
Immediately after inflation, the energy density is dominated by the inflaton component
with an equation of state $w>1/3$, which gradually dilutes relative to the thermalised
relativistic particles as the universe expands leaving a conventional radiation-dominated
FRW era. If this phase is characterised by a temperature $T \gtrsim M_1$ then our
gravitational leptogenesis mechanism could take place at this time. 

The energy density of the radiation component is $\r_\c = \s T^4$, and we assume $T\sim 1/a(t)$.  
In general reheating scenarios, the temperature may have a different and dynamically sensitive
dependence on the scale factor, as discussed in detail in \cite{Chung:1998rq, Giudice:2000ex}.
In our illustrative model, however, we are assuming the inflaton field is decoupled from the
radiation component, in which case $T \sim 1/a(t)$ indeed follows from the analysis 
in \cite{Chung:1998rq, Giudice:2000ex}.   
From the conservation equation (\ref{dd4}), the energy density 
of the exotic component varies with temperature as $\r_w = \s_w T^{3(1+w)}$, where the constant 
$\s_w$ sets its overall magnitude. We can trade this for the critical temperature $T_*$ by setting
$\s_w T_*^{3(1+w)} = \s T_*^4$. It follows immediately that the ratio of the energy densities of the
exotic and radiation components is
\begin{equation}
\frac{\r_w}{\r_\c} \,=\, \left(\frac{z}{z_*}\right)^{1-3w} \ .
\label{f14}
\end{equation}
The total energy density $\r = \r_w + \r_\c$. 

To see how the various terms in the Boltzmann equation are modified in this new background,
we track back through their derivation writing the curvatures and Hubble constant in terms of
the new energy density $\r$. For the decay rate $D(z)$ and $W(z)$, the only change is due to the
factor $1/H$ in their derivations, so we find
\begin{align}
D(z) \,&=\, D_\c(z) \,\left(1+ \Big(\frac{z}{z_*}\Big)^{1-3w}\right)^{-1/2} \nonumber \\
W(z)\,&=\, W_\c(z) \,\left(1+ \Big(\frac{z}{z_*}\Big)^{1-3w}\right)^{-1/2}  \ ,
\label{f15}
\end{align}
where here we define $D_\c$, $W_\c$ {\it etc.} as the formulae given above for a pure 
radiation-dominated background. For the gravitational terms, there is also a direct dependence 
on the curvature, and here we find
\begin{equation}
\mathcal{W}(z)\,=\, \mathcal{W}_\c(z)\, \left(1 + \frac{(1+w)(1+3w)}{(1+w_\c)(1+3w_\c)}\,
\Big(\frac{z}{z_*}\Big)^{1-3w}\right) \ ,
\label{f16}
\end{equation}
and
\begin{equation}
N_L^{eq}(z)\,=\, N_{L\,\c}^{eq}(z) \, \left(1 + \frac{(1+w)(1-3w)}{(1+w_\c)(1-3w_\c)}\,
\Big(\frac{z}{z_*}\Big)^{\frac{3}{2}(1-3w)} \right) \ .
\label{f17}
\end{equation}
where $w_\c =0.3$.

\begin{figure}[h!]
\vskip0.5cm
\centering{
\includegraphics[scale=0.82]{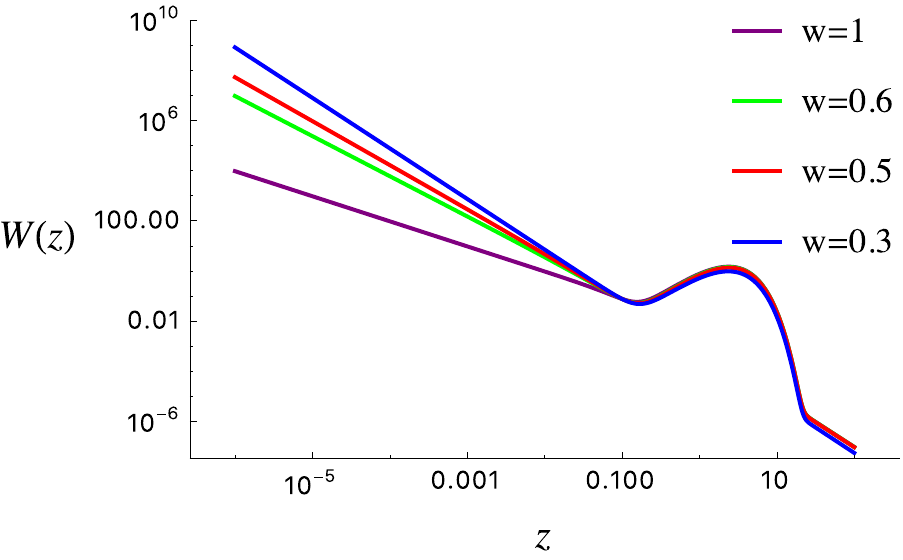}
\hskip0.3cm
\includegraphics[scale=0.82]{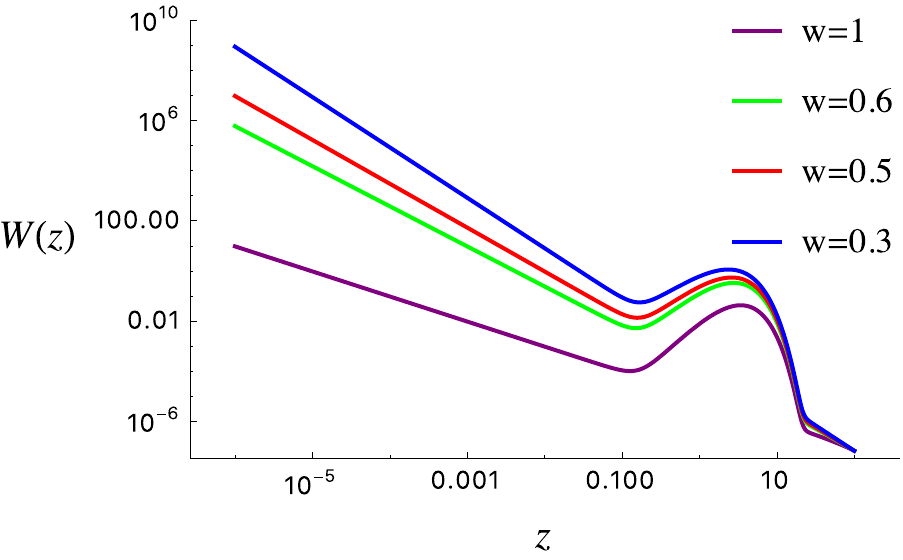}
}
\caption{This shows the lepton number violating rate term $W(z)$ in the Boltzmann equation
with the effective equation of state parameter $w=0.3$, $0.5$, $0.6$ and $1$, and
$z_* = 0.1$ (left-hand figure) and $z_* = 100$ (right-hand figure).
The other parameters here are $M_1=10^{10}\,{\rm GeV}$ and $K=5$.}
\label{Fig 6.6}
\end{figure}
The dependence of $W(z)$ on the parameter $w$ is plotted in Fig.~\ref{Fig 6.6} for the two cases
where $T_*$ is greater or less than $M_1$.  The key feature is that for $z<z_*$, $W(z)$ 
becomes smaller as $w$ is increased from its radiation-dominance value $0.3$ towards
the kination limit $w=1$.

\begin{figure}[h!]
\vskip0.5cm
\centering{
\includegraphics[scale=0.82]{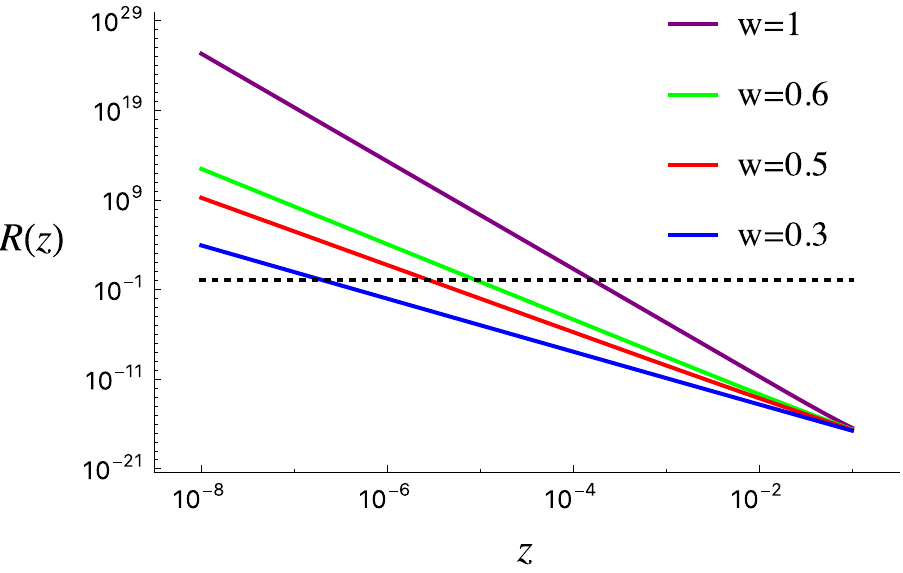}
\hskip0.3cm
\includegraphics[scale=0.82]{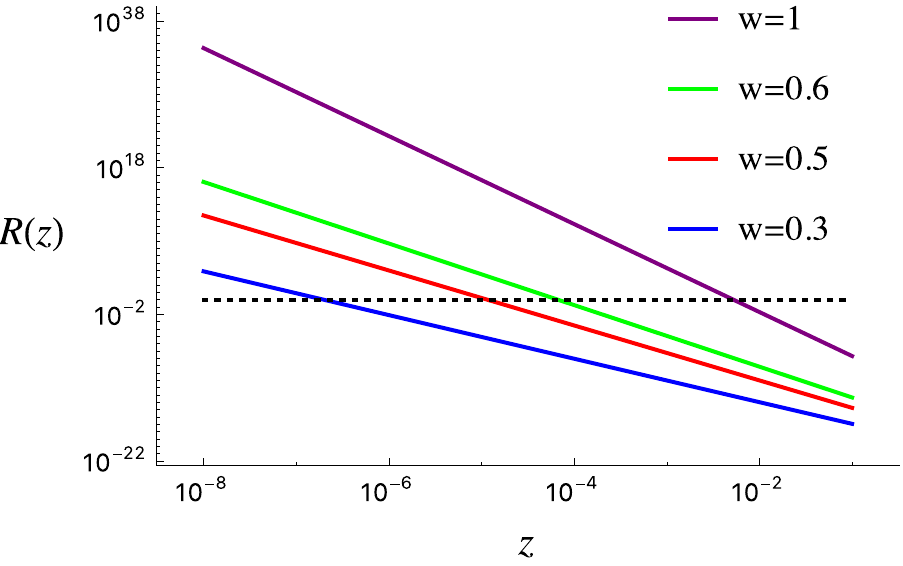}
}
\caption{This figure shows the ratio $R(z)$ of the evolution term $\mathcal{W}(z)$ to the 
lepton number violating term $W(z)$ in the Boltzmann equation for $w=0.3$, $0.5$, $0.6$ 
and $1$, with $z_* = 0.1$ (left-hand figure) and $z_* = 100$ (right-hand figure). 
Note the crossover point where $R(z) = 1$ (dotted line) is shifted to higher values of 
$z$ as $w$ is increased. The parameters are as in Fig.~\ref{Fig 6.6}.
}
\label{Fig 6.7}
\end{figure}
We can also show how the relative strength of the evolution factor $\mathcal{W}(z)$ and 
$W(z)$ depends on $w$. From Fig.~\ref{Fig 6.7}, we see how the crossover point
where $\mathcal{W}(z) \simeq W(z)$ occurs at successively lower temperatures as
$w$ is increased.

\begin{figure}[h!]
\vskip0.5cm
\centering{
\includegraphics[scale=1.1]{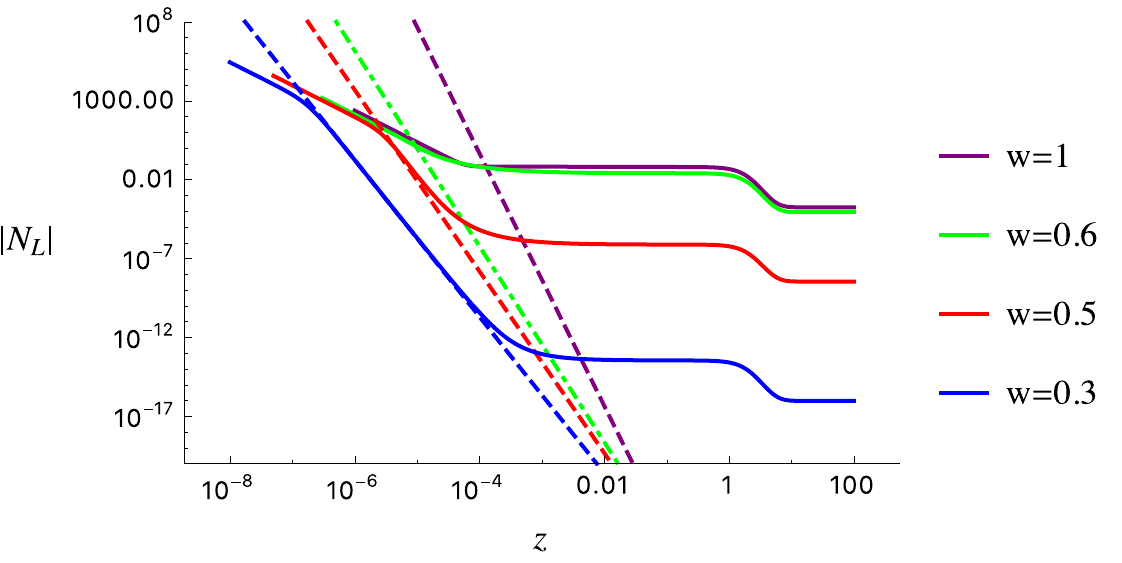}
}
\caption{Dynamical evolution of the gravitationally-induced lepton number asymmetry $N_L(z)$ 
for effective equation of state parameters $w=0.3$, $0.5$, $0.6$ and $1$ with $z_* = 0.1$. 
The corresponding equilibrium values $N_L^{eq}$ are shown as the dot-dashed lines.
Note the universal $N_L(z) \sim 1/z^2$ behaviour at ultra-high temperatures where the
evolution term $\mathcal{W}(z)$ dominates in the Boltzmann equation. Decoupling from
$N_L^{eq}$ occurs earlier for higher values of $w$, resulting in a larger final asymmetry.
The parameters here are $M_1=10^{10}\,{\rm GeV}$, $K=5$, with $M_2= 10^{12}\,{\rm GeV}$,
${\rm Im}K_{12}/(4\pi)^2 = 5 \times  10^{-4}$ and no hierarchy enhancement, $p=0$.}
\label{Fig 6.8}
\end{figure}
These features explain the evolution of the lepton asymmetry $N_L(z)$ shown in 
Fig.~\ref{Fig 6.8}. The qualitative features are the same as described at the end of section 5,
and exact analytic expressions for $N_L(z)$ in the different regions are derived for
arbitrary $w$ in Appendix \ref{Appendix A}. At ultra-high temperatures the evolution 
term $\mathcal{W}(z)$ dominates in the Boltzmann equation and gives the universal behaviour
$N_L(z) \sim 1/z^2$ for all $w$. Beyond the crossover point in Fig,~\ref{Fig 6.7}, $N_L(z)$ 
is driven to the equilibrium value $N_L^{eq}$, which in itself increases with $w$, by the
lepton number violating rate term $W(z)$. However, since $W(z)$ is weaker for successively 
bigger values of $w$, the asymmetry $N_L(z)$ decouples sooner from $N_L^{eq}$ 
and consequently at a higher value. Indeed, as we approach the kination limit, $N_L(z)$ is
so weakly driven to $N_L^{eq}$ that after the initial ultra-high temperature phase
when $\mathcal{W}(z) > W(z)$, it remains essentially constant and the whole Boltzmann 
equation is dominated by the new evolution term $\mathcal{W}$ (see Appendix \ref{Appendix A}).
Overall, the point of decoupling and the final asymmetry are clearly very sensitive to the 
value of $w$.

We see, therefore, that for the same BSM parameters, the final asymmetry produced
by the gravitational leptogenesis mechanism increases with the equation of state
parameter $w$. An optimal case is $N_L(z)$ with $w=0.5$ in Fig.~\ref{Fig 6.8},
yielding a final asymmetry $N_L \simeq 10^{-8}$. 
This shows how with physically reasonable
BSM parameters, the observed baryon asymmetry $\eta$ can be entirely generated
by radiatively-induced gravitational leptogenesis, even without a hierarchy enhancement, 
in a FRW spacetime characterised by an effective equation of state $w \simeq 0.5$.
This could readily be realised in the post-inflationary scenario of \cite{Ford:1986sy,Peebles:1998qn}.

\section{Summary and Outlook}\label{sect 7}

In this final section, we highlight some of the key features of radiatively induced
gravitational leptogenesis and discuss potential future developments.

First, we again emphasise the generality of the leptogenesis mechanism, independent of its 
realisation in the particular BSM theory analysed here. Whatever its origin, the effective 
action (\ref{c1}) summarises the gravitational interactions of the light neutrinos and implies
the picture of the origin and evolution of the lepton number density described above.
We saw there was a clear distinction between the roles of the CP odd and CP even 
operators in (\ref{c1}). The coupling $\partial_\m R\, \overline{\n_L} \c^\m \n_L$ to the
CP odd neutrino operator modifies the light neutrino dispersion relation and gives rise 
to an effective chemical potential for lepton number. This is the origin of the non-vanishing 
equilibrium lepton number density $n_L^{eq}$. On the other hand, the gravitational
couplings to the CP even operators, especially 
$R_{\m\n} \,\overline{\n_L} \c^\m \overleftrightarrow{D}^\n \n_L$,
modify the dynamical evolution of the lepton number density at early times.

To complete the gravitational leptogenesis mechanism, these two tidal curvature effects
encoded in the effective action need to be augmented by a lepton number violating 
reaction to drive the lepton number density $n_L$ towards its equilibrium value $n_L^{eq}$
and maintain it at this value until decoupling.  These non-gravitational $\D L \neq 0$
reactions are required to satisfy the first Sakharov condition and are provided naturally
here by the fundamental BSM theory.

Together, these effects give rise to the generalised Boltzmann equation (\ref{e24})
which implies the picture of the dynamical evolution of the lepton asymmetry summarised 
in Fig.~\ref{Fig 5.1}. An initially zero lepton-to-photon ratio $N_L$
is driven rapidly towards $N_L^{eq}$,
its approach moderated by the new factor $\mathcal{W}(z)$ in the Boltzmann equation.
$N_L$ then tracks $N_L^{eq}$ until the $\D L \neq 0$ reaction rate drops towards the Hubble
expansion rate and can no longer maintain $N_L$ in quasi-equilibrium. At this point,
$N_L$ decouples giving (up to the final dip in region (d) of Fig.~\ref{Fig 5.1})
the final gravitationally-induced lepton asymmetry.

As we have seen, the final value for the lepton-to-photon ratio $N_L$ depends very
sensitively on the temperature at which decoupling from equilibrium occurs.
This is because $N_L^{eq}(z)$ is falling very sharply with temperature, as $1/z^5$
for the radiation-dominated background. This behaviour is inherited from the 
original curvature dependence of $n_L^{eq}$ in (\ref{e14}), from which 
$n_L^{eq} \sim \r\, H \,T^2$. In turn, the point of decoupling is determined by $W(z)$,
so depends on the Hubble parameter (as exploited in the $w > 1/3$ backgrounds
described in section \ref{sect 6.2}) but also on the strength of the $\D L \neq 0$
reactions, which is dependent on the parameters of the BSM theory.

Turning now to the features specific to the see-saw BSM theory we have used here
to illustrate RIGL, note that it has three closely-linked properties relevant to the
analysis of leptogenesis. First, it exhibits the lepton number violating interactions
in Fig.~\ref{Fig 2.1}, which depend on the exchange of virtual $\n_R$ neutrino propagators
with $S^\times$ propagators. Reorienting these diagrams and replacing the Higgs fields
by their VEV gives the see-saw mass generation mechanism for the light neutrinos.
This explains the parameter constraint linking the magnitude of $W(z)$ to the observed
neutrino masses. Third, the same interaction vertex necessarily allows the 
$\n_R \rightarrow \n_L \, H$ and $\n_R \rightarrow \n_L^{\,c}\, H$ decays of real sterile
neutrinos. At one-loop level, these decays exhibit CP violation as described in 
Fig.~\ref{Fig 6D}. As they can occur when the $\n_R$ are out of equilibrium
around $T\sim M_1$ (see Fig.~\ref{Fig 6.2}), this realises the Fukugita-Yanagida leptogenesis
mechanism \cite{Fukugita:1986hr}.  
We noted that the dynamical weighting of the CP-violating combination of Yukawa
couplings at $O(\l^4)$ governing these decays is similar, but not identical, to the
combination arising in the gravitationally-induced lepton number density $n_L^{eq}$.
This difference allows us to choose parameters such that one or other mechanism
for leptogenesis dominates the final asymmetry. Both mechanisms are however necessarily
present in the see-saw model, or any related BSM theory with $L$-violation occurring 
through Feynman diagrams of the type in Fig.~\ref{Fig 2.1}.

It would be interesting to find alternative BSM theories (see \cite{Samanta:2020tcl} for recent work
in this direction)  in which this tight link between 
$\D L \neq 0$ reactions, out-of-equilibrium $\n_R$ decays, and neutrino masses
could be relaxed, giving a greater flexibility in the parameter choices controlling
the magnitude and evolution of the lepton asymmetry in RIGL.

Further exploration of the evolution of the lepton number density $n_L$ at very early
times with $T \gg M_1$ requires going beyond the `low-energy' approximation used in
the effective action. Recall that this is constructed only to leading order in derivatives,
and implies the restriction $T \sqrt{\RR}/M_1^2 < 1$ on its range of validity. To complete
the description of the evolution of $n_L$ beyond this region, we need more powerful
techniques to evaluate the one and two-loop self-energy diagrams in section \ref{sect 2}
beyond the small-momentum expansion. This involves working directly in the BSM model itself.

At one loop, the required methods have been developed in the series of papers
\cite{Hollowood:2007ku, Hollowood:2008kq, Hollowood:2009qz}
where the high-energy behaviour of photon vacuum polarisation diagrams in QED has 
been analysed. The motivation in these papers was to examine fundamental issues
involving causality, analyticity and unitarity in quantum field theories in curved spacetime.
However, the same methods, which exploit the Penrose limit of the background spacetime 
to allow the derivation of explicit non-perturbative formulae for the energy dependence
of the vacuum polarisation,
may readily be adapted to the one-loop self-energy diagrams here, using the generalisation
to fermions described in \cite{Hollowood:2009qz}. So in principle the evolution term 
$\mathcal{W}(z)$ can be found for arbitrarily high temperatures, provided only 
that $\RR/M_1^2 < 1$. 
The remaining technical difficulty is in extending these methods to the two-loop
diagrams necessary to find $n_L^{eq}$. Two loops seems to be necessary since, at least
in the BSM theory considered here, CP violation requires going to fourth order in the
Yukawa couplings. Clearly, the $\sim 1/z^5$ rise in $N_L^{eq}$ will not continue indefinitely
for very small $z$, and the evidence from QED vacuum polarisation strongly suggests
that all these gravitational loop effects revert to zero in the ultra-high energy limit.
This was essential in reconciling causality with the apparent superluminal propagation
in the low-energy effective action for QED in curved spacetime. It would be very interesting
either to overcome this difficulty (which seems unlikely given the complexity in evaluating
all the required two-loop diagrams even in the low-energy approximation) or to find
an alternative BSM theory in which $n_L^{eq}$ is generated at one loop so that the entire
dynamical evolution of $n_L(z)$ from the earliest times could be quantitatively traced.

In conclusion, in this paper we have extended the theory of radiatively-induced 
gravitational leptogenesis developed in 
\cite{McDonald:2015ooa, McDonald:2015iwt, McDonald:2016ehm, McDonald:2014yfg}
to give a fuller picture of the dynamical evolution of the lepton number asymmetry in the
early universe. 
Whether the final value of the lepton asymmetry leading to the observed baryon-to-photon
ratio $\eta$ is determined by the conventional out-of-equilibrium sterile neutrino
decay mechanism, or entirely through the generation of the equilibrium asymmetry
$n_L^{eq}$ by tidal curvature effects at the quantum loop level, we have seen how at 
earlier times the universe will experience a phase with a significantly higher 
lepton-antilepton asymmetry determined purely by gravitational effects, 
as illustrated in Fig.~\ref{Fig 5.1}. It remains an open and interesting question what
other physical consequnces may follow from the existence of such a non-vanishing
matter-antimatter asymmetry in these very early times in the evolution of the
universe.

\vskip0.5cm

\noindent {\bf Acknowledgements}
\vskip0cm

The work of GMS is supported in part by the STFC theoretical particle physics grant
ST/P00055X/1. 
JIM acknowledges support from the Alexander von Humboldt foundation.

\newpage

\appendix

\section{Analytic results for the Boltzmann equation}\label{Appendix A}

The Boltzmann equation (\ref{e24}) for gravitational leptogenesis admits simple analytic
solutions in the temperature regions (a), (b), (c) in Fig.~\ref{Fig 5.1}. This provides additional
insight into the numerical plots for the lepton number asymmetry $N_L(z)$ presented
in the text.

Keeping only the terms relevant to the gravitational mechanism, {\it i.e.}~neglecting the
contribution from $\n_R$ decays, the Boltzmann equation is
\begin{equation}
N_L'(z)\,=\, - W(z) \big(N_L(z) \,-\,N_L^{eq}(z)\big) \,-\,\mathcal{W}(z) N_L(z) \ .
\label{A1}
\end{equation}
In the region $z<1$ (and also $z<z_*$ in the cosmological scenario in section \ref{sect 6.2}), 
this is well-approximated by simple power-law behaviours of the coefficients and we can
write,
\begin{equation}
N_L'(z) \,=\, - \frac{\a}{z^p} \big(N_L(z) \,-\, \frac{\b}{z^q}\big) \,-\, \frac{\c}{z^r} N_L(z) \ .
\label{A2}
\end{equation}

Consider first the radiation-dominated era. Here, $p=2$, $q=5$, $r=5$ and the coefficients
$\a$, $\b$ and $\c$ can be read off from (\ref{f8}), (\ref{f10}) and (\ref{f12}).
It is straightforward to show that the general solution of (\ref{A2}) is 
\begin{equation}
N_L(z)\,=\, f(z) \,-\, f(z_0) \, e^{\left(\a/z + \c/4z^4 - \a/z_0 - \c/4z_0^4\right)}  \ ,
\label{A3}
\end{equation}
with
\begin{equation}
f(z)\,=\, \a\b\, e^{\left(\a/z + \c/4z^4\right)} \, 
\int_0^z dt\, \frac{1}{t^7} e^{-\left(\a/t +\c/4t^4\right)} \ .
\label{A4}
\end{equation}
Here, we have imposed the boundary condition $N_L(z_0)=0$ at some very early time
with temperature given by $z_0$.\footnote{It is clear that with this explicit solution,
$N_L(z)$ rises almost instantly towards its value in region (a) of Fig.~\ref{Fig 5.1},
driven by the large initial value of the $W N_L^{eq}$ term in the Boltzmann equation
for $N_L'(z)$. Note, however, that this is an artifact of using the effective Lagrangian 
for the Boltzmann coefficients at extremely early times outside its range of validity.
In reality, we expect from previous experience with QFT loop calculations in curved 
spacetime that for sufficiently high temperatures, where $T\sqrt{\RR}/M_1^2 \gg 1$,
the loop effects will revert smoothly to zero,
so the transition of $N_L(z)$ from its initial zero value at $z=z_0$ towards its value 
in region (a) is expected to be gradual.}

Now, in region (a) we have $W(z) \ll \mathcal{W}(z)$, so $N_L(z)$ is essentially the solution
of the simpler equation,
\begin{equation}
N_L'(z) \,=\, - \frac{\c}{z^5} N_L(z) \,+\, \frac{\a\b}{z^7} \ .
\label{A5}
\end{equation}
The solution (\ref{A3}) then holds with $\a\rta 0$ in the exponent, and $f(z)$ can be written
explicitly as
\begin{equation}
f(z) \,=\, \frac{\a\b}{\c} \left[ \frac{1}{z^2} \,+\, \frac{\sqrt{\pi}}{\sqrt{\c}} \,e^{\c/4z^4}
\, {\rm erfc}\left(\sqrt{\c}/2z^2\right) \right] \ .
\label{A6}
\end{equation}
For small $z$, {\it i.e.}~such that $\c/z^4 \gg 1$, we can approximate this using the 
asymptotic expansion 
${\rm erfc}(x) \simeq e^{(-x^2)}\left(1/{\sqrt{\pi} x}\right) \left(1 + O(1/x^2)\right)$
at large $x$, so we find
\begin{equation}
N_L(z) \,=\, \frac{\a\b}{\c} \, \frac{1}{z^2} \ , ~~~~~~~~~~~~~~~~~~~~ (z < \c^{1/4})\ .
\label{A7}
\end{equation}
This explains the slower rise of the lepton asymmetry in region (a) compared to the
sharp $1/z^5$ rise of the equilibrium value $N_L^{eq}$.

This matches smoothly to $N_L(z) \sim \b/z^5$ in region (b) where $N_L(z)$ follows
its equilibrium value. From (\ref{A7}), the crossover occurs at $z^3 \simeq \c/\a$, which 
of course matches the point where $W(z) \simeq \mathcal{W}(z)$.

To find the constant value of $N_L(z)$ in region (c) after decoupling, we instead need the 
solution of the Boltzmann equation where $W(z) \gg \mathcal{W}(z)$, {\it i.e.}
\begin{equation}
N_L'(z) \, =\, - \frac{\a}{z^2} N_L(z) \,+\, \frac{\a\b}{z^7} \ .
\label{A8}
\end{equation}
In this case, performing the integral in (\ref{A4}) gives,
\begin{equation}
f(z) \,=\, 5!\, \frac{\b}{\a^5} \, \sum_{n=0}^5 \frac{1}{n!} 
\left(\frac{\a}{z}\right)^n \ .
\label{A9}
\end{equation}
This matches the smooth evolution of $N_L(z)$ from region (b) through decoupling and into
region (c). The dominant term in region (c) is given by the $n=0$ term in the sum,
leaving \cite{McDonald:2016ehm}\footnote{For the `no hierarchy' choice of parameters used in
the text for the radiation-dominated era, we have $\a \simeq 9 \times 10^{-4}$,
$\,\,\b \simeq 4 \times 10^{-32}$ and $\,\c \simeq 5 \times 10^{-24}$.}
\begin{equation}
N_L(z)\, \simeq\, 120 \frac{\b}{\a^5} \ , ~~~~~~~~~~~~~~~~~~~~( z>\a)   \ .
\label{A10}
\end{equation}
 
The same methods can be applied to the pre-radiation cosmological scenario in 
section \ref{sect 6.2} using the expressions (\ref{f15}), (\ref{f17}) and (\ref{f16}) for
$W(z)$, $N_L^{eq}(z)$ and $\mathcal{W}(z)$ approximated for $z \ll z_*$.
In this case, the Boltzmann equation (\ref{A2}) has the power-law dependence
$p= \tfrac{1}{2}(5-3w)$, $\,q=\tfrac{1}{2}(7+9w)$ and $\,r=4 + 3w$.
The coefficients are related to the radiation-dominance ones above by
\begin{align}
\a_w \,&=\, \a\, z_*^{(1-3w)/2} \ , ~~~~~~~~~~~~~~~~~~~~~~~~
\b_w \,=\, \b\, \frac{(1+w)(1-3w)}{(1+w_\c)(1-3w_\c)}\, z_*^{3(3w-1)/2} \ ,
\nonumber \\
\c_w \,&=\, \c\,  \frac{(1+w)(1+3w)}{(1+w_\c)(1+3w_\c)}\, z_*^{3w-1} \ .
\label{A11}
\end{align}

Considering first region (a), we solve the equivalent Boltzmann equation to (\ref{A5}) and find
the obvious generalisation of (\ref{A3}) with 
\begin{equation}
f(z) \,=\, \frac{\a_w \b_w}{\c_w} \, e^{\c_w z^{1-r}/(r-1)}\, 
\left(\frac{r-1}{\c_w}\right)^{(p+q-r)/(r-1)} \, 
\Gamma\left(\frac{p+q-1}{r-1}\, , \,\frac{\c_w z^{1-r}}{r-1}\right) \ ,
\label{A12}
\end{equation}
in terms of the incomplete Gamma function. In the ultra-high temperature region
$\c_w/z^{r-1} \gg 1$, we use the large $x$ asymptotic expansion of the incomplete
Gamma function 
$\Gamma(a,x) \simeq e^{-x} x^{a-1}$ to find
\begin{align}
N_L(z) \,&\simeq\, \frac{\a_w \b_w}{\c_w} \, \frac{1}{z^{p+q-r}}  \ ,~~~~~~~~~~~~~~~~~~~~
(z \ll \c_w^{1/(r-1)}\,)  \nonumber \\
&=\, \frac{\a_w \b_w}{\c_w} \, \frac{1}{z^2} \ .
\label{A13}
\end{align}
Notice that this shows a universal $1/z^2$ behaviour independent of $w$,
as observed in Fig.~\ref{Fig 6.8}.

The behaviour of $N_L(z)$ for larger values of $z$ is not so straightforward for general $w$.
As evident from Fig.~\ref{Fig 6.8}, for values of $w \lesssim 0.5$ the asymmetry $N_L(z)$
is driven temporarily to its equilibrium value, region (b), due to the dominance of $W(z)$
over $\mathcal{W}(z)$ in this region. So as in the radiation-dominance case above, it is 
a good approximation to solve the Boltzmann equation (\ref{A2}) neglecting the evolution
term $\c_w N_L/z^r$. Using the small $x$ behaviour of the incomplete Gamma function
$\Gamma(a,x) \simeq \Gamma(a)$, we then find
\begin{equation}
N_L(z) \,\simeq\, \b_w \left(\frac{p-1}{\a_w}\right)^{q/(p-1)}\, 
\Gamma\left(\frac{p+q-1}{p-1}\right) \ , ~~~~~~~~~~~~~~~~~~ (z \gg \a_w^{1/(p-1)}\,) 
\label{A14}
\end{equation}
generalising (\ref{A10}).

However, for larger values of the equation of state parameter $w \gtrsim 0.7$,
$W(z)$ is too weak and the entire evolution of $N_L(z)$ is dominated by the
evolution term, with the solution given above in (\ref{A12}).  For these values of $w$,
which includes the kination limit $w=1$, $\,N_L(z)$ never holds to its equilibrium
value and its final constant value is given by the large $z$ expansion of (\ref{A12}), {\it viz.}
\begin{equation}
N_L(z)\,\simeq\, \frac{\a_w\b_w}{\c_w}\, \left(\frac{r-1}{\c_w}\right)^{(p+q-r)/(r-1)}
\, \Gamma\left(\frac{p+q-1}{(r-1)}\right) \ .
\label{A15}
\end{equation}

For intermediate values of $w$, we need to keep the full Boltzmann equation (\ref{A2})
balancing both the $W(z)$ and $\mathcal{W}(z)$ terms, with the numerical solutions
for the lepton number asymmetry shown in Fig.~\ref{Fig 6.8}.


\newpage



\begin{thebibliography}{99}


\bibitem{McDonald:2015ooa}
  J.~I.~McDonald and G.~M.~Shore,
 Phys.\ Lett.\ B {\bf 751} (2015) 469,
  [arXiv:1508.04119 [hep-ph]].

\bibitem{McDonald:2015iwt}
  J.~I.~McDonald and G.~M.~Shore,
 JHEP {\bf 1604} (2016) 030, 
  [arXiv:1512.02238 [hep-ph]].

\bibitem{McDonald:2016ehm}
  J.~I.~McDonald and G.~M.~Shore,
 Phys.\ Lett.\ B {\bf 766} (2017) 162,
  [arXiv:1604.08213 [hep-ph]].

\bibitem{Sakharov:1967dj}
 A.~Sakharov,
 Sov. Phys. Usp. \textbf{34} (1991) no.5, 392-393.

\bibitem{Davoudiasl:2004gf}
 H.~Davoudiasl, R.~Kitano, G.~D.~Kribs, H.~Murayama and P.~J.~Steinhardt,
 Phys. Rev. Lett. \textbf{93} (2004) 201301, 
 [arXiv:hep-ph/0403019 [hep-ph]].

\bibitem{Khlebnikov:1988sr}
 S.~Khlebnikov and M.~Shaposhnikov,
 Nucl. Phys. B \textbf{308} (1988) 885.

\bibitem{McDonald:2014yfg}
  J.~I.~McDonald and G.~M.~Shore,
 JHEP {\bf 1502} (2015) 076,
  [arXiv:1411.3669 [hep-th]].

\bibitem{Drummond:1979pp}
 I.~Drummond and S.~Hathrell,
 Phys. Rev. D \textbf{22} (1980) 343.

\bibitem{Ohkuwa:1980jx}
 Y.~Ohkuwa,
 Prog. Theor. Phys. \textbf{65} (1981) 1058.

\bibitem{Antunes:2019phe}
 V.~Antunes, I.~Bediaga and M.~Novello,
 JCAP \textbf{10} (2019) 076,
 [arXiv:1909.03034 [gr-qc]].

\bibitem{Buchmuller:2004nz}
 W.~Buchmuller, P.~Di Bari and M.~Plumacher,
 Annals Phys. \textbf{315} (2005) 305,
 [arXiv:hep-ph/0401240 [hep-ph]].

\bibitem{Fukugita:1986hr}
  M.~Fukugita and T.~Yanagida,
  Phys.\ Lett.\ B {\bf 174} (1986) 45.

\bibitem{Hobson:2006se}
 M.~Hobson, G.~Efstathiou and A.~Lasenby,
``\textit{General relativity: An introduction for physicists}'',
 Cambridge University Press, 2006.

\bibitem{Peskin:1995ev}
 M.~E.~Peskin and D.~V.~Schroeder,
``\textit{An Introduction to quantum field theory},''
 CRC Press, 1995.

\bibitem{Shore:2002gn}
 G.~M.~Shore,
 Nucl. Phys. B \textbf{633} (2002) 271,
 [arXiv:gr-qc/0203034 [gr-qc]].

\bibitem{Hollowood:2007ku}
 T.~J.~Hollowood and G.~M.~Shore,
 Nucl. Phys. B \textbf{795} (2008) 138,
 [arXiv:0707.2303 [hep-th]].

\bibitem{Hollowood:2008kq}
 T.~J.~Hollowood and G.~M.~Shore,
 JHEP \textbf{12} (2008) 091,
 [arXiv:0806.1019 [hep-th]].

\bibitem{Hollowood:2009qz}
 T.~J.~Hollowood, G.~M.~Shore and R.~J.~Stanley,
 JHEP \textbf{08} (2009) 089,
 [arXiv:0905.0771 [hep-th]].

\bibitem{Hollowood:2011yh}
 T.~J.~Hollowood and G.~M.~Shore,
 JHEP \textbf{02} (2012) 120, 
 [arXiv:1111.3174 [hep-th]].

\bibitem{Hollowood:2015elj}
 T.~J.~Hollowood and G.~M.~Shore,
 JHEP \textbf{03} (2016) 129,
 [arXiv:1512.04952 [hep-th]].

\bibitem{deRham:2019ctd}
 C.~de Rham and A.~J.~Tolley,
 Phys. Rev. D \textbf{101} (2020) 063518,
 [arXiv:1909.00881 [hep-th]].

\bibitem{Kostelecky:2000mm}
 V.~Kostelecky and R.~Lehnert,
 Phys. Rev. D \textbf{63} (2001) 065008,
 [arXiv:hep-th/0012060 [hep-th]].

\bibitem{Colladay:1998fq}
 D.~Colladay and V.~Kostelecky,
 Phys. Rev. D \textbf{58} (1998) 116002,
 [arXiv:hep-ph/9809521 [hep-ph]].

\bibitem{Audretsch:1981wf}
 J.~Audretsch,
 J. Phys. A \textbf{14} (1981) 411.

\bibitem{Dodelson:2003ft}
 S.~Dodelson,
``\textit{Modern Cosmology},''
 Academic Press, 2003.

\bibitem{Baumann}
D.~Baumann,
``\textit{Cosmology}'', Cambridge University Lecture Notes, 2014.

\bibitem{Klinkhamer:1984di}
 F.~R.~Klinkhamer and N.~Manton,
 Phys. Rev. D \textbf{30} (1984) 2212.

\bibitem{Buchmuller:2002zs}
 W.~Buchmuller,
``\textit{Baryo- and leptogenesis (brief summary)},''
 ICTP Lect. Notes Ser. \textbf{14} (2003) 41.

\bibitem{Buchmuller:2003gz}
 W.~Buchmuller, P.~Di Bari and M.~Plumacher,
 Nucl. Phys. B \textbf{665} (2003) 445,
 [arXiv:hep-ph/0302092 [hep-ph]].

\bibitem{Turner:1983he}
 M.~S.~Turner,
 Phys. Rev. D \textbf{28} (1983) 1243.

\bibitem{Ford:1986sy}
 L.~Ford,
 Phys. Rev. D \textbf{35} (1987) 2955.

\bibitem{Peebles:1998qn}
 P.~Peebles and A.~Vilenkin,
 Phys. Rev. D \textbf{59} (1999) 063505,
 [arXiv:astro-ph/9810509 [astro-ph]].

\bibitem{Chung:1998rq}
 D.~J.~H.~Chung, E.~W.~Kolb and A.~Riotto,
 Phys. Rev. D \textbf{60} (1999) 063504,
 [arXiv:hep-ph/9809453 [hep-ph]].

\bibitem{Giudice:2000ex}
 G.~F.~Giudice, E.~W.~Kolb and A.~Riotto,
 Phys. Rev. D \textbf{64} (2001) 023508,
 [arXiv:hep-ph/0005123 [hep-ph]].

\bibitem{Samanta:2020tcl}
 R.~Samanta and S.~Datta,
 ``{\it Flavour effects in gravitational leptogenesis},''
 [arXiv:2007.11725 [hep-ph]].



\end{thebibliography}
\end{document}